\documentclass[aps,twocolumn,prb,amsmath,amssymb,notitlepage,superscriptaddress]{revtex4-1}   	
\usepackage{graphicx}				
\usepackage{bm}
\usepackage{array,makecell}
\usepackage{multirow}
\usepackage{color}

\allowdisplaybreaks

\begin{document}

\title{Spin transfer torques and spin-dependent transport in a metallic F/AF/N tunneling junction}

\date{\today }

\author{Kei Yamamoto}
\email{kyamamoto1@ua.edu}
\affiliation{Department of Physics and Astronomy, The University of Alabama, Alabama 35487, USA \\
and Center for Materials for Information Technology (MINT), The University of Alabama, Alabama 35401, USA}
\affiliation{Institut f\"ur Physik, Johannes Gutenberg Universit\"at Mainz, 55128 Mainz, Germany}
\affiliation{Advanced Science Research Center, Japan Atomic Energy Agency, Tokai 319-1195, Japan}
\author{Olena Gomonay}
\affiliation{Institut f\"ur Physik, Johannes Gutenberg Universit\"at Mainz, 55128 Mainz, Germany}
\author{Jairo Sinova}
\affiliation{Institut f\"ur Physik, Johannes Gutenberg Universit\"at Mainz, 55128 Mainz, Germany}
\affiliation{Institute of Physics ASCR, v.v.i. Cukrovarnicka, 10, 16253 Praha 6, Czech Republic}
\author{Georg Schwiete}
\affiliation{Department of Physics and Astronomy, The University of Alabama, Alabama 35487, USA \\
and Center for Materials for Information Technology (MINT), The University of Alabama, Alabama 35401, USA}

\begin{abstract}
We study spin-dependent electron transport through a ferromagnetic-antiferromagnetic-normal metal tunneling junction subject to a voltage or temperature bias, in the absence of spin-orbit coupling. We derive microscopic formulas for various types of spin torque acting on the antiferromagnet as well as for charge and spin currents flowing through the junction. The obtained results are applicable in the limit of slow magnetization dynamics. We identify a parameter regime in which an unconventional damping-like torque can become comparable in magnitude to the equivalent of the conventional Slonczewski's torque generalized to antiferromagnets. Moreover, we show that the antiferromagnetic sublattice structure opens up a channel of electron transport which does not have a ferromagnetic analogue and that this mechanism leads to a pronounced field-like torque. Both charge conductance and spin current transmission through the junction depend on the relative orientation of the ferromagnetic and the antiferromagnetic vectors (order parameters). The obtained formulas for charge and spin currents allow us to identify the microscopic mechanisms responsible for this angular dependence and to assess the efficiency of an antiferromagnetic metal acting as a spin current polarizer. \end{abstract}

\maketitle

\section{Introduction}\label{sec:introduction}

The last few years have witnessed a growing interest in the use of antiferromagnets as active elements in spintronic devices.\cite{Gomonay2014,Jungwirth2016,Jungwirth2018} Antiferromagnets are an attractive platform for novel magnetic recording devices due to their large typical resonance frequency in the THz regime, robustness against magnetic perturbations and the absence of stray magnetic fields. Recent experiments have further revealed that spin transport is strongly affected by antiferromagnetic order. Specifically, precision measurements of magnetoresistance, spin current absorption, and its transmission have proven to be powerful tools for studying antiferromagnetic order in thin film multilayer structures. \cite{Mewes2010,Park2011b,Merodio2014,Merodio2014b,Kriegner2016,Kravets2016,Wang2014d,Wang2015i,Moriyama2015,Prakash2016,Qiu2016a,Wang2017a} Manipulation and switching of the antiferromagnetic order parameter, the N\'eel vector, are possible via current-induced spin torques. In particular, the effectiveness of relativistic (N\'eel) spin-orbit torques, first proposed theoretically in Ref.~\onlinecite{Zelezny2014}, has been demonstrated in several recent experimental works.
\cite{Wadley2016,Roy2016,Grzybowski2017,Bodnar2018} The relativistic N\'eel spin-orbit torque, however, requires significant spin-orbit coupling and a rather particular crystalline structure. It is therefore of interest to understand generic properties of antiferromagnetic metals that persist even in the non-relativistic limit, i.e., for negligible spin-orbit coupling. This is the main aim of this paper, where a minimal microscopic model will be employed to analytically study dynamics and transport in antiferromagnetic nanostructures. Our study complements existing theoretical approaches,\cite{Nunez2006,Haney2008a,Xu2008a,Hals2011,Gomonay2012,Cheng2014c,Saidaoui2014,Zelezny2014,Yamane2016a,Saidaoui2016a,Manchon2016b} which have been predominantly phenomenological\cite{Nunez2006,Haney2008a,Hals2011,Gomonay2012,Cheng2014c,Yamane2016a} or relied on extensive numerical computations.\cite{Xu2008a,Zelezny2014,Saidaoui2014,Saidaoui2016a} Our approach is in a similar spirit to the work of Stiles and Zangwill\cite{Stiles2002} on ferromagnetic spin transfer torques and aims at revealing the anatomy of antiferromagnetic spin-transfer torques.

\begin{figure}[t]
\begin{center}
\includegraphics[width=0.9\linewidth]{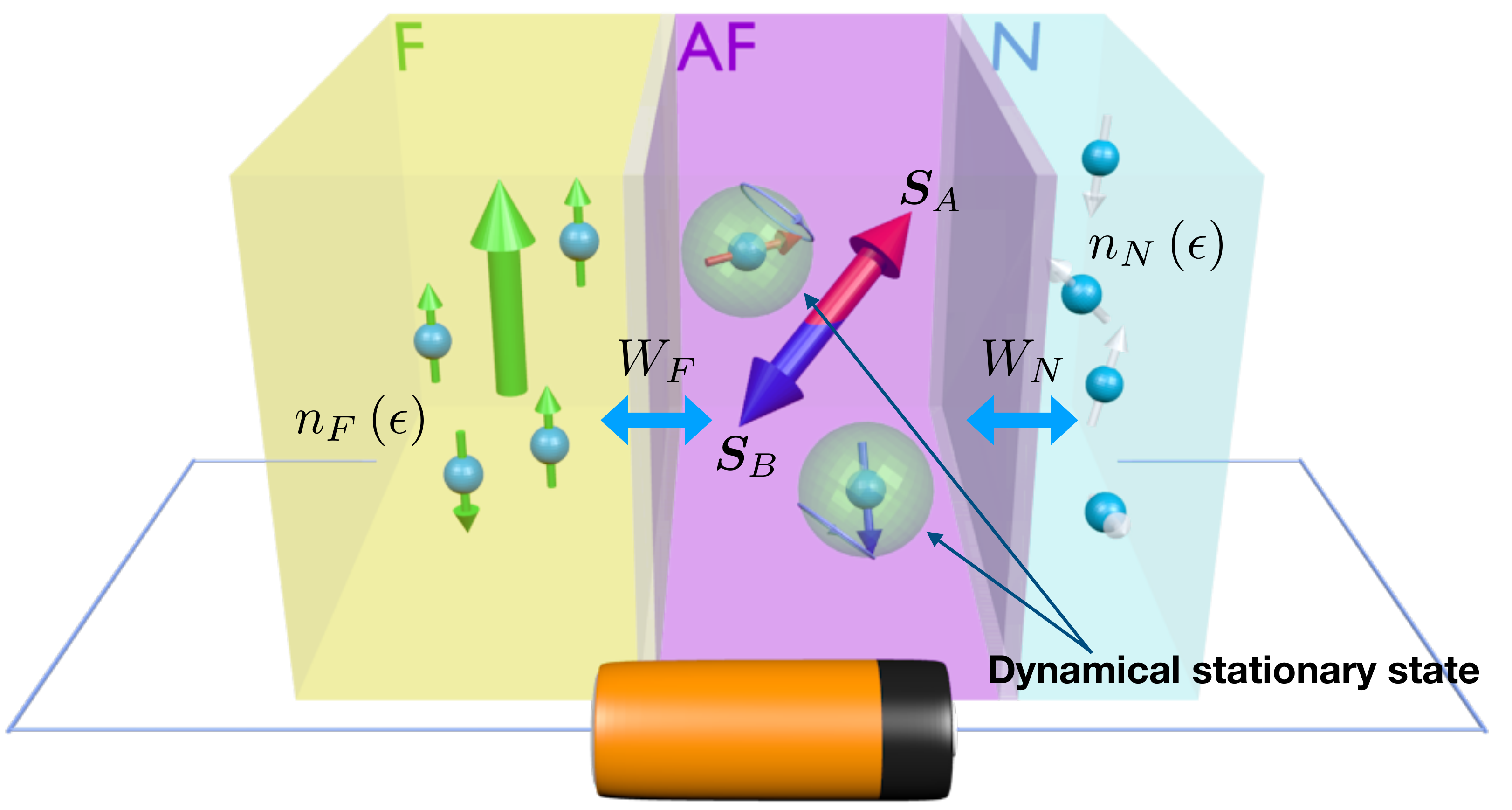}
\caption{Schematic description of the theoretical model considered in the present work. The ferromagnetic (F) and normal metal (N) leads are assumed to have fixed electron distribution functions $n_{F,N} \left( \epsilon \right)$, respectively. A finite difference $n_F -n_N $ drives the antiferromagnet (AF) out of equilibrium and it eventually settles down to a dynamical stationary state. The layers are separated by barriers with tunneling amplitudes $W_{F,N}$.}
\label{fig:setup}
\end{center}
\end{figure}
We consider an antiferromagnetic metal (AF), tunneling coupled to two leads (Fig.~\ref{fig:setup}). The leads are made of ferromagnetic metal (F), and normal metal (N), respectively. We aim at exploring basic transport characteristics of F/AF/N junctions that do not require specific material properties. Spin-orbit coupling is therefore neglected. The tunneling barriers are chosen as spin-conserving and the antiferromagnetic order is established on a bipartite lattice. When a bias voltage or temperature difference is applied between the leads, charge and spin currents flow and generate spin transfer torques acting on the localized spins $\bm{S}_{A,B}$ at the two sublattices in the AF. The main role of the ferromagnetic lead in this setup is to generate a finite spin-polarization. The normal metal lead, in turn, provides an explicit channel of relaxation for the electrons on the AF, a crucial ingredient for a number of phenomena discussed below. 

We derive microscopic formulas for the four symmetry-allowed spin transfer torques and identify physical mechanisms responsible for each of them. We find that in addition to Slonczewski's damping-like torque, familiar from ferromagnetic multilayers, two unconventional types of torque can become relevant in certain parameter regimes. We also obtain analytical expressions for the dependence of the charge and spin currents on the angle between the ferromagnetic and antiferromagnetic order parameters. The spin current transmission depends strongly on this angle, indicating that spin-valve applications are plausible for antiferromagnetic metals. The predictions on the torques and currents may be experimentally tested against each other since they are all given in terms of a common set of control parameters.

Before summarizing our findings in more detail, it is instructive to first review the phenomenology of magnetization dynamics\cite{Gomonay2014} in bipartite antiferromagnets. This phenomenology is based on the observation that the strong exchange interaction responsible for the antiferromagnetic order between $\bm{S}_A $ and $\bm{S}_B$ implies that the normalized total spin angular momentum $\bm{m} =(\bm{S}_A + \bm{S}_B )/2S$ is much smaller than the normalized N\'eel order parameter $\bm{n} = (\bm{S}_A -\bm{S}_B )/2S $ where $S =\left| \bm{S}_A \right| = \left| \bm{S}_B \right| $. Ignoring quantities of order $\left| \bm{m} \right| / \left| \bm{n} \right| \ll 1$, one obtains $\left| \bm{n} \right| \approx 1 $ and $\bm{m}\cdot \bm{n} = 0$. Then the symmetry-permitted terms of spin torque at the leading order in $ \bm{m} $ in the dynamical equations of $\bm{m}$ and $\bm{n}$ are given by
\begin{eqnarray}
\frac{d\bm{m}}{dt} \bigg| _{\rm current-induced} & \approx & \frac{-\bm{n}}{2S} \times \left( \Gamma ^m_{\rm fl} \hat{\bm{z}}_F +  \bm{n}  \times \Gamma ^m_{\rm dl} \hat{\bm{z}}_F  \right) , \label{eq:LLGM} \\
\frac{d\bm{n}}{dt} \bigg| _{\rm current-induced} & \approx &  \frac{-\bm{n}}{2S}  \times \left( \Gamma ^n_{\rm fl} \hat{\bm{z}}_F + \bm{n} \times \Gamma ^n_{\rm dl} \hat{\bm{z}}_F \right) , \label{eq:LLGN}
\end{eqnarray}
where $\hat{\bm{z}}_F $ is a unit vector along the polarization vector of the injected spin current. The subscripts ${\rm fl}$ and ${\rm dl}$ stand for field-like and anti-damping-like respectively. $\Gamma ^m_{\rm dl}$ represents the conventional Slonczewski's spin transfer torque and $\Gamma ^n_{\rm fl}$ acts like an external magnetic field. The so-called N\'eel spin-orbit torque, which can be effective in systems with spin-orbit coupling and is therefore outside the scope of this paper, would enter the equations as a contribution to $\Gamma ^m_{\rm fl}$. The remaining $\Gamma ^n_{\rm dl}$ has not been discussed very much so far in the literature. All four torques are allowed by symmetries and could therefore be introduced on purely phenomenological grounds. In this work, we will go one step further and discuss their relative strengths and microscopic origin for the case of the F/AF/N junction. In general, the dynamical equations for $\bm{m}$ and $\bm{n}$ do not only include torques, but additional terms accounting for damping and noise. While these can be discussed within the formalism described below, they are beyond the scope of the present work.

It is also instructive to interpret the various types of torque in the two-ferromagnet picture, in which the overlap between electronic orbitals located at $A$ and $B$ sublattice sites is negligibly small. In this limit, one may consider AF as a pair of oppositely oriented ferromagents as far as electron transport is concerned. For a ferromagnet, it is well known that quantum mechanical dephasing \cite{Stiles2002} leads to a strong Slonczewski's damping-like torque \cite{Slonczewski1996,Slonczewski2002} and a negligibly small field-like torque. Assuming that the spin torque acting on each individual "ferromagnetic" moment $\bm{S}_{A,B}$ is of the Slonczewski type then corresponds to $\Gamma ^m_{\rm dl} \approx J_F /2 , \Gamma ^m_{\rm fl} \approx \Gamma ^n_{\rm fl} \approx \Gamma ^n_{\rm dl} \approx 0$ where $J_F$ is the total spin current flowing into AF. We note that these estimates are based on a model in which the injected spin current does not appreciably change the state of AF. It therefore implicitly assumes the presence of {\it a relaxation mechanism} faster than the rate of tunneling at the F/AF interface so that it quickly wipes out any influence of the injected current. In this sense, the above estimates apply to a regime of weak F/AF coupling. 

Our calculations generalize the two-ferromagnetic result by including the intersublattice overlap along with N as the source of the relaxation that dissipates the injected current. We show that the intersublattice overlap opens up an additional channel of electron transport in which the dephasing can be avoided and the transverse spin is conserved. This results in a novel contribution to the field-like torque $\Gamma ^n_{\rm fl}$ proportional to the square of the overlap amplitude. The inclusion of N turns out to be crucial here as $\Gamma ^n_{\rm fl}$ is inversely proportional to the relaxation rate. The finite relaxation rate also allows us to explore the regime of strong tunneling at F/AF interface. We find that the antiferromagnetic state modified by the ferromagnetic current generates nonvanishing $\Gamma ^m_{\rm fl}$ and $\Gamma ^n_{\rm dl}$, of which the latter may reach a magnitude comparable to that of $\Gamma ^m_{\rm dl}$. 

Similarly based on the two-ferromagnet picture, each sublattice contributes to the conductance a ferromagnetic angular dependent term, \cite{Landauer1970,Inoue1996} proportional to $\cos \theta _{A,B} = \hat{\bm{z}}_F \cdot \bm{S}_{A,B} /S $ respectively. This angular dependence cancels in the total conductance, however, because of the antiferromagentic order $\bm{S}_B \approx -\bm{S}_A \Rightarrow \cos \theta _B  \approx -\cos \theta _A  $. Phenomenologically this is a consequence of the symmetry between the $A$ and $B$ sublattices and the {\it charge current} $I$ is predicted to be a function of $\cos ^2 \theta $ in the leading order cylindrical harmonics expansion. Still staying within the picture of two superimposed ferromagnets, the spin current flowing at AF/N ($J_N $) is expected to be given approximately by $ J_F \hat{\bm{z}}_F \cdot \bm{n} $ since Slonczewski's spin torque arises from absorption of the transverse components of electron spin by the localized moments via dephasing \cite{Stiles2002} and effectively projects out the spin current polarization onto the N\'eel vector. In contrast, the longitudinal spin is conserved in the absence of spin-orbit interaction. In short, an antiferromagnetic metal should act as an ideal spin-valve according to the two-ferromagnetic picture. We compute the charge and spin currents at both F/AF and AF/N interfaces and explicitly confirm the $\cos ^2 \theta $ dependence of the charge current and the spin-valve-like behaviour of the spin current. Again it is essential to include the two leads since the $\theta $ dependent charge current is second order in F/AF tunneling while the spin transmission requires spin currents at the two interfaces.

The rest of the paper is organized as follows. In Sec.~\ref{sec:model}, after introducing the model Hamiltonian, we give an overview of the physical properties of F/AF/N junction. We explain meanings of all the parameters appearing in the final results of torques and currents. Sections~\ref{sec:torque} and \ref{sec:current} contain the main results for the spin torques and the charge and spin currents in the stationary state. We conclude with discussions of our results and their connection to the previous studies in view of experiments and applications in Sec.~\ref{sec:discussion}. In Appendix~\ref{sec:formalism} and \ref{sec:trace}, we describe our theoretical approach and show details of the calculations. In order to facilitate the comparison with previous studies, Appendix~\ref{sec:scattering} develops a scattering theory approach and Appendix~\ref{sec:AFMlead} discusses F/F/N and AF/AF/N junctions within our framework.

\section{The model}\label{sec:model}

\begin{figure}[t]
\begin{center}
\includegraphics[width=0.95\linewidth]{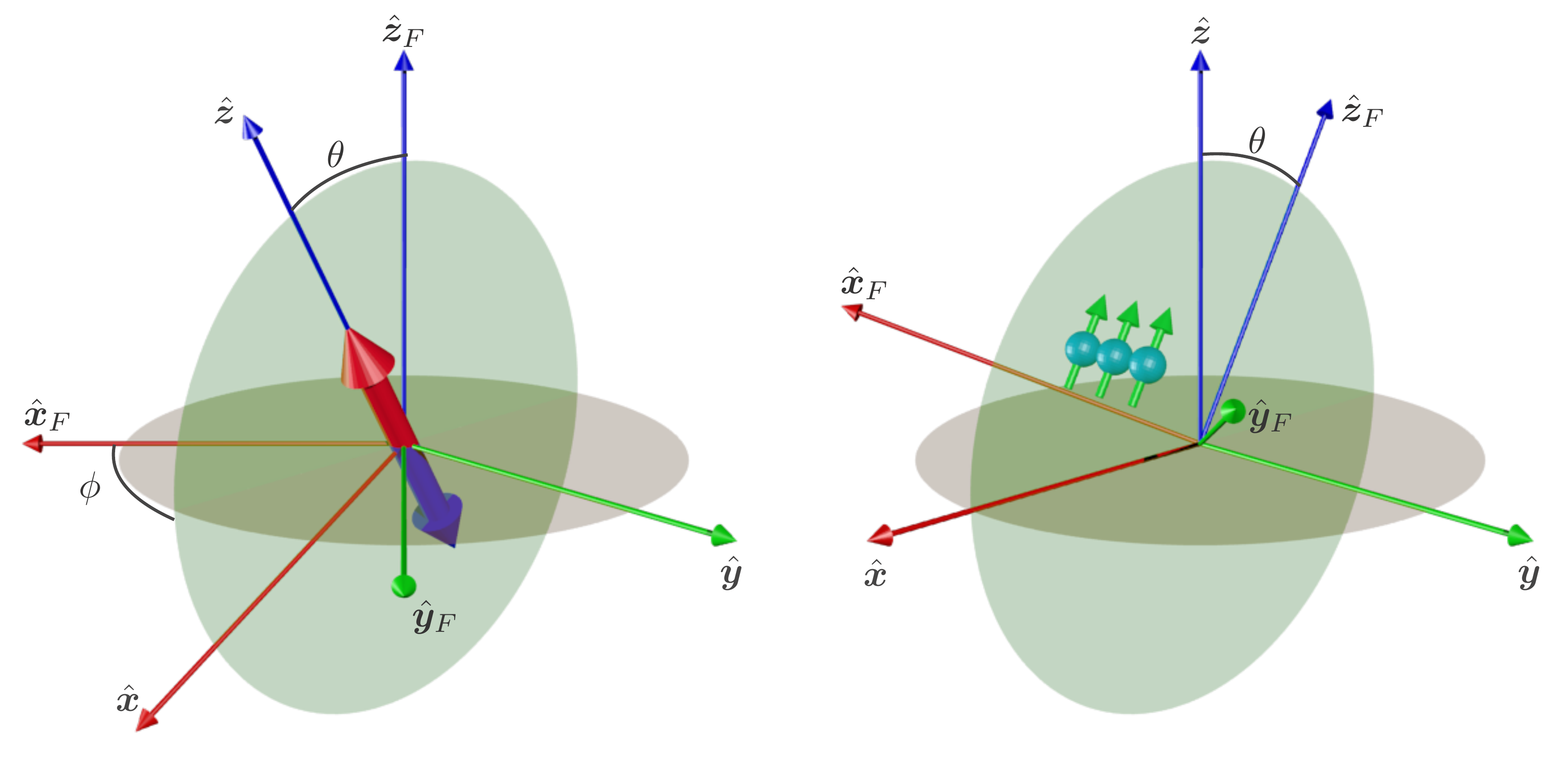}
\caption{The choice of coordinate axes and the definitions of the azimuthal and polar angles $\theta ,\phi $ seen in the ferromagnetic frame spanned by $\hat{\bm{x}}_F ,\hat{\bm{y}}_F , \hat{\bm{z}}_F $ (Left) and in the antiferromagnetic frame spanned by $\hat{\bm{x}} , \hat{\bm{y}} ,\hat{\bm{z}}$ (Right). The N\'eel vector in the ferromagnetic frame is parameterized by the usual spherical coordinates. The incoming spin current, or equivalently the ferromagnetic moment, in the antiferromagnetic frame has $x$ and $z$ components only; $\hat{\bm{z}}_F = \hat{\bm{z}}\cos \theta - \hat{\bm{x}} \sin \theta $.}
\label{fig:axes}
\end{center}
\end{figure}

We consider a system depicted in Fig. \ref{fig:setup} where an antiferromagnetic metal (AF) is connected to a ferromagnetic left lead (F) and a normal metallic right lead (N) by respective tunneling barriers. The model Hamiltonian consists of five distinct parts;
\begin{equation}
H = \sum _{k,\sigma }\epsilon ^F_{k\sigma }c_{k\sigma }^{F \dagger }c^F_{k\sigma } + \sum _{m,\sigma }\epsilon ^N_m c^{N\dagger }_{m\sigma }c^N_{m\sigma } +H_0+ H_F + H_N . \label{eq:total_Hamiltonian}
\end{equation}
The first two terms describe conduction electrons in F and N. Only the electrons in F have spin-dependent energy eigenvalues. Labels $k$ and $m$ are used exclusively for the orbital degrees of freedom of ferromagnetic $c^F$ and normal metallic $c^N$ electrons. $H_0$ represents AF and is given by
\begin{eqnarray}
H_0 &=& \sum _{l,\sigma }\begin{pmatrix}
	a^{\dagger }_{l\sigma } & b_{l\sigma }^{\dagger } \\
	\end{pmatrix} \begin{pmatrix}
	\epsilon _l & \overline{t_l} \\
	t_l & \epsilon _l \\
	\end{pmatrix}_{SL} \begin{pmatrix}
	a_{l\sigma } \\
	b_{l\sigma } \\
	\end{pmatrix} \nonumber \\
&& - \frac{\Delta _{\rm ex}}{S} \sum _{l,\sigma \sigma ^{\prime }} \bm{\sigma }^{\sigma \sigma ^{\prime }}\cdot \left( \bm{S}_A a_{l\sigma }^{\dagger } a_{l\sigma ^{\prime }} + \bm{S}_B b_{l\sigma }^{\dagger } b_{l\sigma ^{\prime }} \right) . \label{eq:AFM_Hamiltonian}
\end{eqnarray}
Here, $a_{l\sigma }$ and $b_{l\sigma }$ annihilate the energy eigenstates with the eigenvalues $\epsilon _l$ residing in the $A$ and $B$ sublattice respectively, in the absence of intersublattice overlap of the atomic orbitals and also of the exchange interaction with the localized spins $\bm{S}_{A,B}$. The overlap amplitudes $t_l $ are assumed to be diagonal in this basis mainly for the ease of implementing the sublattice symmetry ($\overline{t_l}$ denotes the complex conjugate of $t_l$). We note that this form is generic in the case where $l$ denotes the crystalline momentum. \cite{Yamane2016a} Specifically for a checker board structure, $t_l$ and $\epsilon _l$ correspond to the Fourier transforms of the nearest-neighbor and next-nearest-neighbor hopping amplitudes respectively. $2\Delta _{\rm ex}$ is the exchange split of the antiferromagnetic electrons. $\bm{\sigma } =\left( \sigma _1 ,\sigma _2 ,\sigma _3 \right) $ are the Pauli matrices in spin space and we set $\hbar =1$. The Hilbert space for a given $l$ is four dimensional; two for the spin and two for the sublattice. Two by two matrices in the spin space and the sublattice space are distinguished by subscripts $SP$ and $SL$ except for the Pauli matrices for which we use $\tau _{1,2,3}$ in the sublattice space. $H_F $ and $H_N $ represent spin-conserving tunneling processes to and from F and N respectively;
\begin{eqnarray}
H_F &=& \sum _{kl,\sigma \sigma ^{\prime }} \left[   c_{k\sigma }^{F\dagger } \left( W_F \right) ^{\sigma \sigma ^{\prime }}_{kl} \begin{pmatrix}
	a_{l\sigma ^{\prime }} \\
	b_{l\sigma ^{\prime }} \\
	\end{pmatrix}  + {\rm h.c.} \right] , \label{eq:Hamiltonianc} \\
H_N &=& \sum _{lm,\sigma }\left[  c_{m\sigma }^{N\dagger }  \left( W_N \right) _{ml} \begin{pmatrix}
	a_{l\sigma } \\
	b_{l\sigma } \\
	\end{pmatrix} + {\rm h.c.} \right] . \label{eq:Hamiltoniand}
\end{eqnarray} 
The tunneling matrices $W_{F,N}$ are vectors in the sublattice space, defined by
\begin{eqnarray}
\left( W_F \right) _{kl}^{\sigma \sigma ^{\prime }} &=&R^{\sigma \sigma ^{\prime }}_{SP}  \begin{pmatrix}
	\left( W^A_F \right) _{kl} & \left( W^B_F \right) _{kl} \\
	\end{pmatrix} ,  \\
\left( W_N \right) _{ml} &=& \begin{pmatrix}
	\left( W^A_N \right) _{ml} & \left( W^B_N \right) _{ml} \\
	\end{pmatrix} .
\end{eqnarray}
The spin rotation matrix $R_{SP}$ reads
\begin{equation}
R_{SP}  = \begin{pmatrix}
	e^{-i\phi /2}\cos \left( \theta /2 \right) & -e^{-i\phi /2}\sin \left( \theta /2\right) \\
	e^{i\phi /2} \sin \left( \theta /2\right) & e^{i\phi /2} \cos \left( \theta /2\right) \\
	\end{pmatrix} ,
\end{equation}
encoding the difference in the reference frame between the ferromagnet and the antiferromagnet, as indicated in Fig.~\ref{fig:axes}.
They have been chosen such that $\bm{n} =  \left( \sin \theta \cos \phi ,\sin \theta \sin \phi ,\cos \theta \right) $ in the ferromagnetic frame. 

We remark on the dynamics of $\bm{S}_{A,B}$. The Hamiltonian should be augmented by terms which do not involve the electrons, e.g. the antiferromagnetic exchange and crystalline anisotropy. However, the details of these terms will not affect our computation as long as the resulting magnetization dynamics is sufficiently slow. This condition will be quantified in the followings. We also assume throughout that any deviation from the collinearity relation $\bm{S}_B = -\bm{S}_A $ is due to time-dependent dynamical part of $\bm{S}_{A,B}$. This allows us to specify a common spin quantization axis for the two sublattices. In reality, there can be some effective fields (torques to be derived) generated through the tunneling to F that will induce a nonzero total spin $\bm{m}$ in equilibrium or stationary state. It implies that there will be an additional requirement that the antiferromagnetic exchange be much stronger than the spin torques, which is expected to be safely satisfied in most circumstances of interest.

\subsection{Band structure}

\begin{figure}[t]
\begin{center}
\includegraphics[width=1\linewidth]{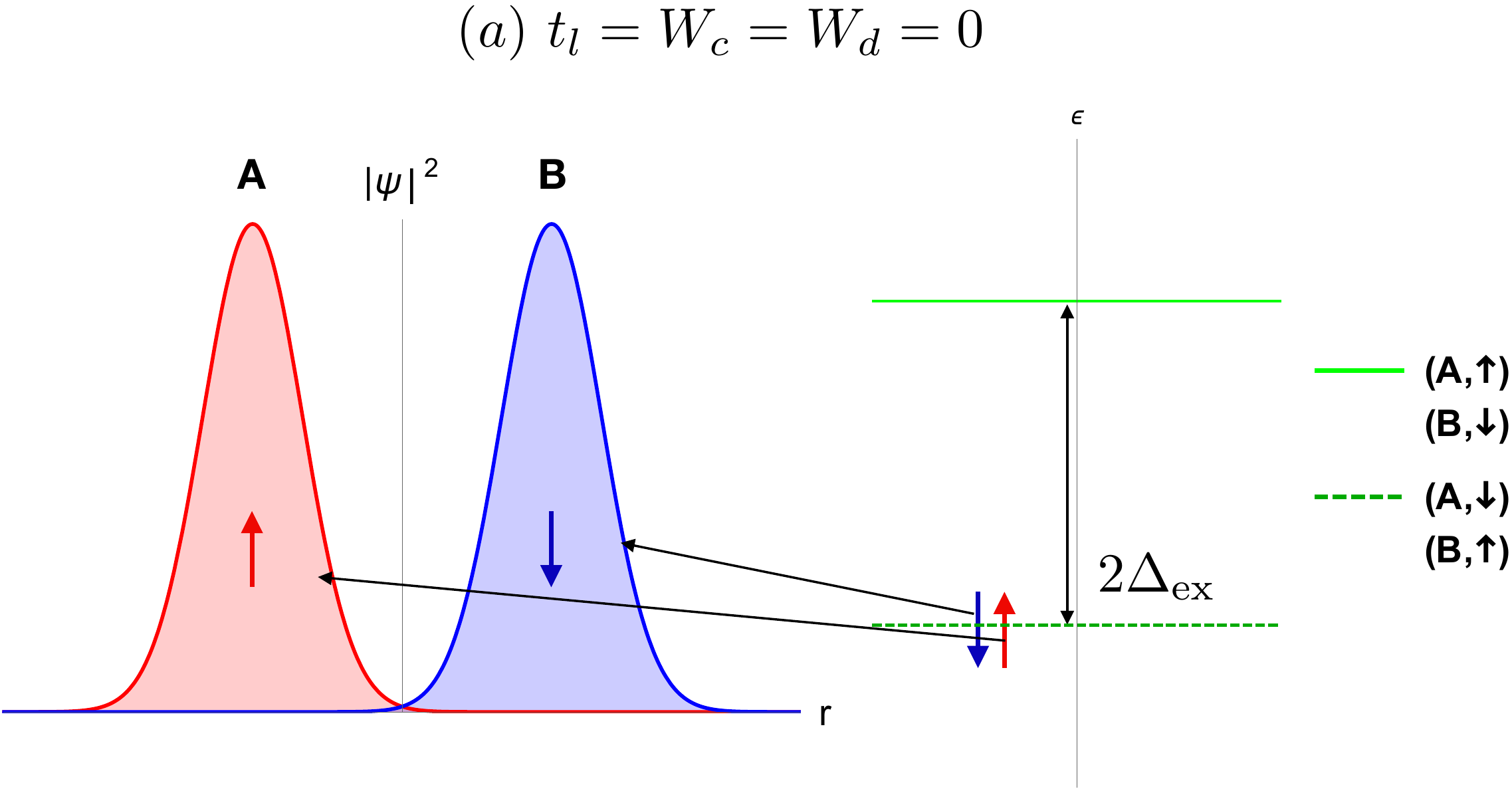}
\includegraphics[width=1\linewidth]{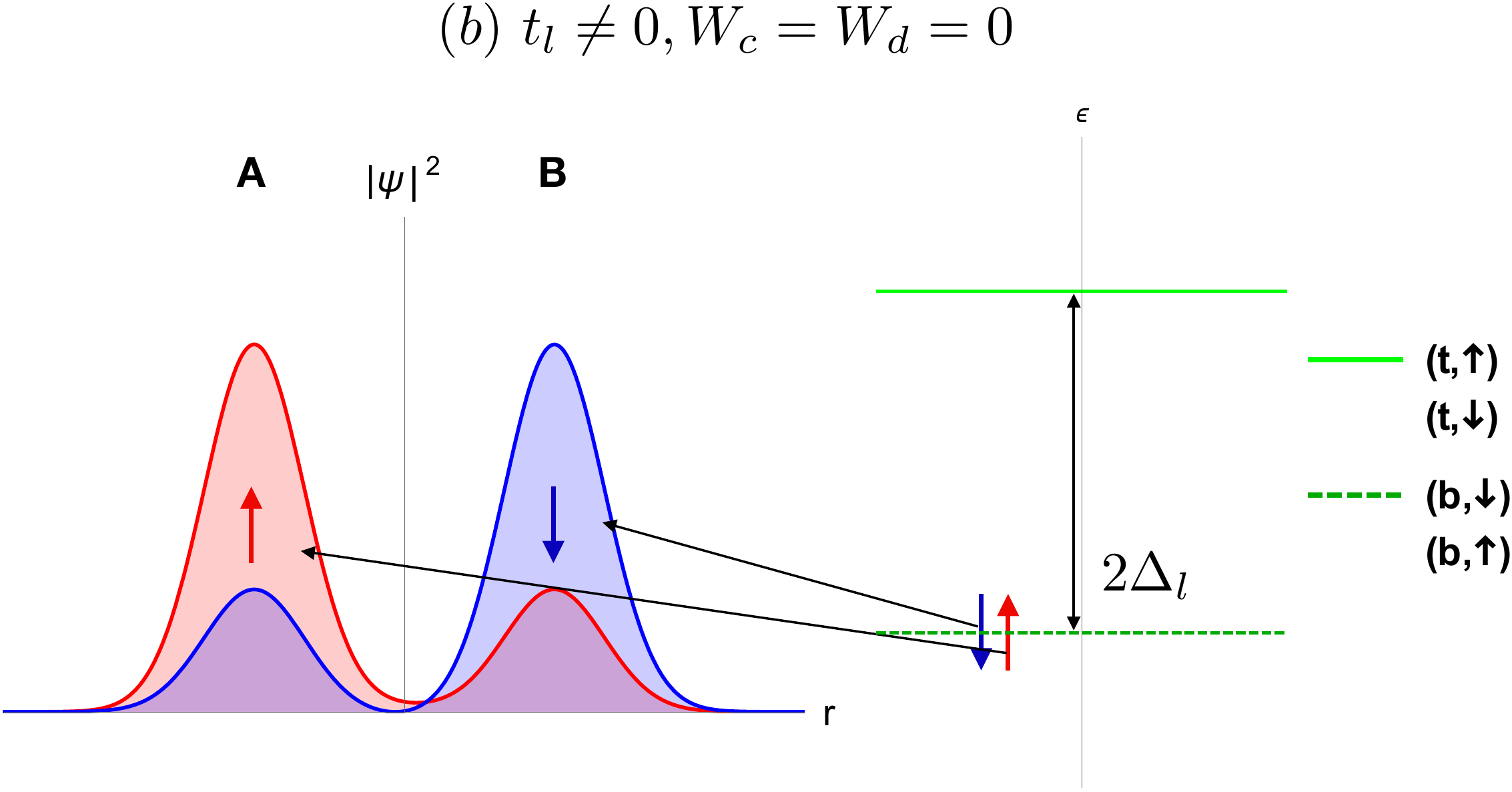}
\includegraphics[width=1\linewidth]{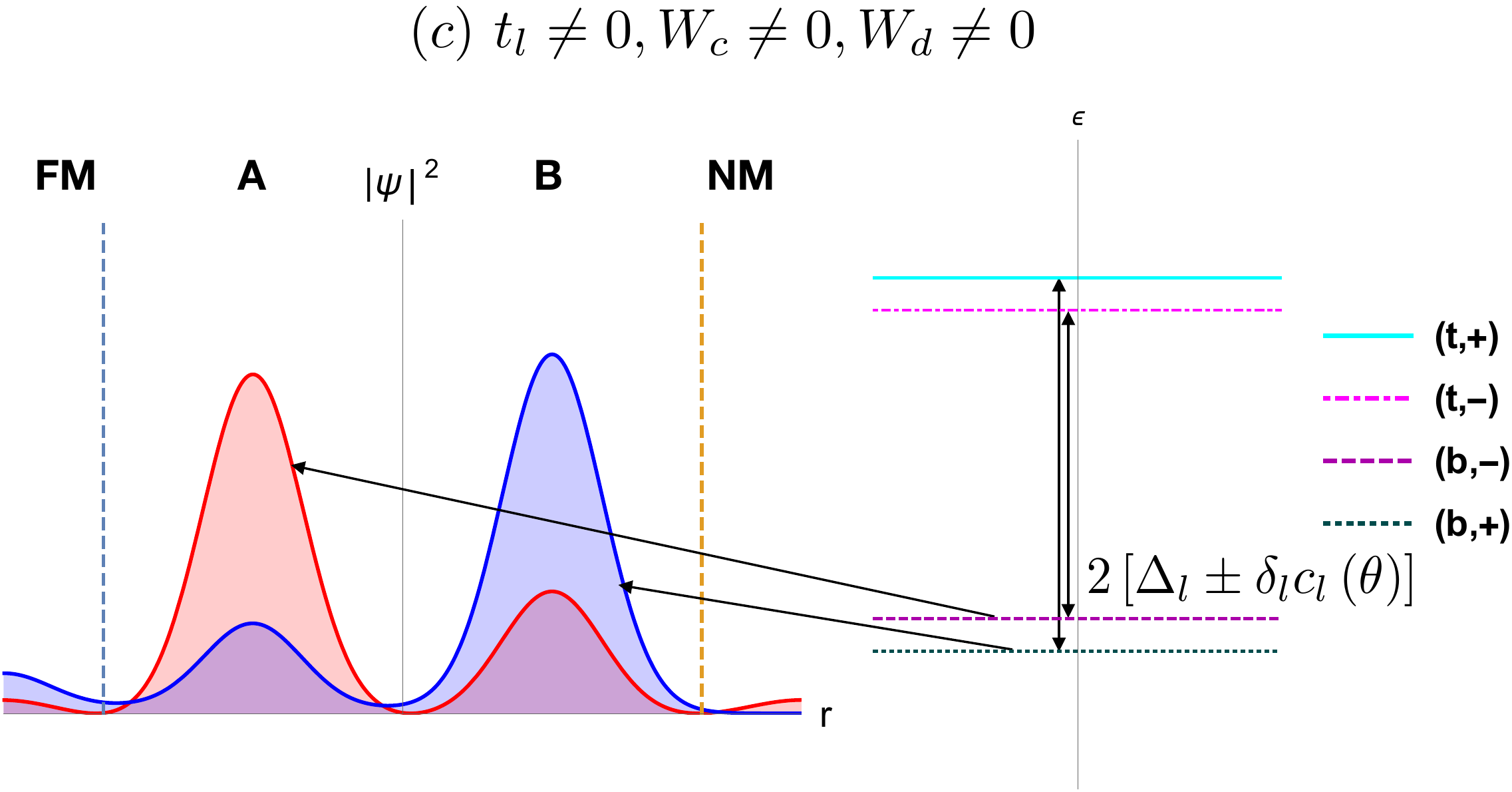}
\caption{Spatial variation of wave functions (Left) and the corresponding energy spectrum (Right) $(a)$ with neither intersublattice overlap nor tunneling, $(b)$ including intersublattice overlap without tunneling and $(c)$ for full electron eigenstates of the structure under consideration. ${\rm t}$ and ${\rm b}$ denote the top and bottom bands split by the large gap $2\Delta _l$. Spin is a good quantum number in an isolated bipartite antiferromagnet as for $(a)$ and $(b)$. Since the mixing with ferromagnetic electrons breaks rotational symmetry completely unless $\hat{\bm{z}}_F \parallel \hat{\bm{z}}$, spin cannot be used to label the states in $(c)$. The ferromagnet also breaks the symmetry between $A$ and $B$ sites, reflected on the asymmetry in the wavefunctions.}\label{fig:band}
\end{center}
\end{figure}

Figure \ref{fig:band} shows how the electron wave functions and the energy spectrum change as we turn on the intersublattice overlap $t_l $ and tunneling to the leads $W_{F,N}$. In the two-ferromagnet limit $t_l = W_F = W_N =0$, the strong $sd$ exchange interaction $\Delta _{\rm ex}$ splits the energy of up and down spins within each sublattice. Since $\bm{S}_B = -\bm{S}_A $ in the equilibrium, the spins of upper and lower energy states are swapped between $A$ and $B$ sublattices and the bands are doubly degenerate in spin. Thus we call them top $({\rm t},\uparrow \downarrow )$ and bottom $({\rm b},\uparrow  \downarrow )$ bands, even though the energy gap $2\Delta _{\rm ex}$ is essentially the exchange spin splitting. The introduction of $t_l $ does not qualitatively change the band structure as it preserves the sublattice symmetry and conservation of the $z$ component of spin. As we shall see, however, the nonzero overlap between the $A$ and $B$ wave functions opens up new channels of transport and alters some observables qualitatively. The energy gap $2\Delta _l =2 \sqrt{ \Delta _{\rm ex}^2 + \left| t_l \right| ^2 }$ is taken to be the largest relevant energy scale of the problem. When the coupling to F is taken into account, the degeneracy between the sublattices is lifted and spin ceases to be a good quantum number in general (Fig. \ref{fig:band}$(c)$). The energy split originating from the tunneling is denoted by $2\delta _l c_l ( \theta) $, where
\begin{eqnarray}
\delta _l &=& \sum _{\substack{k \\ \alpha =A,B }} \frac{\left| \left( W^{\alpha }_F \right) _{kl}\right| ^2  }{4} \mathcal{P}\left( \frac{1}{\epsilon -\epsilon ^F_{k\uparrow }} -\frac{1}{\epsilon -\epsilon ^F_{k\downarrow }} \right) , \\
c_l \left( \theta \right) &=& \sqrt{1- \frac{\Delta _{\rm ex}^2 }{\Delta _l^2 } \sin ^2 \theta } .  \label{eq:cl}
\end{eqnarray}
The corresponding split bands are labelled by $\pm $. This splitting plays an important role in interpreting $\Gamma ^m_{\rm fl} $ component of spin torque. Although the band structures of F and N are also modified by the influence of AF, these modifications are neglected by the assumption that the leads are much greater in spatial dimension and electron density of states than AF.

\subsection{Relaxation rates}
\label{sec:relax}
In considering transport of electrons, the lifetime of the electronic eigenstates, whose inverse we call {\it relaxation rate}, plays a crucial role alongside with the band structure. In our model, this occurs for the electrons in AF only through a tunneling into either of the two leads as we have not included other sources of scatterings such as disorder, electron-electron collisions, or phonons. First of all, the relaxation rate associated with the tunneling into N is independent of spin and sublattice by assumption; we denote it by $1/\tau _l^N $. The relaxation into F, in contrast, depends on both spin and sublattice, and is also a function of $\theta $. As derived in Sec.~\ref{sec:formalism}, the bands $({\rm t},+)$ and $({\rm b},-)$ have the same ferromagnetic relaxation rate $1/\tau _{l+}^F $ while the other two $({\rm t},-)$ and $({\rm b},+)$ decay at a different rate $1/\tau _{l-}^F $. The origin of the difference between $1/\tau _{l\pm }^F $ is the spin-dependent tunneling into F. We define the isotropic and anisotropic parts of the ferromagnetic tunneling rates, $1/\tau_l^F$ and $1/\tau_l^a$ respectively, as
\begin{eqnarray}
\frac{1}{\tau _l^F}=\frac{1}{2} \left( \frac{1}{\tau ^F_{l+}}+ \frac{1}{\tau _{l-}^F} \right), \quad \frac{c_l \left( \theta \right) }{\tau _l^a} =\frac{1}{2} \left( \frac{1}{\tau ^F_{l+}} - \frac{1}{\tau _{l-}^F} \right). \label{eq:rates}
\end{eqnarray} 
The quantity $c_l(\theta)$ which also governs the angle-dependence of the energy split of the $\pm$ bands has been introduced in Eq. (\ref{eq:cl}). The relaxation rates $1/\tau _l^{F,a}$ defined in Eq.~\eqref{eq:rates}  are independent of $\theta $. Microscopic expressions for $1/\tau _l^{F,N,a}$ in terms of the tunneling matrix elements are given by
\begin{eqnarray}
\frac{1}{\tau _l^N} &=& \sum _{\substack{m \\ \alpha =A,B}} \left| \left( W^{\alpha }_N \right) _{ml} \right| ^2 \delta \left( \epsilon -\epsilon ^N_m \right) , \label{eq:relax_N} \\
\frac{1}{\tau _l^F} &=& \sum _{\substack{k \\ \alpha =A,B}} \left| \left( W^{\alpha }_F \right) _{kl} \right| ^2 \frac{ \delta \left( \epsilon -\epsilon ^F_{k\uparrow } \right) + \delta \left( \epsilon -\epsilon ^F_{k\downarrow } \right)}{2} , \label{eq:relax_F}  \\
\frac{1}{\tau _l^a} &=&  \sum _{\substack{k \\ \alpha =A,B}} \left| \left( W^{\alpha }_F \right) _{kl} \right| ^2 \frac{ \delta \left( \epsilon -\epsilon ^F_{k\uparrow } \right) - \delta \left( \epsilon -\epsilon ^F_{k\downarrow } \right)}{2} . \label{eq:relax_a}
\end{eqnarray}
They can be estimated if one assumes that $\left| \left( W^{A,B}_{F/N} \right) _{k/ml}\right| ^2 \approx {\rm const} \equiv w_{F/N}^2  $. This leads to
\begin{eqnarray}
&&\frac{1}{\tau _l^F} \sim w_F^2  \frac{D_{F\uparrow } + D_{F\downarrow } }{2} ,\quad \frac{1}{\tau _l^a}  \sim  w_F^2 \frac{D _{F\uparrow } -D_{F\downarrow }}{2}, \label{eq:esttcta}\\
&&\qquad \qquad \qquad \frac{1}{\tau _l^N} \sim w_N^2 D _N, 
\end{eqnarray}
where $D _{F\uparrow ,\downarrow } , D _N $ are the respective densities of states at the Fermi energies of F and N. In contrast to $1/\tau_l^F$ and $1/\tau_l^a$, $\delta_l$ is determined by states with a wide range of energies $\epsilon_k$ in F. Therefore, the dependence of $\delta_l$ on characteristic energy scales is more difficult to estimate. As an example, for a $3d$ ferromagnet with quadratic dispersion one finds that the dimensionless product $\delta_l \tau_l^a$ scales with $\sqrt{\mu_F/\Lambda}$, where $\Lambda$ is the bandwidth. Therefore, one may expect the inequality $\left| \delta _l \right| < 1/\left| \tau _l^a \right| $ to hold. 

We can now be more precise about the approximation we have made, namely the slowness of the dynamics of $\bm{S}_{A,B}$. Its characteristic frequency $\omega $ must satisfy $\left| \omega  \right| \ll 1/\tau_l $, so that the electron dynamics, characterized by the typical dwell time $\tau_l$, is much faster than the magnetization dynamics.  

\subsection{Nonequilibrium stationary state}

In the remainder of the paper, we consider dynamics of AF driven by fixed electron distribution functions $n_{F/N}\left( \epsilon \right) = \left\{ \exp \left[ \left( \epsilon -\mu _{F/N} \right) /T_{F/N} \right] +1\right\} ^{-1}$ for F and N respectively. The externally applied bias voltage $\mu _F -\mu _N $ or temperature difference $T_F - T_N $ induces charge and spin currents in AF. Consequently a nonequilibrium spin accumulation develops in AF, which generates spin transfer torques via the $sd$ exchange interaction. The nonequilibrium state is characterized by the lesser Green's function $\left( G^<_a \right) _{ij}  = i \left\langle \psi ^{\dagger }_j \psi _i \right\rangle $ where $\psi _{i,j} $ are any of $a_{l\sigma },b_{l\sigma }$ and the expectation value is taken over the nonequilibrium probability distribution of the quantum mechanical states. We focus on a {\it stationary state} in which all the macroscopic observables such as currents and torques are independent of time. The stationary state is fully determined by $n_{F,N}$ and the instantaneous magnitude and orientation of the localized spins in AF. The calculation of $G^<_{a}$ is relegated to Appendix~\ref{sec:formalism}. Once $G^<_{a}$ is known, the torques and currents are readily computed as explained in the next sections. 

It is helpful to compare our setup with the weak tunneling regime discussed in Sec.~\ref{sec:introduction}. The latter concerns a situation where AF is in a prescribed equilibrium state $n_0 \left( \epsilon \right) $ and a weak contact with either F or N is introduced adiabatically. As long as AF stays close to the equilibrium state $n_0  $, the tunneling charge currents that flow across the interfaces AF/F ($I_F$) and AF/N ($I_N$) are given by
\begin{eqnarray}
I_F &=& 2e \sum _{l,\pm } \frac{n_F \left( \epsilon _l \pm \Delta _l \right) - n_0 \left( \epsilon _l \pm \Delta _l \right) }{\tau _l^F} ,  \label{eq:tunnel_current_c} \\
 I_N &=& 2e \sum _{l,\pm } \frac{n_N \left( \epsilon _l \pm \Delta _l \right)  - n_0 \left( \epsilon _l \pm \Delta _l \right) }{\tau _l^N} . \label{eq:tunnel_current_d}
\end{eqnarray}
The factor $2$ takes account of the spin degrees of freedom. The summation over $\pm $ corresponds to the top and bottom bands. Similarly, one can compute the tunneling spin current leaving F ($J_F^{z_F}) $ as
\begin{equation}
J^{z_F}_F =  \sum _{l,\pm } \frac{n_F \left( \epsilon _l \pm \Delta _l \right) -n_0 \left( \epsilon _l \pm \Delta _l \right) }{\tau _l^a} . \label{eq:tunnel_spin}
\end{equation}
Note that here the spin polarization is along the ferromagnetic axis $\hat{\bm{z}}_F$ and the current does not depend on $\theta $. In the normalization used in this manuscript, the tunneling spin current equals twice the Slonczewski's spin transfer torque in the weak tunneling regime,\cite{Chudnovskiy2008} setting a reference time scale for the magnetization dynamics. Under a bias voltage $V$, one can estimate $\omega \sim J^{z_F}_F \sim eV D_{AF} /\tau _l^a $ with the antiferromagnetic density of states $D_{AF}$. Thus the aforementioned slowness condition is self-consistent as long as $\left| eVD_{AF} \right| \ll 1$. We also remark that the spin current measured in AF is given by $ J^z_{\rm AF} = J^{z_F}_F \cos \theta $, where the polarization is along the N\'eel vector $\bm{n} =\hat{\bm{z}}$. The spin current at AF/N is zero in this weak tunneling limit. Therefore, one can interpret $1/\tau _l^{F,N}$ and $1/\tau _l^a $ roughly as charge and spin currents per antiferromagnetic electronic state. Note that the weak tunneling regime cannot be a stationary state unless a strong relaxation mechanism maintains AF in the state $n_0 $. We use a full nonequilibrium electron distribution of AF self-consistently determined by the tunneling processes to compute the currents. The results reduce to (\ref{eq:tunnel_current_c}) and (\ref{eq:tunnel_spin}) for $1/\tau _l^F \ll 1/\tau _l^N $ where N acts as the relaxation source against the currents from F. Yet the currents and torques in the strong tunneling regime are also expressed in terms of those same relaxation rates $1/\tau _l^{N,F,a}$ as we shall see.

Our formalism is formally valid for arbitrary values of $1/\tau _l^{N,F,a} $ and $\mu _F -\mu _N , T_F -T_N $ as long as the interfaces are in the tunneling regime;
\begin{equation}
\frac{D_{F\uparrow }+D_{F\downarrow }}{2\tau _l^F} \ll N_F , \quad \frac{D_N}{\tau ^N_l} \ll N_N .
\end{equation}
Here $N_{F,N}$ are the number of conduction channels for F and N and assumed to be large integers. These conditions can be interpreted that the conductivity of each channel must be small, i.e. they are nonmetallic tunneling contacts. Nevertheless, the total current can be large as there can be many channels. Although the band gap $\Delta _l $ can be arbitrary, we focus on the regime $|\delta _l | /\Delta _l , 1/|\tau _l^a |\Delta _l \ll 1$ for which physics can be discussed in the language of the antiferromagnetic band structure. The general results can be found in the Appendix.

Finally, it is worth repeating that our model system does not include spin-orbit interactions. The inclusion of electron-electron-interactions or interactions with phonons is also beyond the scope of this work. Relaxation is therefore modeled entirely through the coupling to the leads.

\section{Four types of spin torque}\label{sec:torque}
From the lesser Green's function Eq.~(\ref{eq:lesser}), one can readily compute the spin torques. The Heisenberg equations of motion for the averaged spin $\bm{S}_{A,B}$ for $H_0$ are given by
\begin{eqnarray}
\frac{d\bm{S}_A}{dt} &= & \frac{1}{i}\left[ \bm{S}_A ,H_0 \right] = \bm{S}_A \times \frac{\Delta _{\rm ex}}{S} \sum _l a_l^{\dagger } \bm{\sigma } a_l , \label{eq:LLGA} \\
\frac{d\bm{S}_B}{dt} &= & \frac{1}{i} \left[ \bm{S}_B ,H_0 \right] = \bm{S}_B \times \frac{\Delta _{\rm ex}}{S} \sum _l b_l^{\dagger } \bm{\sigma } b_l . \label{eq:LLGB}
\end{eqnarray}
In the antiferromagnetic frame, $\bm{S}_{A,B}$ are pointing in the $\hat{\bm{z}}$ and $-\hat{\bm{z}}$ directions respectively so that $d\bm{S}_{A,B}/dt$ have only $x$ and $y$ components. The slowness of the magnetization dynamics implies one can replace the electron operators by their expectation values. Rearranging (\ref{eq:LLGA}) and (\ref{eq:LLGB}) and discarding terms proportional to $\bm{m}$ on the right-hand-sides yield (\ref{eq:LLGM}) and (\ref{eq:LLGN}) with the coefficients identified to be
\begin{eqnarray}
\Gamma ^m_{\rm fl} \sin \theta &=&  -i\Delta _{\rm ex} \int \frac{d\epsilon }{2\pi } {\rm tr}\left[ \sigma _1 \otimes \tau _3 G^<_a \right] , \label{eq:torquemf} \\
\Gamma ^m_{\rm dl} \sin \theta &=& -i\Delta _{\rm ex} \int \frac{d\epsilon }{2\pi }{\rm tr}\left[ \sigma _2 \otimes \tau _3 G^<_a \right]  , \label{eq:torquemd} \\
\Gamma ^n_{\rm fl} \sin \theta &=& -i\Delta _{\rm ex} \int \frac{d\epsilon }{2\pi } {\rm tr}\left[ \sigma _1 \otimes 1_{SL} G^<_a \right]  , \label{eq:torquenf} \\
\Gamma ^n_{\rm dl} \sin \theta &=& -i\Delta _{\rm ex} \int \frac{d\epsilon }{2\pi } {\rm tr}\left[ \sigma _2 \otimes 1_{SL} G^<_a \right]  . \label{eq:torquend} 
\end{eqnarray}
We have noted
\begin{eqnarray}
i\sum _l \left\langle a_l^{\dagger }\bm{\sigma }a_l \right\rangle &=& \int \frac{d\epsilon }{2\pi } {\rm tr}\left[ \bm{\sigma } \otimes \frac{1_{SL}+\tau _3}{2} G^<_a \right] , \\
i \sum _l \left\langle b_l^{\dagger } \bm{\sigma }b_l \right\rangle &=& \int \frac{d\epsilon }{2\pi } {\rm tr}\left[ \bm{\sigma } \otimes \frac{1_{SL} - \tau _3}{2} G^<_a \right] .
\end{eqnarray}
One can observe that torques appearing in the equation for $\bm{n}$ are expectation values of the total electron spin while those driving $\bm{m}$ come from the staggered spin ($\bm{\sigma }\otimes \tau _3 $) expectation values.
\begin{figure}
\caption{\label{fig:torques} Four types of spin torque. They are classified according to the effective field orientations illustrated by the electron spin accumulations on the two sublattices. If the effective field lies in the plane spanned by the N\'eel vector $\bm{n}$ and the spin polarization $\hat{\bm{z}}_F$, the torque is field-like. An out-of-plane field represents an anti-damping-like torque. Each category has two staggered and non-staggered varieties based on the relative sign between the effective fields at the two sublattice sites. The direction of the torques themselves are indicated by planer arrows.}
\includegraphics[width=1\linewidth]{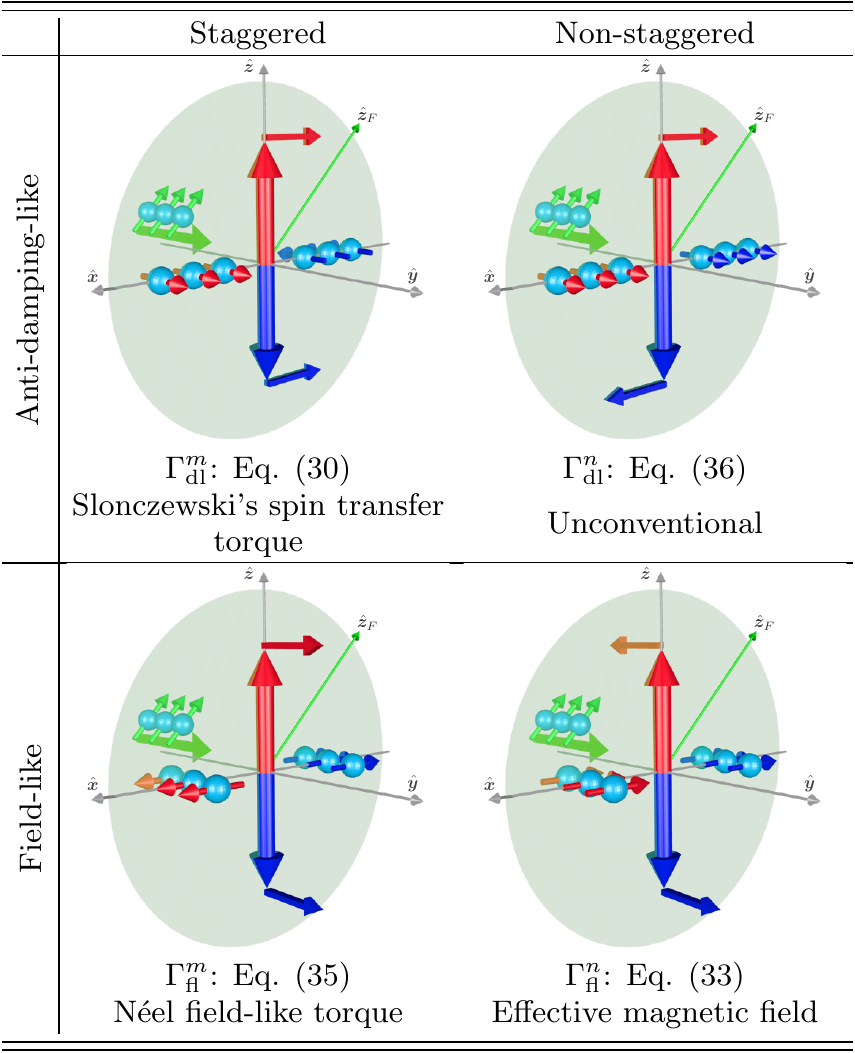}
\end{figure}
Based on this observation, we call $\Gamma ^m_{\rm fl,dl}$ {\it staggered} torques and $\Gamma ^n_{\rm fl,dl}$ {\it non-staggered} torques. When the torques are expressed as effective magnetic fields, the field direction coincides with the direction of the electron spin accumulation. Confusingly, if an effective field is staggered, the direction of the corresponding torque is non-staggered since the torque is a product of the field and the local magnetization, which is itself staggered. Our designation of {\it staggered} and {\it non-staggered} refers to the effective fields, not the torques. The situation is summarized in Fig.~\ref{fig:torques}. Note that in the antiferromagnetic frame, $\hat{\bm{z}}_F$ has $z$ and $x$ components only (Fig. \ref{fig:axes}). Therefore, the field-like torques are related to the injected transverse spin (i.e. spin perpendicular to $\bm{n}$) and the damping-like torques require the spin expectation value that is orthogonal to the polarization of the injected spin current. From these considerations alone, one can anticipate that the field-like torques can arise from mechanisms that conserve the transverse spin inside the antiferromagnet while the non-conservation of transverse spin is essential in generating the damping-like torques. Carrying out the traces in Eqs. (\ref{eq:torquemf}) - (\ref{eq:torquend}) is a straightforward matter as presented in Appendix \ref{sec:trace}. Below we discuss each of the four components in detail. 

\subsection{Slonczewski's spin transfer torque}\label{sec:dlm}
We start from the contribution which is most familiar in the ferromagnetic dynamics, which turns out to be $\Gamma ^m_{\rm dl} $ in our notation;
\begin{eqnarray}
\Gamma ^m_{\rm dl} &=& \int \frac{d\epsilon }{2\pi } \left( n_F -n_N \right)  \sum _l \frac{1}{\tau _l^a} \left( \frac{\Delta _{\rm ex}}{\Delta _l} \right) ^2 \nonumber \\
&& \times \frac{1}{\tau _l \tau _l^N} \left[ \frac{1}{\tau _l^2} -\left( \frac{c_l \left( \theta \right)}{\tau _l^a} \right) ^2 \right] ^{-1}  \left( A_{{\rm t}l}+A_{{\rm b}l} \right) , \label{eq:torquemd2}
\end{eqnarray}
where the spectral functions $A_{{\rm t,b}l}$ are given by
\begin{equation}
A_{{\rm t,b}l} = \frac{1}{2\tau _l} \frac{1 }{\left( \epsilon - \epsilon _l \mp \Delta _l \right) ^2 + 1/4\tau _l^2} .
\end{equation}
In our terminology, it can also be called {\it staggered anti-damping-like torque}. Note that at the leading order in $|\delta _l |/\Delta _l , 1/| \tau _l^a |\Delta _l $, the distinction between the $\pm $ bands has disappeared from the final expression. To identify it with the spin transfer torque, we take the limit $t_l =0 , 1/\tau _l^F \ll 1/\tau _l^N \ll \Delta _l $ and obtain (${\rm sgn}\left( \uparrow /\downarrow \right) = \pm 1$)
\begin{eqnarray}
\Gamma ^m_{\rm dl}  &\approx & \pi \sum _{kl, \sigma \pm} \frac{n_F \left(  \epsilon ^F_{k\sigma } \right) -n_N \left( \epsilon _l \pm \Delta _l \right) }{2} \nonumber \\
&& \times {\rm sgn}\left( \sigma \right) \left| \left( W_F \right) _{kl} \right| ^2 \delta \left( \epsilon ^F_{k\sigma } -\epsilon _l \mp \Delta _l \right)  . \label{eq:torquemd3}
\end{eqnarray}
The right-hand-side is precisely the spin current per sublattice in the leading order tunneling approximation (\ref{eq:tunnel_spin}) with assumption $n_0 = n_N $.\cite{Chudnovskiy2008} Even though physical interpretation of this formula has been well discussed in many places, \cite{Stiles2002,Slonczewski2002} we repeat the argument here in the context of the two-ferromagnetic description of antiferromagnet. Ignoring the intersublattice overlap $t_l$, an electron in F can only tunnel into a superposition of up and down spin states in one sublattice, which have different de Broglie wavelengths. Accordingly they dephase as they propagate and induce precession of the transverse spin component. The precession frequency differs for different orbital indices $l$. Upon averaging over $l$, the transverse component of the injected spin current is rapidly lost and absorbed into the magnetizations $\bm{S}_{A,B}$ as required by the overall spin conservation, resulting in the torque.

\begin{figure}[t]
\begin{center}
\includegraphics[width=1\linewidth]{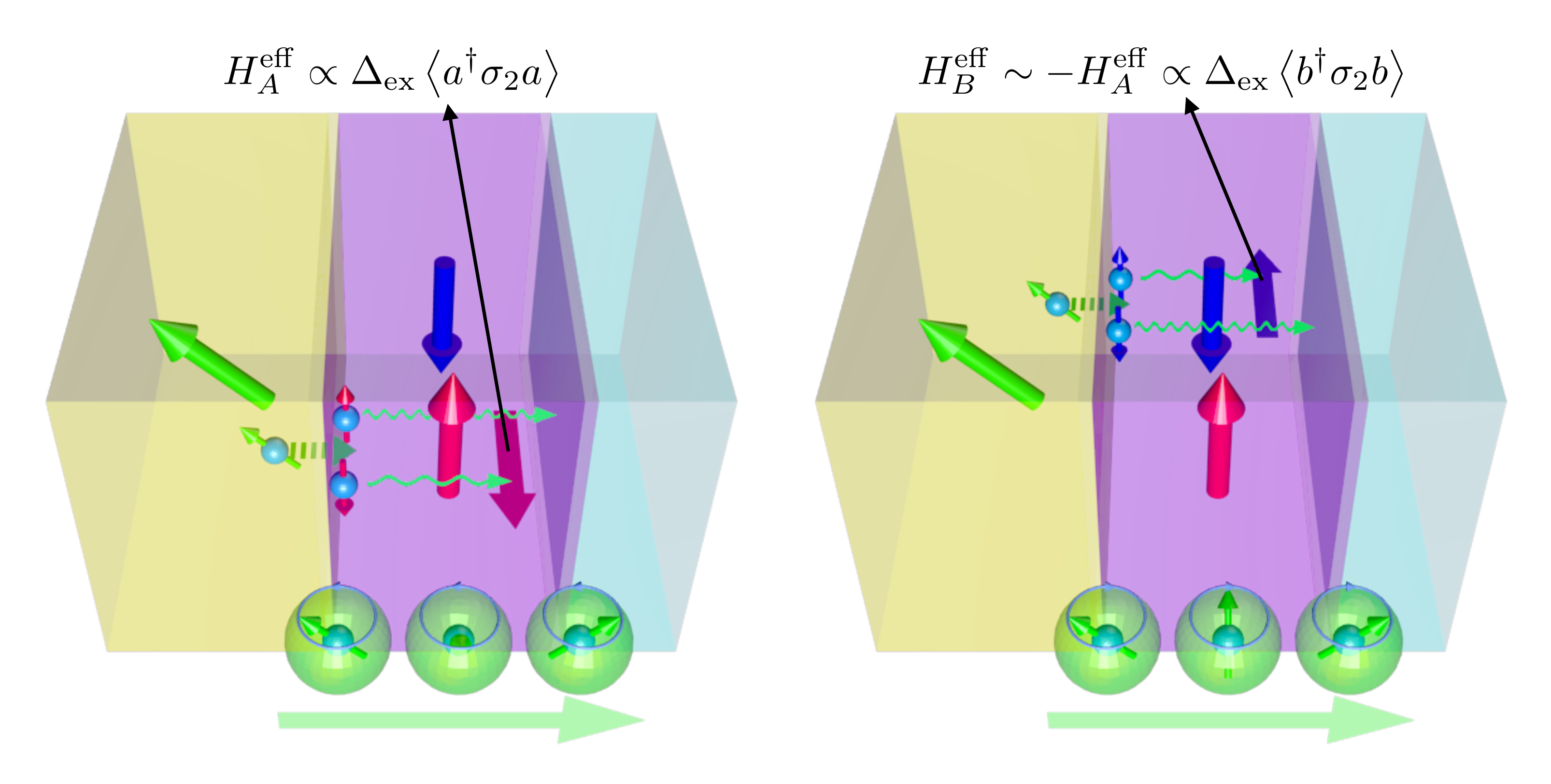}
\caption{Illustration of dephasing processes through $A$ and $B$ sublattices. An electron tunnels into a superposition of up and down states due to the difference in the quantization axes in F and AF. The dephasing leads to precession of the electron spin, whose chirality is opposite for the two sublattices. This intrasublattice process is the only channel of electron transport through AF if there is no intersublattice overlap $t_l =0$.}
\label{fig:dephase}
\end{center}
\end{figure}

Note that the torque $\Gamma^m_{\rm dl}$ appears as the expectation value of the staggered spin operator $\sigma _2 \otimes \tau _3 $. It is due to the opposite handedness of the dephasing-induced precession in the two sublattices as depicted in Fig. \ref{fig:dephase}. Our generalized expression (\ref{eq:torquemd2}) shows that the spin transfer torque in antiferromagnets is as effective as in ferromagnets even when $t_l \neq 0$ and multiple tunneling processes are taken into account. 

\subsection{Non-staggered field-like torque}

As stated above, the transverse component of spin is rapidly lost upon entering the antiferromagnet according to the two-ferromagnet description. Next we discuss the fate of transverse spin conservation in the presence of $t_l \neq 0$ by looking at $\Gamma ^n_{\rm fl} $, which is essentially the expectation value of the $x$ component of the spin $\sigma _1 \otimes 1_{SL}$ and given by
\begin{eqnarray}
\Gamma ^n_{\rm fl} &=& - \int \frac{d\epsilon }{2\pi } \left( n_F -n_N \right) \sum _l \frac{2\Delta _{\rm ex}}{\tau _l^N} \nonumber \\
&& \times \left[ \frac{1}{\tau _l^2 }-\left( \frac{c_l \left( \theta \right) }{\tau _l^a} \right) ^2 \right] ^{-1} \Bigg[ \frac{\left| t_l \right| ^2 }{\tau _l^a \Delta _l^2} \left( A_{{\rm t}l}+A_{{\rm b}l}\right) \nonumber \\
&& -\frac{\delta _l}{\tau _l \Delta _l} \left( \frac{\Delta _{\rm ex}}{\Delta _l} \right) ^2 \left( A_{{\rm t}l}-A_{{\rm b}l} \right) \Bigg] . \label{eq:torquenf2}
\end{eqnarray}
Note that the second term in the second line is one order higher in $\delta _l /\Delta _l$ compared to the first term. It has been kept, nevertheless, since the leading order term is proportional to $\left| t_l \right| ^2 $, which implies that this contribution would have been absent in the two-ferromagnet model and is unique to antiferromagnets. 

\begin{figure}[t]
\begin{center}
\includegraphics[width=1\linewidth]{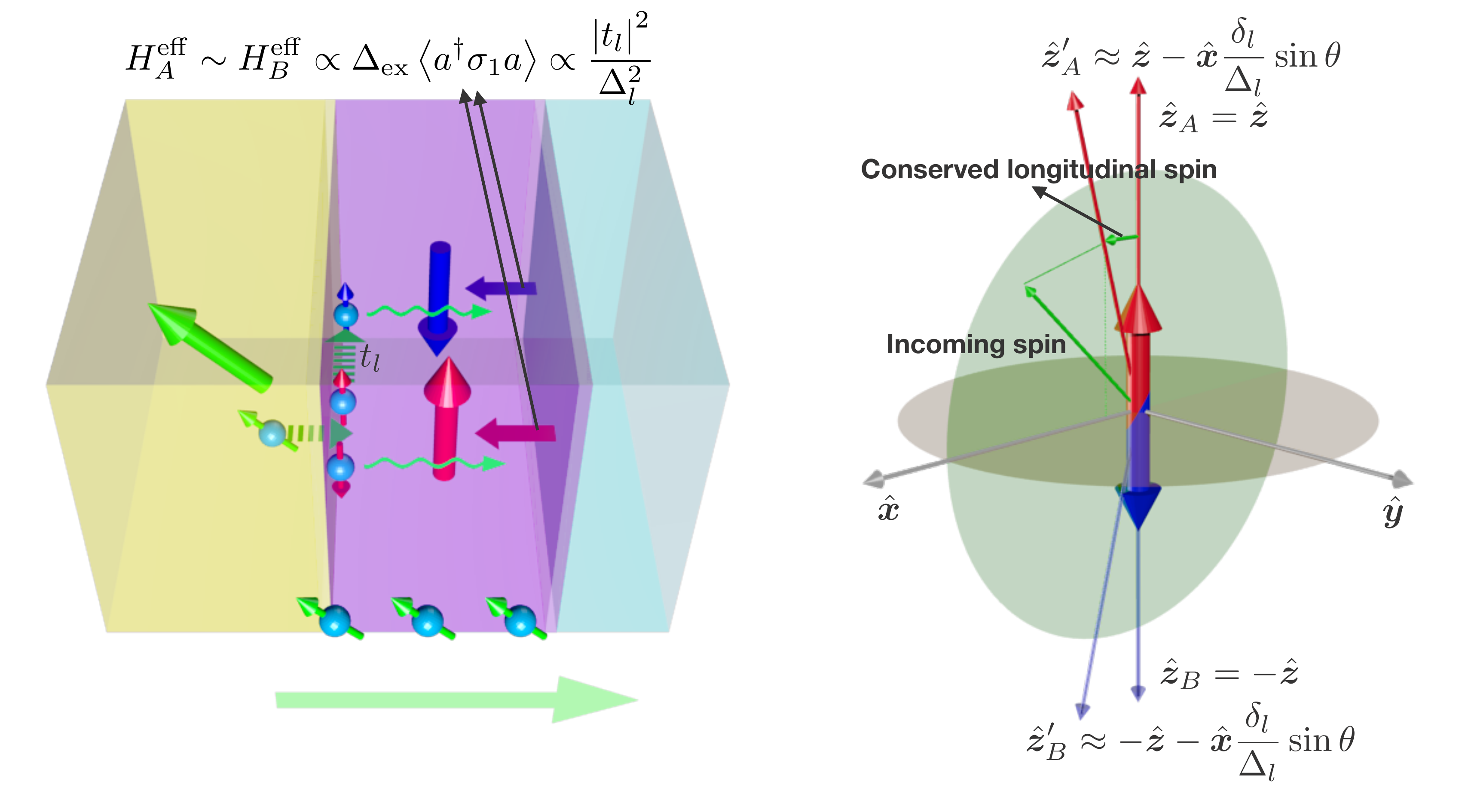}
\caption{Two mechanisms of transverse spin conservation in AF. Left: When $t_l \neq 0$, a ferromagnetic electron may propagate through AF as a superposition of up and down states that have exactly the same energy and wavelength. In the perturbative picture, an electron that tunneled into the $A$ site is initially a superposition of up and down with different wavelength. Subsequently, one of the electron states may hop onto a state in the $B$ site via $t_l$, which has the same wavelength as the state remaining at $A$. The phases of the two states evolve at the same rate, thus avoiding dephasing. Right: The tilt of the quantization axes . At the first order in $\delta _l /\Delta _l $, the axes for both $A$ and $B$ sites change by the same amount. At each sublattice, the $z^{\prime }_{A,B}$ component of the spin is conserved, which has a finite $x$ component.}
\label{fig:field-like}
\end{center}
\end{figure}

In AF with non-vanishing $t_l$, all the four bands $({\rm t}/{\rm b},\uparrow /\downarrow )$ have a nonzero amplitude at both $A$ and $B$ sublattice sites as shown in Fig. \ref{fig:band}(b). Thus a ferromagnetic electron may tunnel into a superposition of up and down states of exactly the same energy and wavelength, say $({\rm t},\uparrow )$ and $({\rm t},\downarrow )$. Alternatively, if one treats $t_l$ as a perturbation, an electron tunnels into a superposition of, e.g. $(A,\uparrow )$ and $(A,\downarrow )$, then via $t_l$, the state $(A,\downarrow )$ hops onto $(B,\downarrow )$ that has exactly the same wavelength as $(A,\uparrow )$ due to the sublattice symmetry. Either way, after the tunneling, the two electron states, representing a single electron, propagate with exactly the same phase evolution, dephasing is thus avoided, and the transverse spin is conserved (Fig. \ref{fig:field-like}). We reiterate that this is a consequence of the complete sublattice symmetry assumed in our model. Consequently, there will be a nonvanishing expectation value of the $x$ component of spin proportional to the fraction of electrons undergoing the intersublattice hopping $\left| t_l \right| ^2 /\Delta _l^2 $, which is represented by the first term in (\ref{eq:torquenf2}). This also explains the factor $ \left( \Delta _{\rm ex}/\Delta _l \right) ^2 $ in Eq. (\ref{eq:torquemd2}), which coincides with the fraction of electrons propagating with different wavelengths and affected by the dephasing. 
We note that a related mechanism was discussed in the context of antisymmetric F/N/F spin valves.\cite{Brataas2003} 

The physics behind the second term should then be related to intrasublattice processes as it also comes with the factor of $\left( \Delta _{\rm ex} /\Delta _l \right) ^2 $. It represents the residual $x$ component of spin that has managed to survive the dephasing. One way to interpret this term is to consider the tilt of electron quantization axes in AF due to the influence of F. It should not be confused with the tilt of $\bm{S}_{A,B}$ as they are assumed fixed in the electron time scale. The part of self-energy proportional to $\delta _l$ can be considered as an additional Zeeman term in the direction of the ferromagnetic moment $\sigma _z \cos \theta - \sigma _x \sin \theta $. In the leading order approximation in $\delta _l /\Delta _l$, taking it into account yields the direction of the effective magnetic field (preferred quantization axis) $\hat{\bm{z}}^{\prime } \sim \hat{\bm{z}} - \left( \delta _l \sin \theta /\Delta _l \right) \hat{\bm{x}}$ for the antiferromagentic electrons (Fig. \ref{fig:field-like}). While the spin transverse to the quantization axis is lost by dephasing, the longitudinal component is conserved by definition. Thus the fraction $\delta _l \sin \theta /\Delta _l$ of the injected spin $x$ component will be conserved and contribute to the field-like torque.

It is helpful to write down $\Gamma ^n_{\rm fl} $ in the weak ferromagnetic tunneling limit $1/\tau _l^F \ll 1/\tau _l^N $, yielding
\begin{eqnarray}
\Gamma ^n_{\rm fl} & \approx & -  2\int \frac{d\epsilon }{2\pi } \left( n_F -n_N \right) \sum_l\Bigg[ \Delta _{\rm ex} \frac{\tau _l^N}{\tau _l^a} \frac{\left| t_l \right| ^2 }{\Delta _l^2} \left( A_{{\rm t}l}+A_{{\rm b}l}\right) \nonumber \\
&& - \delta _l \left( \frac{\Delta _{\rm ex}}{\Delta _l} \right) ^3 \left( A_{{\rm t}l}-A_{{\rm b}l} \right) \Bigg] .\label{eq:Gnfl}
\end{eqnarray}
As one can see, the intersublattice contribution (the first term) has the relaxation time in the numerator. When this first term dominates $\Gamma ^n_{\rm fl}$, its relative magnitude compared to the spin transfer torque $\Gamma ^m_{\rm dl}$ is given by $\sim \tau _l^N  \left| t_l \right| ^2 /2\Delta _{\rm ex} $. Therefore $\Gamma ^n_{\rm fl} $ could in principle greatly exceed $\Gamma ^m_{\rm dl} $ depending on how $1/\tau _l^N \sim w_N^2 D_N $ and $t_l $ compare with $\Delta _{\rm ex}$. This is one of the reasons for specifying the origin of the relaxation in the present work. The intrasublattice contribution (the second term) is always suppressed by a factor of $\left| \delta _l \tau _l^a \right|  \ll 1$ with respect to $\Gamma ^m_{\rm dl}$. This contribution exists for ferromagnets as well, though mostly neglected because it is always subdominant. 

\subsection{N\'eel field-like torque}
As we have explained, the $x$ component of the spin expectation value is related to the conservation of the injected spin current. In the picture given above that is based on first order tunneling processes, the generated expectation value is the same for the two sublattices, which therefore leads to a non-vanishing $\Gamma ^n_{\rm fl} $. If the higher order effects of the tunneling from F are considered, however, they break the symmetry between the sublattices and result in different expectation values of $s_x^A = \left\langle a_l^{\dagger }\sigma _1 a_l \right\rangle $ and $s_x^B =\left\langle b_l^{\dagger } \sigma _1 b_l \right\rangle $. This breaking of the sublattice symmetry manifests itself through two different parameters. The first is the energy split $\delta _l /\Delta _l$, whose first order effect gives an equal magnitude of $s_x^A $ and $s_x^B$ as we have just seen, but at the second order leads to an asymmetric change of quantization axis for $A$ and $B$ sublattices, and in turn to different values of $s_x^{A,B}$. The other parameter is $\tau _l^N /\tau _l^a $, which represents the different exit rates to F for the antiferromagnetic electrons in $A$ and $B$ sublattices. This is also expected to give an asymmetric correction to the conserved $s_x^{A,B}$, given that $s_x^{A,B}$ have a finite expectation value, since if electrons in one sublattice escape faster than those in the other, it should reduce the spin expectation value for the former compared to the latter.

In our notation, the asymmetric part of the conserved $x$ component corresponds to $\Gamma ^m_{\rm fl} $ and is given in terms of the expectation value of $\sigma _1 \otimes \tau _3 $, which reads
\begin{eqnarray}
\Gamma ^m_{\rm fl} &= & \int \frac{d\epsilon }{2\pi } \left( n_F -n_N \right) \sum _l \frac{2\Delta _{\rm ex}}{\tau _l^N}\left[ \frac{1}{\tau _l^2} - \left( \frac{c_l \left( \theta \right) }{\tau _l^a} \right) ^2 \right] ^{-1} \nonumber \\
&& \times \frac{\cos \theta }{\tau _l^a} \frac{\delta _l}{\Delta _l} \frac{\Delta _{\rm ex}}{\Delta _l} \left( A_{{\rm t}l} + A_{{\rm b}l} \right) . \label{eq:torquemf2}
\end{eqnarray}
The similarity of this formula to the intrasublattice contribution of $\Gamma ^n_{\rm fl}$ is apparent; the differences are the factor $\tau _l^N \cos \theta /\tau _l^a $, the different power of $\Delta _{\rm ex}/\Delta _l$, and the sign in front of $A_{{\rm b}l}$. The latter two features are related to the fact that $\Gamma ^m_{\rm fl} $ is an order higher also in $\Delta _{\rm ex}$ than $\Gamma ^n_{\rm fl}$. Hence it is reasonable to interpret this correction as arising from the escape rate difference. Since the same estimate as for the intrasublattice term of $\Gamma ^n_{\rm fl} $ applies, $\Gamma ^m_{\rm fl} $ is expected to be subdominant in the entire parameter space and we do not go deeper in its physical interpretation.

\subsection{Unconventional anti-damping-like torque}
Similarly to the field-like torques, the leading order effect of the ferromagnetic tunneling gives a sublattice symmetric contribution to $s_y^A =\left\langle a_l^{\dagger } \sigma _2 a_l \right\rangle $ and  $s_y^B =\left\langle b^{\dagger }_l \sigma _2 b_l \right\rangle $, albeit the expectation values are staggered $s_y^B = -s_y^A$. Then the symmetry breaking effects discussed in the previous subsection should generate a non-staggered component $s_y^A + s_y^B$, which corresponds to $\Gamma ^n_{\rm dl} $. To the leading order in $\delta _l /\Delta _l , 1/\tau _l^a \Delta _l $, one derives
\begin{eqnarray}
\Gamma _{\rm dl}^n &=&  \int \frac{d\epsilon }{2\pi } \left( n_F -n_N \right) \sum _l \frac{1}{\tau _l^a} \left( \frac{\Delta _{\rm ex}}{\Delta _l} \right) ^3 \cos \theta  \nonumber \\
&& \times \frac{1}{\tau _l^N \tau _l^a} \left[ \frac{1}{\tau _l^2}-\left( \frac{c_l \left( \theta \right) }{\tau _l^a} \right) ^2 \right] ^{-1} \left( A_{{\rm t}l}-A_{{\rm b}l} \right), \label{eq:torquend2}
\end{eqnarray}
whose close connection to (\ref{eq:torquemd2}) is clear. This contribution can be interpreted as caused by the different electron escape rates between $A$ and $B$ sublattices limiting the expectation value of the $y$ component of the spin. The $\cos \theta $ factor is also reasonable as there is no symmetry breaking when the ferromagnetic moment is perpendicular to the N\'eel vector.

\subsection{Comparison of the torques}

The expressions for the four torques given in  Eqs.~(\ref{eq:torquemd2}), (\ref{eq:torquenf2}), (\ref{eq:torquemf2}), (\ref{eq:torquend2}) can be evaluated explicitly once the parameters of the model Hamiltonian such as the spectra of the leads and the antiferromagnetic island and the tunneling amplitudes are specified. Here, we aim to make general statements about the relative importance of the different kinds of spin torque. 

A brief look at Eqs.~\eqref{eq:torquemd2}, (\ref{eq:torquenf2}), (\ref{eq:torquemf2}), (\ref{eq:torquend2}) reveals that the summation in $l$ involves terms that are even in $1/\tau_l^a$ for $\Gamma_{\rm dl}^n$, Eq.~\eqref{eq:torquend2} and odd in $1/\tau_l^a$ for the remaining three torques. Since $1/\tau^a_l$ can take both positive and negative values, cancellations may in principle occur when evaluating $\Gamma_{\rm dl}^m$, $\Gamma_{\rm fl}^n$ and $\Gamma_{\rm fl}^m$. This fact makes general statements about their magnitude difficult. For the further discussion, we therefore assume that $1/\tau_l^a$ does not change sign within the relevant interval of energies.

First, we will be concerned with the comparison of the two damping-like torques. For this case, one arrives at the inequality $\left| \Gamma ^n_{\rm dl} \right| < \left| \Gamma ^m_{\rm dl} \right| $, which is a direct consequence of the following hierarchy of relaxation rates, $1/\left| \tau _l^a \right| < 1/\tau _l^F <1/\tau _l $. Let us discuss the ratio $r_{\rm d}=|\Gamma ^n_{\rm dl} /\Gamma ^m_{\rm dl}| $ in two limiting cases. In the limit of weak ferromagnetic tunneling, one has $r_{\rm d} \sim \tau _l^N /\left| \tau _l^a \right| \ll 1 $. This implies that when the first order tunneling approximation is justified, $\Gamma ^n_{\rm dl} $ is likely negligible compared to the conventional spin transfer torque $\Gamma^m_{\rm dl}$. In the opposite regime, i.e. $1/\tau _l^N \ll 1/\tau _l^F $, the ratio can be estimated as $r_d\sim\tau_l^F/|\tau_l^a|$. The estimate \eqref{eq:esttcta} for the relaxation rates suggests that in this case $r_{\rm d}\sim |D_{F\uparrow}-D_{F\downarrow}|/(D_{F\uparrow}+D_{F\downarrow})\sim \Delta_F/\mu_F<1$, where $\Delta_F$ is the exchange splitting in the ferromagnetic lead.

Next, we will be concerned with the comparison of the torques entering Eq.~\eqref{eq:LLGM} for $d{\bf m}/dt$, $\Gamma ^m_{\rm dl} $ and $\Gamma ^m_{\rm fl} $. The relevant control parameter is $|\Gamma^m_{\rm fl}/\Gamma^m_{\rm dl}|\sim  \left| \delta _l \right| \tau _l < \left| \delta _l \tau _l^a \right| < 1$. This estimate suggests that the N\'eel field-like torque in the absence of spin-orbit coupling is insignificant in almost all circumstances. 

We will now turn to the two torques entering Eq.~\eqref{eq:LLGN} for $d{\bf n}/dt$, $\Gamma ^n_{\rm dl} $ and $\Gamma ^n_{\rm fl} $. Here, the control parameter is $|\Gamma^n_{\rm fl}/\Gamma^n_{\rm dl}|\sim \Delta_l|\tau_l^a|\times |t_l|^2/\Delta _{\rm ex}^2$ (assuming that the first term dominates in expression for $\Gamma_{\rm fl}^n$, Eq.~\eqref{eq:Gnfl}). This ratio is proportional to $|\tau^a|$ and therefore enhanced for weak coupling to F. Further, the ratio is strongly affected by the value of $t_l $, which is difficult to estimate from microscopic considerations. Since $\Gamma^n_{\rm fl}$ is likely negligible when $\left| t_l \right| \ll \Delta _{\rm ex} $, the detection of a significant $\Gamma ^n_{\rm fl} $ might be used as an experimental probe into the extent of microscopic intersublattice wave function overlap, although it could be practically difficult to eliminate the possible field-like contributions from spin-orbit interactions. 

For the comparison of $\Gamma ^n_{\rm dl} $ and $\Gamma ^m_{\rm fl} $, one can probably assume $\left| \Gamma ^m_{\rm fl} \right| < \left| \Gamma ^n_{\rm dl} \right| $ as $| \Gamma ^m_{\rm fl}/  \Gamma ^n_{\rm dl}|\sim \Delta |\tau_l^a|\times \delta_l/\Delta _{\rm ex}$.

In summary, the discussion presented above suggests the hierarchy $
|\Gamma_{\rm dl}^m|> |\Gamma_{\rm dl}^n|>| \Gamma ^m_{\rm fl}|$. As for the non-staggered field-like torque, a sizeable $\Gamma^{n}_{\rm fl}$ requires large intersublattice overlap amplitudes $|t_l|$. Indeed, its relative magnitude compared to $\Gamma^m_{\rm dl}$, for which one estimates $|\Gamma^{n}_{\rm fl}/\Gamma^m_{\rm dl}|\sim |t_l|/\Delta _{\rm ex} \times |t_l|\tau_l$ and compared to $\Gamma^n_{\rm dl}$, for which $|\Gamma^{n}_{\rm fl}|/\Gamma^n_{\rm dl}|\sim \Delta_l\tau_l^a\times |t_l|^2/\Delta _{\rm ex}^2$, depends crucially on the relation between $|t_l|$ and $\Delta _{\rm ex}$ as well as the relaxation rates $1/\tau_l^a$ and $1/\tau_l$. When the overlap is ignored $t_l =0$, one obtains $\left| \Gamma ^n_{\rm fl} /\Gamma ^m_{\rm dl} \right| \sim \left| \delta _l \tau _l^a \right| \sim \sqrt{\mu _F / \Lambda }$ where $\Lambda $ is the bandwidth of F. Thus one would typically expect $\left| \Gamma ^n_{\rm fl}/\Gamma ^m_{\rm dl} \right| \ll 1$, as was also found for an interface between a normal metal and an antiferromagnetic insulator.\cite{Cheng2014c}

\subsection{Threshold current in the presence of $\Gamma ^n_{\rm dl} $}

In contrast to Slonczewski's spin-transfer torque $\Gamma ^m_{\rm dl} $ as well as the conventional field-like torque $ \Gamma ^n_{\rm fl} $\cite{Gomonay2012,Cheng2014} and the N\'eel field-like torque $\Gamma ^m_{\rm fl} $\cite{Zelezny2014}, which have all been studied in the literature in some form, dynamical consequences of the unconventional anti-damping-like torque $\Gamma ^n_{\rm dl} $ have hardly been discussed so far in the spintronics literature. As we have found that the magnitude of $\Gamma ^n_{\rm dl} $ can be a sizable fraction ($\sim \Delta_F/\mu_F$) of the usually dominant $\Gamma ^m_{\rm dl} $, it is of a practical interest to explore how $\Gamma ^n_{\rm dl} $ manifests itself in the antiferromagnetic dynamics. Here, we compute the threshold current that destabilizes a ground state configuration in the presence of both $\Gamma ^m_{\rm dl} $ and $\Gamma ^n_{\rm dl} $ for a uniaxial antiferromagnet. To simplify the discussions, we ignore the generically small $\Gamma ^m_{\rm fl} $ and also assume $\left| t_l \right| \ll \Delta _{\rm ex} $ so that $\Gamma ^n_{\rm fl} $ is negligible. 

The Landau-Lifshitz-Gilbert equations for a uniaxial bipartite antiferromagnet in the macrospin approximation are given by
\begin{eqnarray}
\frac{d\bm{m}}{dt} &=& - \bm{n} \times \left( \omega _A n^z \hat{\bm{z}} + \frac{\Gamma ^m_{\rm dl}}{2S} \bm{n} \times \hat{\bm{z}}_F \right)  \nonumber \\
&& +\alpha \bm{n} \times \frac{d\bm{n}}{dt} , \\
\frac{d\bm{n}}{dt} &=&  2\omega _E \bm{n} \times \bm{m} -\frac{\Gamma ^n_{\rm dl}}{2S} \bm{n} \times \left( \bm{n} \times \hat{\bm{z}}_F \right) \nonumber \\
&& + \alpha \bm{n} \times \frac{d\bm{m}}{dt} . \label{eq:LLGn2}
\end{eqnarray}
Here $\omega _A $ and $\omega _E $ are the crystalline anisotropy and exchange field in the unit of frequency respectively, and $\alpha $ is the bulk Gilbert damping constant. Note that $\hat{\bm{z}}_F $ denotes the polarization of the incoming spin current and $\hat{\bm{z}}$ is now fixed in the direction of the anisotropy axis. We dropped the anisotropy term in the equation for $\bm{n}$, Eq. (\ref{eq:LLGn2}), as it is always negligible compared to the exchange term. We pick the ground state configuration $\bm{n}_0 = \hat{\bm{z}} , \bm{m}_0 =0$ and consider the linear perturbation $\bm{n} -\bm{n}_0 = \left( n_x ,n_y , 0 \right) , \bm{m} -\bm{m}_0 = \left( m_x ,m_y ,0 \right) $. Introducing the complex variables $n_+ = n_x + in_y , m_+ = m_x + im_y $, the linearized equations of motion read
\begin{eqnarray}
\frac{dm_+}{dt} &=& i\omega _A n_+ - \frac{\Gamma ^m_{\rm dl}}{2S} n_+ + i\alpha \frac{dn_+}{dt} , \\
\frac{dn_+}{dt} &=&  2i\omega _E  m_+ - \frac{\Gamma ^n_{\rm dl}}{2S} n_+ + i\alpha \frac{dm_+}{dt} .
\end{eqnarray}
Applying the Fourier transform $n_+ = n_{\omega }e^{-i\omega t} , m_+ = m_{\omega }e^{-i\omega t} $, the frequency eigenvalues are determined by
\begin{equation}
\omega \left( \omega + i\frac{\Gamma ^n_{\rm dl}}{2S} \right) = \left( 2\omega _E  - i\alpha \omega \right) \left( \omega _A - i\alpha \omega + i\frac{\Gamma ^m_{\rm dl}}{2S} \right) .
\end{equation}
We discard terms first order in $\omega _A /\omega _E $ and second order in $\alpha , \Gamma ^{m,n}_{\rm dl}$ and obtain two eigenfrequencies
\begin{eqnarray}
\omega _{\pm }&=& \pm \sqrt{ 2\omega _E \omega _A  } - i\alpha  \omega _E  \pm i\sqrt{\frac{2\omega _E }{\omega _A}}\frac{\Gamma ^m_{\rm dl}}{4S} - i \frac{\Gamma ^n_{\rm dl}}{4S} .
\end{eqnarray}
As long as the Gilbert damping $\alpha >0 $ dominates the imaginary part, the ground state is stable and the system stays in the linear regime. The damping-like torques $\Gamma ^{m,n}_{\rm dl}$ can change sign depending on the direction of the bias field and spin polarization so that they can contribute either positive or negative damping. When the current is strong enough so that the imaginary part of one of $\omega _{\pm }$ is positive, the ground state becomes unstable. We note that the way the two damping-like torques $\Gamma ^{m,n}_{\rm dl} $ contribute to the instability is qualitatively different. Because of the $\pm $ sign in front, $\Gamma ^m_{\rm dl}$ destabilizes one mode and stabilizes the other regardless of whether $\Gamma ^m_{\rm dl} \gtrless 0$. In contrast, if $\Gamma ^n_{\rm dl} $ is positive (negative), its effect is always stabilizing (destabilizing) for the both $\omega _{\pm }$ modes. The threshold current at which the instability sets in is determined by
\begin{equation}
\alpha \omega _E  = \sqrt{\frac{2\omega _E}{\omega _A}} \frac{\left| \Gamma ^m_{\rm dl} \right| }{4S}  
 -  \frac{\Gamma ^n_{\rm dl}  }{4S}   .
\end{equation}
Therefore, a finite $\Gamma ^n_{\rm dl} $ results in a threshold current that depends on the signature of $\Gamma ^n_{\rm dl}$. For instance, one can switch the sign of $\Gamma ^n_{\rm dl} $ by reversing the direction of the bias voltage. This asymmetry of the threshold current with respect to the direction of the applied charge current can be used to experimentally study the unconventional damping-like torque $\Gamma ^n_{\rm dl} $.

\section{Conductance and spin transmission}\label{sec:current}

The nonequilibrium stationary state of AF is maintained by continual flows of electrons at F/AF and AF/N interfaces. These charge currents are given by the Meir-Wingreen formula\cite{Meir1992};
\begin{eqnarray}
I_{\beta } &=& e \int \frac{d\epsilon }{2\pi } {\rm tr}\Big[ \left\{ G^<_a + n_{\beta }\left( G^R_a -G^A_a \right) \right\} \nonumber \\
&& \quad \times  W_{\beta }^{\dagger } \left( G^R_{\beta }-G^A_{\beta }\right) W_{\beta } \Big] ,\label{eq:current}
\end{eqnarray}
where $\beta = F,N$ for F/AF and AF/N interfaces respectively and $G^{R,A}_{a,F,N}$ are the retarded and advanced Green's functions of AF, F and N respectively, whose details are given in Appendix \ref{sec:formalism}. Similar expressions can be readily derived for the spin currents;
\begin{eqnarray}
J^{z_F}_F &=& \frac{1}{2}\int \frac{d\epsilon }{2\pi }{\rm tr}\Big[ \left\{ G^<_a  + n_F \left( G^R_a -G^A_a \right) \right\} \nonumber \\
&& \quad \times W_F^{\dagger } \left( G^R_F -G^A_F \right) \sigma _3 W_F \Big] ,  \\
\bm{J}_N &=& \frac{1}{2} \int \frac{d\epsilon }{2\pi } {\rm tr}\Big[ \left\{ G^<_a + n_N \left( G^R_a - G^A_a \right) \right\} \nonumber \\
&& \quad\times  W_N^{\dagger } \left( G^R_N  -G^A_N \right) \bm{\sigma } W_N \Big] . \label{eq:spin_current}
\end{eqnarray}
Note that the Meir-Wingreen formula is based on the conservation of the current in the lead so that $x_F , y_F $ components of the spin current at F/AF are not meaningful.

\subsection{Magnetoresistance}\label{sec:charge}

Since the charge is conserved in the whole system, $I_F = -I_N $ holds for stationary states with the latter being slightly simpler to calculate. Substituting (\ref{eq:lesser}) into (\ref{eq:current}) and taking the leading order term in $\delta _l /\Delta _l ,1/\tau _l^a \Delta _l $ lead to
\begin{eqnarray}
I_N &=& \int \frac{d\epsilon }{2\pi } \sum _l \frac{2e}{\tau _l^N} \left( n_F -n_N \right) \left( A_{{\rm t}l} + A_{{\rm b}l} \right) \nonumber \\
&& \times \left\{ 1- \frac{1}{\tau _l \tau _l^N} \left[ \frac{1}{\tau _l^2} - \left( \frac{c_l \left( \theta \right) }{\tau _l^a} \right) ^2 \right] ^{-1} \right\} . \label{eq:charge}
\end{eqnarray}
The first line is the tunneling current into the normal metallic lead (\ref{eq:tunnel_current_d}), which indeed $I_N $ reduces to in the limit of $1/\tau _l^N \ll 1/\tau _l^F $. The magnetoresistance manifesting itself through the $\theta $-dependence of $I_N $ thoroughly comes from the second line. As $c_l \left( \theta \right) $ is a monotonically decreasing function of $\sin ^2 \theta $, $I_N $ is maximum when $\theta =\pi /2 $ and minimum for $\theta =0,\pi $. 

In order to infer physical processes responsible for the angular dependence, we again appeal to the weak ferromagnetic tunneling approximation $1/\tau _l^F \ll 1/\tau _l^N $ to rewrite (\ref{eq:charge}) as
\begin{equation}
I_N \approx \int \frac{d\epsilon }{2\pi }e \left( n_F -n_N \right) \sum _{l,\pm } \left[ \frac{1}{\tau ^F_{l\pm }}  -\frac{\tau _l^N}{\left( \tau ^F_{l\pm } \right) ^2} \right] \left( A_{{\rm t}l}+A_{{\rm b}l} \right) , \label{eq:charge2}
\end{equation}
where it has been expressed in terms of the escape rates to F for the $({\rm t},\pm )$ eigenstates $1/\tau ^F_{l\pm }= 1/\tau _l^F \pm c_l (\theta ) /\tau _l^a $. One can see that the angular dependences from $1/\tau ^F_{l+ } $ and $1/\tau ^F_{l-} $ cancel each other at the first order (i.e. the first term in the square bracket in Eq. (\ref{eq:charge2})) in accordance with the two-ferromagnet picture. The sign of the second order term (proportional to $1/(\tau ^F_{l\pm })^2 $) is the opposite of the first order term, which implies a backflow of the current from AF to F. It microscopically represents simple second-order processes where an electron enters from the lead at the rate $1/\tau ^F_{l\pm }$, then exits at the same rate $1/\tau ^F_{l\pm }$. This escape rate should be compared with the escape rate to the other lead $1/\tau _l^N $, the only other source of relaxation, which accounts for the factor of $\tau _l^N $. Taking this backflow into account reduces the total current by the fraction $\tau _l^N / \tau ^F_{l\pm } $ in the case of positive bias.  

\begin{figure}[tbp]
\begin{center}
\includegraphics[width=1\linewidth]{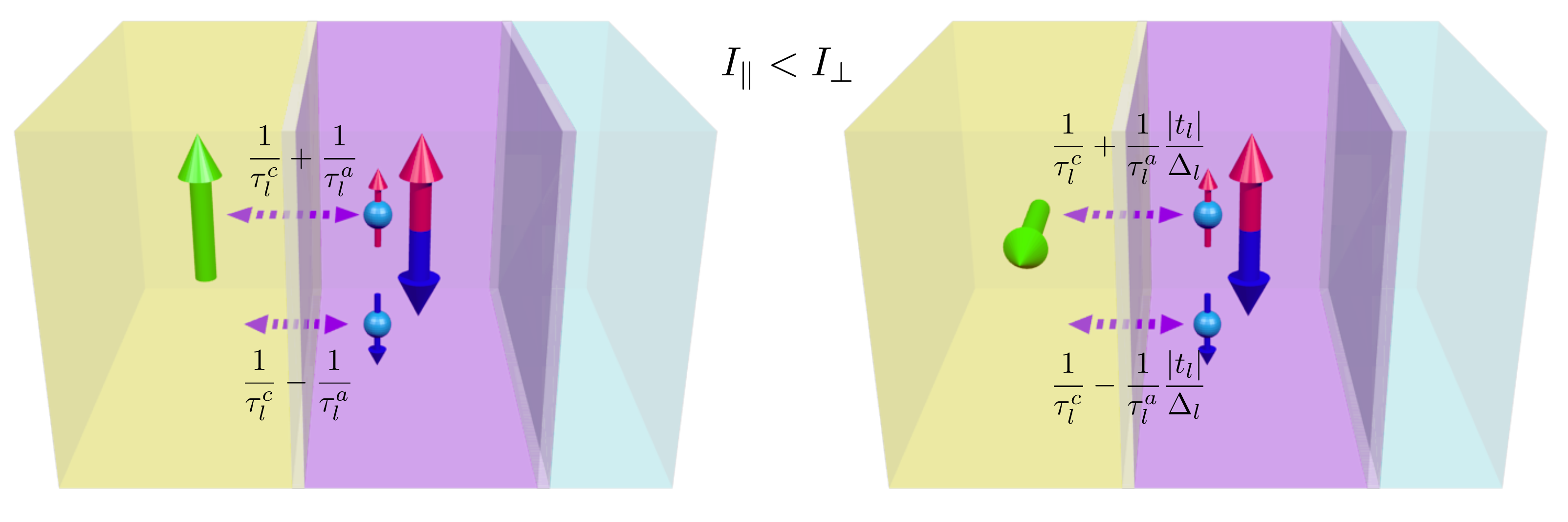}
\caption{Mechanism of the magnetoresistance. In the perturbation in the influence of F on AF $\sigma _l^R /\Delta _l $, $1/\tau _{l\pm }^F$ can be considered to be the rate of particle exchange between F and AF for up and down electrons respectively. At the leading order, the charge current is proportional to $1/\tau _{l+}^F + 1/\tau _{l-}^F $ and independent of $\theta $. The second order effect is a backflow proportional to $-1/(\tau _{l+}^F )^2 - 1/(\tau _{l-}^F )^2$. The sum of the squares is $\theta $-dependent and it is greatest for the collinear configuration (Left) and smallest for the perpendicular configuration (Right). The backflow is accordingly strongest for the collinear configuration.}
\label{fig:resistance}
\end{center}
\end{figure}

The monotonic increase of the current as a function of $\sin ^2 \theta $ arises from the fact that the angular dependent splitting of the escape rates is symmetric between the $\pm $ states (Fig. \ref{fig:resistance}).
Therefore, when they are added at the linear order, the angular dependence disappears: $1/\tau _{l+}^F + 1/\tau _{l-}^F = 2/\tau ^F_l $. When they are added after being squared, it always increases as the result of a simple algebraic identity; $1/(\tau _{l+}^F )^2 + 1/(\tau _{l-}^F )^2 = 2/(\tau ^F_l ) ^2 + 2 c_l (\theta )^2 / (\tau _l^a )^2 > 2/(\tau ^F_l )^2 $. The amount of the increase is proportional to the square modulus of the split $c_l (\theta ) /\tau _l^a $, and the split is clearly smallest when the ferromagnet is perpendicular to the antiferromagnet. 

Finally, one can rephrase the result in terms of the conductance in the linear response regime by expanding $n_F -n _N $ in terms of $\mu _F -\mu _N$;
\begin{eqnarray}
G &=& \frac{2e^2 }{h} \sideset{}{'}\sum _l \frac{1}{\tau _l^N} \left( A_{{\rm t}l} + A_{{\rm b}l}\right) \nonumber \\
&& \times \left\{ 1 - \frac{1}{\tau _l \tau _l^N} \left[ \frac{1}{\tau _l^2 } -\left( \frac{c_l \left( \theta \right) }{\tau _l^a} \right) ^2 \right] ^{-1} \right\} ,
\end{eqnarray}
where we have restored $\hbar $ and $\sum\nolimits'$ denotes summation over states on the Fermi surface. 

\subsection{Input spin current}

\begin{figure}[tbp]
\begin{center}
\includegraphics[width=1\linewidth]{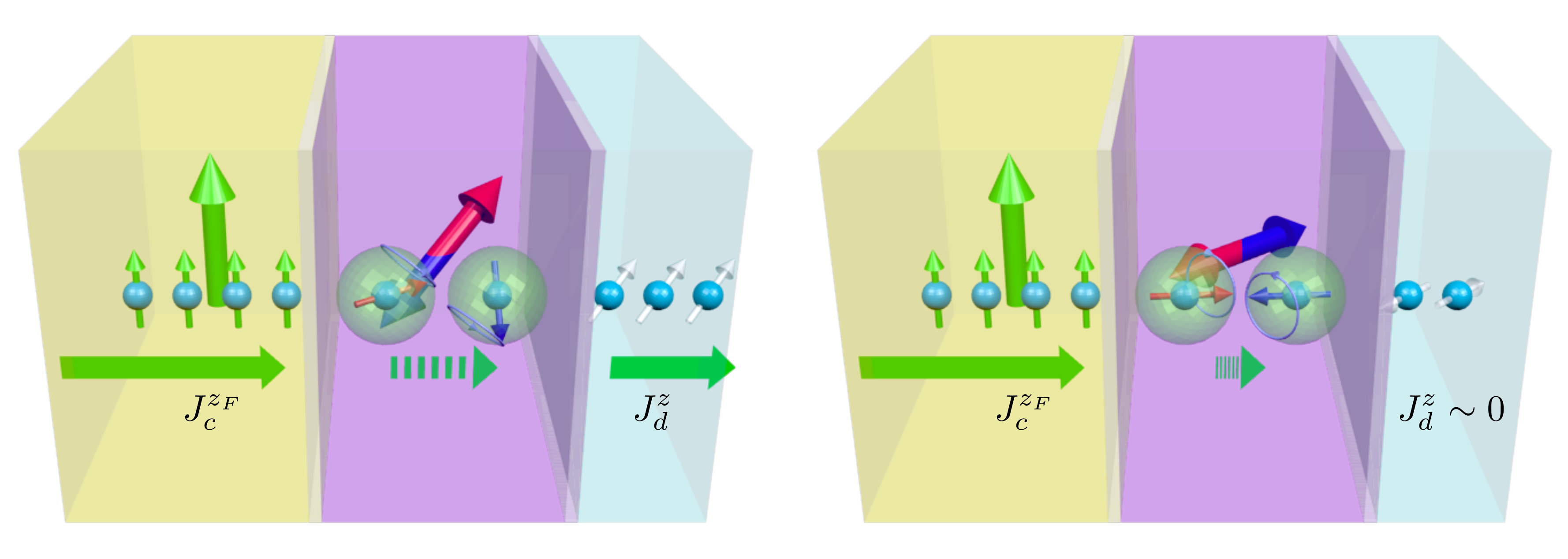}
\caption{Dependence of the spin current transmission on the mutual orientation of the incoming spin current polarization and the N\'eel vector. Left: Operation as a spin current polarizer. When $\hat{\bm{z}}_F \cdot \bm{N} \neq 0$, the injected spin current $J^{z_F}_F $ is projected onto the direction of the N\'eel vector $\hat{\bm{z}}$. The transverse ($x,y$) components are projected out and the outgoing spin current is mainly polarized along the N\'eel vector; $\bm{J}_N \approx J^z_N \hat{\bm{z}}$. Right: Operation as a spin current filter. When $\hat{\bm{z}}_F $ and $\bm{N}$ are perpendicular, the output spin current is effectively zero. This is a direct consequence of transverse spin absorption in AF by the dephasing mechanism.}
\label{fig:spin_transmission}
\end{center}
\end{figure}

Next we turn our attention to the spin current transmission. The presence of the spin transfer torque is due to the efficient absorption of the transverse spin current by AF. In contrast, the longitudinal component of spin should be conserved inside AF, at the leading order approximation in the tunneling, resulting in unhindered transmission of the longitudinal spin current. Therefore, AF is expected to act as an effective spin current polarizer (Fig.~\ref{fig:spin_transmission}). 
Although ferromagnets as well possess the above characteristics, they act as a source of spin current by themselves, with which the spin polarizing effect is largely washed out. In addition to the potential for applications as a spin valve, this property may serve as a method for detecting the orientation of the N\'eel vector by measuring the spin current transmission. Here we compute both the spin current injected at F/AF interface and the transmission into N. Note that since our non-perturbative treatment of the coupling to F means no component of spin is conserved in AF, it does not make much sense to talk about spin current inside AF. 

We start with the spin current at F/AF interface. In F, the $z$ component of spin current (in the ferromagnetic frame, namely spin along $\hat{\bm{z}}_F $) is conserved and given by
\begin{eqnarray}
J^{z_F}_F &=& \int \frac{d\epsilon }{2\pi } \sum _l \left( n_F -n_N \right) \left[ \frac{1}{\tau _l^N} + \frac{1}{\tau _l^F} \left( \frac{\Delta _{\rm ex}}{\Delta _l} \right) ^2 \sin ^2 \theta \right] \nonumber \\
&& \times \frac{1}{\tau _l^N \tau _l^a} \left[ \frac{1}{\tau _l^2} -\left( \frac{c_l \left( \theta \right) }{\tau _l^a} \right) ^2 \right] ^{-1} \left( A_{{\rm t}l} + A_{{\rm b}l} \right) . \label{eq:spin_zF}
\end{eqnarray}
Its angular dependence is qualitatively similar to that of the charge current $I_N $, namely taking maximum at $\theta = \pi /2 $. This is reasonable as $I_F = - I_N $ and $J^{z_F}_N $ coincide in the limit of a fully spin-polarized ferromagnetic lead. 

\subsection{Output spin current}
The spin in N is fully conserved and all the components of spin current are well-defined. In the antiferromagnetic frame, the individual components read
\begin{eqnarray}
J^x_N &=& \int \frac{d\epsilon }{2\pi } \sum _l \left( n_F -n_N \right) \frac{\sin \theta }{\tau _l^a}  \frac{\left| t_l \right| ^2}{\Delta _l^2}  \\
&& \times \left( \frac{1}{\tau _l^N} \right) ^2 \left[ \frac{1}{\tau _l^2 } -\left( \frac{c_l \left( \theta \right) }{\tau _l^a} \right) ^2 \right] ^{-1} \left( A_{{\rm t}l} + A_{{\rm b}l} \right) ,\nonumber \\
J^y_N &=& \int \frac{d\epsilon }{2\pi } \sum _l \left( n_F -n_N \right) \frac{\sin \theta }{\tau _l^a} \frac{\cos \theta }{2\tau _l^a \Delta _l} \left( \frac{\Delta _{\rm ex}}{\Delta _l} \right) ^2 \label{eq:spin_y}  \\
&& \times \left( \frac{1}{\tau _l^N} \right) ^2 \left[ \frac{1}{\tau _l^2} -\left( \frac{c_l \left( \theta \right) }{\tau _l^a} \right) ^2 \right] ^{-1} \left( A_{{\rm t}l} - A_{{\rm b}l} \right) , \nonumber \\
J^z_N &=& \int \frac{d\epsilon }{2\pi } \sum _l \left( n_F -n_N \right) \frac{\cos \theta }{\tau _l^a}  \label{eq:spin_z} \\
&& \times \left( \frac{1}{\tau _l^N} \right) ^2 \left[ \frac{1}{\tau _l^2} - \left( \frac{c_l \left( \theta \right) }{\tau _l^a} \right) ^2 \right] ^{-1} \left( A_{{\rm t}l} + A_{{\rm b}l} \right) .\nonumber 
\end{eqnarray}
For AF to work as a spin polarizer, it is first of all necessary that the output spin current is mainly polarized along the N\'eel vector, i.e. $\left| J^z_N \right| \gg \left| J^{x,y}_N \right| $. Comparing (\ref{eq:spin_y}) and (\ref{eq:spin_z}), it is clear that $\left| J^z_N \right| \gg \left| J^y_N \right| $ holds generally as the latter is suppressed by $\sin \theta / 2\tau _l^a \Delta _l $. This is reasonable since the input spin current is polarized along $\hat{\bm{z}}_F = \hat{\bm{z}}\cos \theta - \hat{\bm{x}}\sin \theta $ and does not have a finite $y$ component. The nonzero value of the output $y$ component $\left| J^y_N \right| $ is therefore due to the nonconservation of spin inside AF. $\left| J^y_N \right| $ is also expected to be closely related to the expectation value of electron spin $y$ component in AF. Indeed, the unconventional damping-like torque $\Gamma ^n_{\rm dl} $ given in (\ref{eq:torquend2}) and $\left| J^y_N \right| $ given in (\ref{eq:spin_y}) are very similar, indicating their common origin. More precisely, one can write $2J^y_N = {\rm tr}\left[ \sigma _2 \otimes 1_{SL} G^<_a W_N^{\dagger }\left( G^R_N -G^A_N \right) W_N \right] $ while $i \Gamma ^n_{\rm dl} \sin \theta = \Delta _{\rm ex} {\rm tr}\left[ \sigma _2 \otimes 1_{SL} G^<_a \right] $ (c.f. Eq. (\ref{eq:torquend})). Hence both $\Gamma ^n_{\rm dl} $ and $J^y_N $ arise from the sublattice asymmetric part of the dephasing process.

The comparison between $J^x_N $ and $J^z_N $ depends on the strength of the intersublattice overlap $t_l$. When $\left| t_l \right| \ll \Delta _{\rm ex}$, we have $\left| J^x_N \right| \ll \left| J^z_N \right| $ and AF acts as a good spin current polarizer. The fact that $J^x_N $ is proportional to $\left| t_l \right| ^2 $ can be understood by recalling the discussion on the origin of the field-like torque $\Gamma ^n_{\rm fl} $. In the absence of the intersublattice overlap $t_l$, all the electrons that enter AF undergo the dephasing process and lose their transverse spin, i.e. the $x$ component. By turning on the overlap $t_l$, electrons can avoid the dephasing by hopping onto the other sublattice, generating a finite expectation value for the $x$ component proportional to $\left| t_l \right| ^2$, which is essentially the field-like torque $i \Gamma ^n_{\rm fl} \sin \theta = \Delta _{\rm ex}{\rm tr}\left[ \sigma _1 \otimes 1_{SL} G^<_a \right] $. The finite spin $x$ component escapes towards the normal metal lead at the rate $1/\tau _l^N \propto i W_N^{\dagger }\left( G^R_N -G^A_N \right) W_N $, resulting in the $x$ component of spin current $2J^x_N \sim {\rm tr}\left[ \sigma _1 \otimes 1_{SL} G^<_a W_N^{\dagger }\left( G^R_N -G^A_N \right) W_N \right] $ (this relation holds approximately by neglecting subdominant terms). 

While we have so far clarified that $J^x_N $ and $J^y_N$ share the same physical origins as the torques $\Gamma ^n_{\rm fl} $ and $\Gamma ^n_{\rm dl} $ respectively, the $z$ component of spin current $J^z_N $ does not appear in the discussions of spin torque. Rather, it is determined by the conservation of the longitudinal spin inside AF. If the longitudinal component, i.e. the $z$ component, is fully conserved, the output $z$ component $J^z_N $ should be the projection of the input $J^{z_F}_F $ onto the $z$ axis; $J^z_N = J^{z_F}_F \cos \theta $. The general relation inferred from Eqs. (\ref{eq:spin_zF}) and (\ref{eq:spin_z}) is instead an inequality $\left| J^z_N  \right| \leq \left| J^{z_F}_F \cos \theta \right| $ where the equality is satisfied for $\theta  = 0 ,\pi $. It is again reasonable since when $\sin \theta \neq 0$, the longitudinal spin conservation inside AF is broken by the influence of F and therefore the output should be less than the input. 

In the context of the spin current polarizer, it is useful to look at the same results in the ferromagnetic frame. Taking appropriate linear combinations of $J^x_N $ and $J^z_N $ yields
\begin{eqnarray}
J^{x_F }_N &=& \int \frac{d\epsilon }{2\pi } \sum _l \left( n_F -n_N \right) \frac{\cos \theta \sin \theta }{\tau _l^a} \left( \frac{\Delta _{\rm ex}}{\Delta _l} \right) ^2 \\
&& \times \left( \frac{1}{\tau _l^N} \right) ^2 \left[ \frac{1}{\tau _l^2} - \left( \frac{c_l \left( \theta \right) }{\tau _l^a} \right) ^2 \right] ^{-1} \left( A_{{\rm t}l} + A_{{\rm b}l} \right) ,\nonumber  \\
J^{z_F}_N &=& \int \frac{d\epsilon }{2\pi } \sum _l \left( n_F -n_N \right) \frac{1}{\tau _l^a}  c_l \left( \theta \right) ^2  \\
&& \times \left( \frac{1}{\tau _l^N} \right) ^2 \left[ \frac{1}{\tau _l^2} - \left( \frac{c_l \left( \theta \right) }{\tau _l^a} \right) ^2 \right] ^{-1} \left( A_{{\rm t}l}+A_{{\rm b}l} \right) .\nonumber 
\end{eqnarray}
One can make an estimate of the spin transmission of the $z_F $ component;
\begin{eqnarray}
\frac{J^{z_F }_N}{J^{z_F }_F} &\approx &\left[ 1-\left( \frac{\Delta _{\rm ex}}{\Delta _l}\right) ^2 \sin ^2 \theta \right] \nonumber \\
&& \times  \left[ 1+\frac{\tau _l^N}{\tau _l^F}\left( \frac{\Delta _{\rm ex}}{\Delta _l} \right) ^2 \sin ^2 \theta \right] ^{-1} .
\end{eqnarray}
The contrast is maximized for $t_l =0$ and $1/\tau _l^N \ll 1/\tau _l^F $. Regardless of the relative strengths between the two tunneling barriers, however, one should expect a contrast in spin transmission of order unity unless the exchange splitting $\Delta _{\rm ex} $ is much smaller than the intersublattice overlap $\left| t_l \right| $. From our discussions above, the origin of the angular dependent spin transmission is the same as for Slonczewski's damping-like torque; namely the absorption of the transverse spin via dephasing. Therefore, we expect that the spin polarizer effect of antiferromagnets is as robust as the conventional spin transfer torque. 

\section{Discussions}\label{sec:discussion}

We have theoretically studied a model F/AF/N junction without spin-orbit coupling and systematically derived microscopic expressions for spin transfer torques as well as charge and spin currents in response to a voltage or temperature bias. The results for the torques are discussed in Sec.~\ref{sec:torque} and charge conductance and spin transmission in Sec.~\ref{sec:current}. 

The results strongly support the validity of two commonly made assumptions, the dominance of Slonczewski's damping-like torque and the $\cos ^2 \theta $ dependence of the magnetoresistance. At the same time, we have found several additional features that can be of experimental interest. The intersublattice overlap of antiferromagnetic electron wave functions opens up a new channel of electron transport and results in a contribution to the non-staggered field-like torque that depends sensitively on the overlap and relaxation rates. Higher-order tunneling processes into F generate additional non-staggered damping-like and staggered field-like torques. The magnitude of the non-staggered damping-like torque can become a sizable fraction of Slonczewski's staggered damping-like torque and causes the threshold current for destabilizing the ground state to depend on the direction of bias voltage. The staggered field-like torque, also called N\'eel field-like torque, appears always small in the present model, suggesting that it is difficult to generate a sizable torque of this kind without spin-orbit interactions. The same physical mechanisms responsible for the torques lead to a strong dependence of the spin current transmission on the angle between the spin polarization and the N\'eel vector. In particular, unless the intersublattice overlap overwhelms the exchange splitting, one may expect a metallic bipartite antiferromagnet to act as a spin current polarizer. All of the mentioned features could potentially be used for experimentally probing electronic properties of antiferromagnetic nanostructures.

In comparing our results with experiments, however, we have to stress that this study is based on a specific model and comes with various limitations, some of which are implicit and not easily quantifiable. For instance, restrictions on macroscopic parameters such as temperature or the size of the antiferromagnetic island result from the neglect of electron-electron and electron-phonon interactions and the associated relaxation processes or from limiting the electron dwell time on the island. Clearly, slow relaxation requires low temperatures and short dwell times require small spatial dimensions. The question whether a given real system is well described by the model, however, is not easily answered. We also remark in this context that the antiferromagnetic Hamiltonian (\ref{eq:AFM_Hamiltonian}) is fairly restrictive due to the assumed sublattice symmetry. For disordered antiferromagnets, the symmetry should be implemented only on average, as we have done here for the tunneling matrices. 

It is useful to discuss the connection between the non-equilibrium Green's function approach used in this work and scattering theory methods that have been employed in similar contexts.\cite{Nunez2006,Cheng2014c} In Appendix \ref{sec:scattering}, we present the scattering matrix derived from our model and show that it yields strongly spin-dependent reflection and transmission coefficients. We have also included a more elementary scattering model of an antiferromagnet to illustrate the spin dependance. These two results indicate that transverse spin is generically altered upon transmission through antiferromagnets while the longitudinal spin is conserved. This spin-dependent transmission underlines the potential of antiferromagnets for application as spin current polarizers. 

The formalism presented in this work can be applied to other multilayer models such as F/F/N and AF/AF/N junctions. These are discussed in Appendix \ref{sec:AFMlead}. In particular, we demonstrate that for the model AF/AF/N junction neither spin torques nor spin-dependent transport occur. This conclusion may appear at odds with the studies of current-induced torques in slowly varying antiferromagnetic textures.\cite{Yamane2016a} We clarify the difference between the two approaches in the appendix and argue that there is no contradiction.

Our study may be extended in several different ways. Firstly, knowledge of the non-equilibrium Green's function allows us to compute damping constants and spin current fluctuations induced by the tunneling processes. They play crucial roles in determining dynamics of stochastic magnetic switching. Secondly, it would also be important to develop a theoretical framework that can handle non-equilibrium transport through multilayer structures in the presence of spin-orbit interaction. Finally, a similar theoretical approach can be applied to structures involving insulating magnets.\cite{Zheng2017a} Antiferromagnetism is more common in insulating metal oxides and how torques and spin transport depend on microscopic parameters in insulator nanostructures remains to be explored. 

\begin{acknowledgements}
The authors would like to thank Gerrit Bauer, Dazhi Hou, Jun'ichi Ieda, Hiroshi Kohno, Claudia and Tim Mewes, and Yuta Yamane for discussions and helpful comments. This work was supported by the College of Arts and Sciences at the University of Alabama (K.~Y. and G.~S.), the Alexander von Humboldt Foundation, the ERC Synergy Grant SC2 (No.~610115), the Transregional Collaborative Research Center (SFB/TRR) 173 SPIN+X and Grant Agency of Czech Republic grant No.~14-37427, the DAAD project "MaHoJeRo." (K.~Y.), the EU FET Open RIA Grant No.~766566 (H.~G. and J.~S.) and the DFG project SHARP 397322108 (H~.G.).

\end{acknowledgements}

\appendix

\section{The formalism}\label{sec:formalism}

Starting from the Hamiltonian (\ref{eq:total_Hamiltonian}), we study a non-equilibrium stationary state of AF by employing the Keldysh Green's function technique. As indicated in Fig. \ref{fig:setup}, the non-equilibrium state is driven by the difference in the equilibrium distribution functions $n_{F,N}\left( \epsilon \right) $ of the two leads. We assume that the system is completely relaxed and macroscopic observables are independent of time. Our approach to the problem is based on the formalism developed by Kamenev and his collaborators \cite{Chudnovskiy2008,Swiebodzinski2010,Kamenev2011} for the description of ferromagnetic tunneling junctions. Their two-layer model consists of a ferromagnetic free layer coupled to a single ferromagnetic lead with fixed magnetization direction. In addition to the relaxation into the lead, fast relaxation on the free layer is assumed without specifying the mechanism. Here, we have explicitly included N as the source of relaxation. This allows us to handle the regime where the two relaxation rates are comparable. Our results reduce to those obtained in a two-layer formalism in the limit where the relaxation rate into F is smaller than the other relaxation rate into N. The extension to three layers turns out to be necessary for describing three of the four types of torque $\Gamma ^n_{\rm dl}, \Gamma ^m_{\rm fl} ,\Gamma ^n_{\rm fl}$ as well as the magnetoresistance and the spin current transmission. Our formulation can also be viewed as a leading-order truncation of the non-equilibrium Born-Oppenheimer approximation employed, e.g., in Ref.~\onlinecite{Bode2012a}.

\subsection{Dyson's equation}
According to the Keldysh formalism,\cite{DiVentra2008} observables represented by electronic one-body operators follow from the lesser Green's function $G^<_{ij} = i\left\langle \psi _j^{\dagger } \psi _i \right\rangle $, where $\psi _{i,j}$ are any of $a_{l\sigma },b_{l\sigma },c^F_{k\sigma },c^N_{m\sigma }$ and the expectation value is taken over an arbitrary probability distribution of the quantum mechanical states. Loosely speaking, $G^<_{ij}$ can be considered as a quantum mechanical generalization of the classical distribution function in statistical mechanics, and can be determined by solving the Dyson's equation on the Keldysh time contour, which is translated into the Dyson's equation in the Keldysh matrix space on the real time contour. In our present model, the only time-dependent parameters in the electronic Hamiltonian are $\bm{S}_{A,B}$ and we assume that their dynamics is slow. We apply the Wigner transform in time and neglect all the terms that contain time derivatives (leading order non-equilibrium Born-Oppenheimer approximation). The resulting Dyson's equation reads
\begin{equation}
\hat{G} = \left[ \begin{pmatrix}
	\hat{G}_F^{-1} & 0 & 0\\
	0 & \hat{G}_0^{-1} & 0 \\
	0 & 0 & \hat{G}_N^{-1} \\
	\end{pmatrix} - \begin{pmatrix}
	0 & W_F  & 0\\
	W_F^{\dagger } & 0 & W_N^{\dagger } \\
	0 & W_N & 0 \\
	\end{pmatrix}  \right] ^{-1},  \label{eq:Dyson}
\end{equation}
where the hats indicate nontrivial matrix structures in the $2 \times 2$ Keldysh space and the three blocks are assigned to the ferromagnetic, antiferromagnetic and normal metal electrons respectively and labelled by $f,a,n$ as
\begin{equation*}
\hat{G} = \begin{pmatrix}
	\hat{G}_{f} & \hat{G}_{fa} & \hat{G}_{fn} \\
	\hat{G}_{af} & \hat{G}_{a} & \hat{G}_{an} \\
	\hat{G}_{nf} & \hat{G}_{na} & \hat{G}_{n} \\
	\end{pmatrix} .
\end{equation*}
Our convention is to use subscripts $F,0,N$ for the bare and $f,a,n$ for the corresponding full Green's functions. The bare Green's functions $\hat{G}_{0,F,N}$ have the common matrix structure in the Keldysh space
\begin{equation}
\hat{G}_{0,F,N} = \begin{pmatrix}
	G^R_{0,F,N} & G^<_{0,F,N}\\
	0 & G^A_{0,F,N} \\
	\end{pmatrix} , \label{eq:bare}
\end{equation}
where the retarded and advanced Green's functions $G^{R,A}_{0,F,N}$, which are matrices acting on the respective Hilbert space and contain the information about energy eigenstates, are given by
\begin{eqnarray*}
&&\left( G^R_0 \right) _{ll^{\prime }} = \delta _{ll^{\prime }} \left[ \epsilon + i\delta -1_{SP}\otimes \begin{pmatrix}
	\epsilon _l & \overline{t_l} \\
	t_l & \epsilon _l \\
	\end{pmatrix} +\Delta _{\rm ex} \sigma _3 \otimes \tau _3 \right] ^{-1} , \\
&&\left( G^R_F \right) _{kk^{\prime }} = \delta _{kk^{\prime }} \begin{pmatrix}
	\left( \epsilon - \epsilon ^F_{k\uparrow } +i\delta  \right) ^{-1} & 0 \\
	0 & \left( \epsilon - \epsilon ^F_{k\downarrow } +i\delta  \right) ^{-1} \\
	\end{pmatrix}_{SP} , \\
&& \left( G^R_N \right) _{mm^{\prime }} = \frac{\delta _{mm^{\prime }} 1_{SP} }{\epsilon - \epsilon ^N_m + i\delta } , \quad G^A_{0/F/N} = \left( G^R_{0/F/N} \right) ^{\dagger } ,
\end{eqnarray*} 
with $\delta >0 $ being infinitesimally small. One can formally expand the right-hand-side of Eq. (\ref{eq:Dyson}) in powers of $W_{F,N}$. Elementary algebra leads to
\begin{equation}
\hat{G}_a \equiv \begin{pmatrix}
	G_a^R & G_a^< \\
	0 & G_a^A \\
	\end{pmatrix} = \left( \hat{G}_0^{-1} - \hat{\Sigma }_F - \hat{\Sigma }_N \right) ^{-1} , \label{eq:Dyson_AFM}
\end{equation}
where $\hat{\Sigma }_{F,N}$ are the self-energies arising from the leads given by $\hat{\Sigma }_{F/N} = W_{F/N}^{\dagger } \hat{G}_{F/N} W_{F/N} $.
We then impose the conditions that enforce the non-equilibrium state of AF to be fully determined by the individual equilibrium states of F and N, which amounts to choosing
\begin{eqnarray}
G^<_{F/N} &=& -n_{F/N}\left( \epsilon \right) \left( G^R_{F/N} -G^A_{F/N} \right).  \label{eq:boundary_conditions1} 
\end{eqnarray}
We assume that intrinsic relaxation mechanisms inside AF are negligible so that the lesser component of $\hat{G}_0^{-1}$ is a pure regularization. The Fermi-Dirac distribution functions $n_{F,N}\left( \epsilon \right)$ drive the currents through the differences in chemical potential $\mu _F -\mu _N $ and temperature $T_F -T_N $. Eqs.~(\ref{eq:bare}) - (\ref{eq:boundary_conditions1}) fully determine $\hat{G}_a $ and are consistent if the influence of $W_{F,N}$ on the leads is negligible, in which case the remaining diagonal components of (\ref{eq:Dyson}) trivially read $\hat{G}_f =\hat{G}_F , \hat{G}_n =\hat{G}_N $. The off-diagonal components of (\ref{eq:Dyson}) are needed in deriving the Meir-Wingreen formulae (\ref{eq:current}) - (\ref{eq:spin_current}).

\subsection{Self-enegies}
The retarded components of the self-energies representing the tunneling Hamiltonians (\ref{eq:Hamiltonianc}) and (\ref{eq:Hamiltoniand}) are given by
\begin{eqnarray}
\Sigma ^R_F &=& \sum _k  \Bigg[ \frac{1}{2}\left( \frac{1}{\epsilon -\epsilon _{k\uparrow }+i\delta }+\frac{1}{\epsilon -\epsilon _{k\downarrow }+i\delta } \right) 1_{SP}  \\
&&  + \frac{1}{2}\left( \frac{1}{\epsilon -\epsilon ^F_{k\uparrow }+i\delta } -\frac{1}{\epsilon -\epsilon ^F_{k\downarrow }+i\delta } \right) S_{SP}\left( \theta \right) \Bigg]  \nonumber \\
&& \otimes  \begin{pmatrix}
	\overline{\left( W^A_F \right) _{kl}} \left( W^A_F \right) _{kl^{\prime }} & \overline{\left( W^A_F \right) _{kl}} \left( W^B_F \right) _{kl^{\prime }} \\
	\overline{\left( W^B_F \right) _{kl}} \left( W^A_F \right) _{kl^{\prime }} & \overline{\left( W^B_F \right) _{kl}} \left( W^B_F \right) _{kl^{\prime }} \\
	\end{pmatrix}_{SL} , \nonumber \\
\Sigma ^R_N &=& \frac{1_{SP}}{\epsilon -\epsilon ^N_m +i\delta }  \\
&& \otimes  \begin{pmatrix}
	\overline{\left( W^A_F \right) _{ml}} \left( W^A_F \right) _{ml^{\prime }} & \overline{\left( W^A_F \right) _{ml}} \left( W^B_F \right) _{ml^{\prime }} \\
	\overline{\left( W^B_F \right) _{ml}} \left( W^A_F \right) _{ml^{\prime }} & \overline{\left( W^B_F \right) _{ml}} \left( W^B_F \right) _{ml^{\prime }} \\
	\end{pmatrix}_{SL} . \nonumber
\end{eqnarray}
Solving Eq.~(\ref{eq:Dyson_AFM}) is a matter of matrix inversion. Yet the self-energies have off-diagonal elements between the bare orbital energy eigenstates of AF $l \neq l^{\prime }$, which obstructs making progress analytically. Since we are not interested in properties attributed to a particular realization of the tunneling matrices $W_{F,N}$, we assume that the matrix elements $( W^{A,B}_F ) _{kl} , ( W^{A,B}_N ) _{ml}$ for each fixed $l$ are randomly distributed across different lead channels labelled by $k$ and $m$. In order to enforce the sublattice symmetry on average, we further assume that the matrix elements for $A$ and $B$ sublattices have the same variance. Mathematically, these conditions can be written as
\begin{eqnarray}
\lim _{N_F \rightarrow \infty } \frac{1}{N_F} \sum _k \left( W^{\alpha }_F \right) _{kl}  &=& 0 , \\
\lim _{N_F \rightarrow \infty } \frac{1}{N_F}\sum _k \overline{\left( W^{\alpha }_F \right) _{kl}}\left( W^{\beta }_F \right) _{kl^{\prime }} &=& w_{F}^2  \delta _{\alpha \beta }  \delta _{ll^{\prime }}  , \label{eq:variance_F} \\
\lim _{N_N \rightarrow \infty } \frac{1}{N_N} \sum _m \left( W^{\beta }_N \right) _{ml} &=& 0  , \\
\lim _{N_N \rightarrow \infty }\frac{1}{N_N} \sum _m \overline{\left( W^{\alpha }_N \right) _{ml}}\left( W^{\beta }_N \right) _{ml^{\prime }} &=&  w_{N}^2\delta _{\alpha \beta }  \delta _{ll^{\prime }}  . \label{eq:variance_N}
\end{eqnarray}
Here $N_{F,N}$ denote the number of channels for the ferromagnetic and normal metal leads, $\alpha ,\beta \in \{ A,B \}$, and $ w_{F,N}^2 $ are the variances of the tunneling matrix elements. Note that we have also assumed that $W^A_{F/N}$ and $W^B_{F/N}$ are statistically independent, embodied in $\delta _{\alpha \beta }$ in (\ref{eq:variance_F}) and (\ref{eq:variance_N}). Although we could have let $w_{F,N}$ depend on $l$, here we keep it simple since these variances are used only for order-of-magnitude estimates. These assumptions allow us to ignore the off-diagonal components of $\hat{\Sigma }_{F,N}$ in the orbital eigenstate space ($ll^{\prime } $ indices) \cite{Ludwig2017}, as we demonstrate below. 

Using (\ref{eq:variance_F}) and (\ref{eq:variance_N}), one can estimate the summations over the channel indices $k,m$ for the diagonal components as ($\alpha = A,B$)
\begin{eqnarray}
\sum _{k} \frac{\left| \left( W^{\alpha }_F \right) _{kl} \right| ^2 }{\epsilon -\epsilon ^F_{k\uparrow ,\downarrow }+i\delta }  &\sim & -i N_F  D_{F\uparrow ,\downarrow } w_F^2  ,\\
\sum _m \frac{\left| \left( W^{\alpha }_N \right) _{ml} \right| ^2 }{\epsilon -\epsilon ^N_m +i\delta } & \sim & -i N_N  D_N  w_N^2 .
\end{eqnarray}
When $N_{F,N}$ are large, one can approximate the distributions by normal distributions with the same mean and variance, in which case one can also make the following estimates;
\begin{eqnarray}
&&\sum _k \frac{\overline{\left( W^{\alpha }_F \right) _{kl}} \left( W^{\alpha }_F \right) _{kl^{\prime }}}{\epsilon -\epsilon ^F_{k\uparrow ,\downarrow }+i\delta }  \sim \sqrt{N_F} D_{F\uparrow ,\downarrow } w_F^2  , \quad l\neq l^{\prime } , \\
&&\sum _m \frac{\overline{\left( W^{\alpha }_N \right) _{ml}} \left( W^{\alpha }_N \right) _{ml^{\prime }}}{\epsilon -\epsilon ^N_m +i\delta }  \sim  \sqrt{N_N} D_N w_N^2  , \quad l\neq l^{\prime } , \\
&& \sum _k \frac{\overline{\left( W_F^{\alpha }\right) _{kl} } \left( W^{\beta }_F \right) _{kl}}{\epsilon -\epsilon ^F_{k\uparrow ,\downarrow }+i\delta } \sim   \sqrt{N_F} D_{F\uparrow ,\downarrow }  w_F^2 ,\quad \alpha \neq \beta , \\
&& \sum _m \frac{\overline{\left( W^{\alpha }_N \right) _{ml}}\left( W^{\beta }_N \right) _{ml}}{\epsilon - \epsilon ^N_m + i\delta } \sim \sqrt{N_N}D_N w_n^2 , \quad \alpha \neq \beta , \\
&& \sum _k \frac{\left| \left( W^A_F \right) _{kl}  \right| ^2 - \left| \left( W^B_F \right) _{kl}\right| ^2 }{\epsilon -\epsilon ^F_{k\uparrow ,\downarrow }+i\delta } \sim \sqrt{N_F} D_{F\uparrow ,\downarrow } w_F^2 .
\end{eqnarray}
The latter three rely on the statistical independence of $W^A_F $ and $W^B_F$. They imply that all the off-diagonal components of $\Sigma ^R_{F,N}$ (except in the spin space) are suppressed by a factor of $1/\sqrt{N_{F,N}}$ compared to the diagonal components. The fact that these components can indeed be neglected in computing the torques and currents follows from a perturbative expansion analogous to the one presented in Ref.~\onlinecite{Ludwig2017} and we do not repeat it here. Thus dropping the orbital and sublattice off-diagonal terms and the real parts proportional to $1_{SP} \otimes 1_{SL}$ yields Eqs. (\ref{eq:self_c}) and (\ref{eq:self_d}).

The off-diagonal terms being ignored, the resulting self-energies can be written as
\begin{align}
\hat{\Sigma }_{F/N} &= \begin{pmatrix}
	\Sigma _{F/N}^R & - n_{F/N}\left( \epsilon \right) \left( \Sigma ^R_{F/N} - \Sigma ^A_{F/N} \right) \\
	0 & \Sigma _{F/N}^A \\
	\end{pmatrix} , \\
\left( \Sigma ^R_F \right) _{ll^{\prime }} &=\delta _{ll^{\prime }}\left[  -\frac{i}{2\tau _l^F} 1_{SP} \otimes 1_{SL} + \sigma _l^R S_{SP}\left( \theta \right) \otimes 1_{SL} \right] , \label{eq:self_c}  \\
\left( \Sigma ^R_N \right) _{ll^{\prime }} &= -\delta _{ll^{\prime }} \frac{i}{2\tau _l^N} 1_{SP} \otimes 1_{SL} , \quad \Sigma _{F/N}^A = \left( \Sigma ^R_{F/N}\right) ^{\dagger }. \label{eq:self_d}
\end{align} 
The real parts proportional to $1_{SP} \otimes 1_{SL}$ have been dropped since they only renormalize $\epsilon _l$. The imaginary parts set the characteristic time scales for the tunneling processes that are given by Eqs.~(\ref{eq:relax_N}) and (\ref{eq:relax_F}). As we explained in Sec.~\ref{sec:relax}, these relaxation times largely determine charge currents in the weak tunneling regime.

The angular dependence of tunneling originates from the ferromagnetic spin operator in the antiferromagnetic frame
\begin{equation}
S_{SP}\left( \theta \right) = R^{\dagger }_{SP} \sigma _3 R_{SP} = \begin{pmatrix}
	\cos \theta & -\sin \theta \\
	-\sin \theta & -\cos \theta \\
	\end{pmatrix} .
\end{equation}
Its coefficient $\sigma _l^R $ represents the spin-dependent part of the self-energy;
\begin{eqnarray}
\sigma _l^R &=& \sum _k  \frac{ \left| \left( W^A_F \right) _{kl} \right| ^2 + \left| \left( W^B_F \right) _{kl} \right| ^2 }{2} \nonumber \\
&& \times   \frac{1}{2} \left( \frac{1}{\epsilon -\epsilon ^F_{k\uparrow }+i\delta } - \frac{1}{\epsilon -\epsilon ^F_{k\downarrow } +i\delta } \right) . \label{eq:tau_a}
\end{eqnarray}
The real part of $\sigma _l^R $, denoted $\delta _l $, is essentially the energy split due to the sublattice symmetry breaking while the imaginary part, denoted $-1/2\tau _l^a $, is the difference in the rate of escape into the ferromagnet between the $\pm $ bands split by $\delta _l $.

 \subsection{Retarded Green's function}\label{sec:Green}
 
Fortunately, the matrix inversion in the remaining $4 \times 4$ structures can be carried out analytically. The dressed retarded Green's function of the antiferromagnet now reads
\begin{eqnarray}
\left( G_a^R \right) _{ll^{\prime }}^{-1} &=& \delta _{ll^{\prime }} \Bigg[ \epsilon + \frac{i}{2\tau _l} -\epsilon _l - 1_{SP} \otimes \begin{pmatrix}
	0 & \overline{t_l} \\
	t_l & 0 \\
	\end{pmatrix} _{SL} \nonumber \\
&& + \Delta _{\rm ex} \sigma _3 \otimes \tau _3 - \sigma _l^R S_{SP}\left( \theta \right) \otimes 1_{SL} \Bigg] .
\end{eqnarray}
To carry out the matrix inversion, we first note that the bare part $G^R_0 $ is diagonalized by a unitary matrix chosen to be
\begin{eqnarray}
U_l &=& \begin{pmatrix}
	\cos \left( \theta _l /2 \right) & 0  \\
	0 & \sin \left( \theta _l /2 \right) \\
	\end{pmatrix}_{SP} \otimes e^{-i\phi _l \tau _3 /2}   \\
&& -i \begin{pmatrix}
	\sin \left( \theta _l /2\right) & 0 \\
	0 & \cos \left( \theta _l /2 \right) \\
	\end{pmatrix} _{SP} \otimes \tau _2 e^{i\phi _l \tau _3 /2} ,\nonumber 
\end{eqnarray}
where
\begin{equation}
\cos \theta _l = \frac{\Delta _{\rm ex}}{\Delta _l} , \quad e^{i\phi _l}\sin \theta _l = -\frac{t_l}{\Delta _l} .
\end{equation}
Thus one can write
\begin{align}
&\left( G_a^R  \right) _{ll^{\prime }}=\\
& \quad\delta _{ll^{\prime }}U_l \left[ \epsilon +\frac{i}{2\tau _l} -\epsilon _l + \Delta _l 1_{SP}\otimes \tau _3 -\sigma _l^R L_l \left( \theta \right) \right] ^{-1} U_l^{\dagger } \nonumber
\end{align}
with
\begin{eqnarray}
L_l \left( \theta \right) &=&  U_l^{\dagger } \left[ S_{SP}\left( \theta \right) \otimes 1_{SL} \right] U_l \nonumber \\
&=& \left( \sigma _3 \cos \theta - \sigma _1 s_l \right) \otimes 1_{SL} - \sigma _2 \otimes \tau _2 \sin \theta \cos \theta _l \nonumber \\
&=& \left[ e^{i\chi _l \sigma _2 /2}\otimes 1_{SL} \right] \left( \sigma _3 \otimes 1_{SL}c_l - \sigma _2 \otimes \tau _2 s_l \right)  \nonumber \\
&& \left[ e^{-i\chi _l \sigma _2 /2}\otimes 1_{SL} \right] .
\end{eqnarray}
We have introduced $s_l = \sin \theta \cos \theta _l , c_l =\sqrt{1-s_l^2} $ and
\begin{equation}
\cos \chi _l = \frac{\cos \theta }{c_l} , \quad \sin \chi _l = \frac{\sin \theta \sin \theta _l}{c_l} .
\end{equation}
Hence one can write
\begin{equation}
\left( G_a^R \right) _{ll^{\prime }} = \delta _{ll^{\prime }} V_l \left( \begin{array}{cc}
	G_{l+}^R & 0 \\
	0 & G^R_{l-} \\
	\end{array} \right) _{SL} V_l^{\dagger } \label{eq:retarded}
\end{equation}
where
\begin{equation}
V_l = U_l \left[ e^{i\chi _l \sigma _2 /2 }\otimes 1_{SL} \right] \begin{pmatrix}
	0 & 0 & 1 & 0 \\
	1 & 0 & 0 & 0 \\
	0 & 1 & 0 & 0 \\
	0 & 0 & 0 & 1\\
	\end{pmatrix} , \\
\end{equation}
and the $2 \times 2 $ blocks are given by
\begin{eqnarray}
G^R_{l\pm } &=& \left( \epsilon + \frac{i}{2\tau _l}-\epsilon _l - h_{l\pm }^R \right) ^{-1} , \\
h_{l\pm }^R &=&   -\begin{pmatrix}
	\Delta _l \pm \sigma _l^R c_l & \pm \sigma _l^R s_l \\
	\pm \sigma ^R_l s_l & -\Delta _l \mp \sigma _l^R c_l \\
	\end{pmatrix}_{SP} .
\end{eqnarray}
The blocks $G^R_{l\pm }$ belong to two pairs of bands in Fig. \ref{fig:band}(c). The topmost and bottommost bands form one pair, denoted as ($+$), and the two bands in the middle form the other pair, denoted as ($-$). When the $2 \times 2$ matrices $G^R_{l\pm }$ are diagonalized, which takes a similarity transformation by a non-unitary matrix, the diagonal elements read
\begin{eqnarray}
g^R_{{\rm t}l\pm } &=& \left( \epsilon + \frac{i}{2\tau _l } -\epsilon _l - e_{l\pm }^R \right) ^{-1} , \\
g^R_{{\rm b}l\pm } &=& \left( \epsilon +\frac{i}{2\tau _l} -\epsilon _l  +e_{l\pm }^R \right) ^{-1} ,
\end{eqnarray}
where $1/\tau _l = 1/\tau _l^F + 1/\tau _l^N $ and
\begin{eqnarray}
e_{l\pm }^R &= &\sqrt{\Delta _l^2 + \left( \sigma _l^R \right) ^2 \pm 2\Delta _l \sigma _l^R c_l \left( \theta \right) } ,\\
c_l \left( \theta \right) &=& \sqrt{1-\left( \frac{\Delta _{\rm ex}}{\Delta _l} \right) ^2 \sin ^2 \theta } .
\end{eqnarray}
The principal branch (positive real part) is taken for square roots throughout. Note that $c_l \left( \theta \right) = \left| \cos \theta \right| $ when $t_l =0$. We are mainly concerned with the regime $\left| \sigma _l^R \right| \ll \Delta _l $ where a perturbative picture in terms of the isolated antiferromagnet makes sense. To the leading order in $\sigma _l /\Delta _l $, one obtains
\begin{eqnarray}
g^R_{{\rm t}l\pm} &=& \left( \epsilon - \epsilon _l - \Delta _l \mp \delta _l c_l \left( \theta \right) + \frac{i}{2\tau _l^{\pm }} \right) ^{-1} , \\
g^R_{{\rm b}l\pm } &=& \left( \epsilon -\epsilon _l + \Delta _l \pm \delta _l c_l \left( \theta \right) + \frac{i}{2\tau _l^{\mp }} \right) ^{-1} ,
\end{eqnarray}
where
\begin{equation}
\frac{1}{\tau _l^{\pm }} = \frac{1}{\tau _l^N} + \frac{1}{\tau _l^F} \pm \frac{1}{\tau _l^a} c_l \left( \theta \right) . \label{eq:escape_times}
\end{equation}
Note that $1/\tau _l^{\pm }$ are positive since $1/\tau _l^F \pm 1/\tau _l^a \geq 0$. In this approximation, ignoring $\epsilon $ dependence of $\delta _l , 1/\tau _l^{\pm }$, one obtains the energy dispersion $\epsilon _{{\rm t}l\pm } = \epsilon _l + \Delta _l \pm \delta _l c_l \left( \theta \right) , \epsilon _{{\rm b}l\pm } = \epsilon _l -\Delta _l \mp \delta _l c_l \left( \theta \right) $, and the relaxation rates due to the tunneling processes are given by $1/\tau _l^+$ for $( {\rm t},+ )  , ({\rm b},- )$ bands and $1/\tau _l^- $ for $({\rm t},-) ,({\rm b},+)$. In terms of $g^R_{{\rm b,t}l\pm }$, the blocks $G^R_{l\pm }$ are expressed as
\begin{align}
G^R_{l\pm } &= \frac{g^R_{{\rm t}l \pm }+g^R_{{\rm b}l\pm }}{2} - \frac{g^R_{{\rm t}l\pm } -g^R_{{\rm b}l\pm }}{2} \left( C_{l\pm }^R \sigma _3 + S^R_{l\pm }\sigma _1 \right) , \label{eq:retarded_block} \\
C_{l\pm }^R &= \frac{\Delta _l }{e_{l\pm }^R}  \pm \frac{\sigma _l^R}{e_{l\pm }^R} c_l \left( \theta \right)  , \quad  S_{l\pm }^R = \pm \frac{\sigma _l^R }{e_{l\pm }^R} \frac{\Delta _{\rm ex}}{\Delta _l} \sin \theta .
\end{align}

\subsection{Lesser Green's function}

While the energy spectrum is fully described by the retarded component, non-equilibrium problems concern dynamical occupation of those energy states. This information is contained in the lesser component of Eq. (\ref{eq:Dyson_AFM}), which becomes
\begin{equation}
G_a^< = - G_a^R \left[ n_F \left( \Sigma ^R_F -\Sigma _F^A \right) + n_N \left( \Sigma _N^R - \Sigma _N^A \right) \right] G_a^A . \label{eq:lesser_def} 
\end{equation}
To simplify the expression, we use the identity
\begin{equation}
\Sigma ^R_F - \Sigma ^A_F = \left( G^A_a \right) ^{-1} - \left( G^R_a \right) ^{-1} - \left( \Sigma ^R_N - \Sigma ^A_N \right) .
\end{equation}
Noting that $\Sigma ^R_N - \Sigma ^A_N $ and $G^R_a $ commute (since the orbital off-diagonal terms have been discarded), one obtains
\begin{equation}
G^<_a =  \left( n_F -n_N \right) \left( \Sigma ^R_N - \Sigma ^A_N \right) G^R_a G^A_a  -n_F \left( G^R_a -G^A_a \right) . \label{eq:lesser}
\end{equation}
When $n_F =n_N $ and hence there is no driving force, $G^<$ reduces to the equilibrium form $-n_F \left( G_a^R - G_a^A \right) $. The first term proportional to $n_F -n_N$ can be further reduced by noting
\begin{equation}
\left( G^R_a G^A_a \right) _{ll^{\prime }}  = \delta _{ll^{\prime }} V_l  \begin{pmatrix}
	G^R_{l+ }G^A_{+ } & 0 \\
	0 & G^R_{l-} G^A_{l-} \\
	\end{pmatrix}_{SL} V_l^{\dagger } ,
\end{equation}
and $G^R_{l\pm }G^A_{l\pm } = \sum _{\mu =0}^3 \mathcal{G}_{\mu l\pm }\sigma _{\mu }$ with $\sigma _0 \equiv 1_{SP}$ where
\begin{align}
\mathcal{G}_{0l\pm } =&  \frac{1+\left| C_{l\pm }^R \right| ^2 + \left| S_{l\pm }^R \right| ^2 }{4i\tau _l^N} \left( \left| g_{{\rm t}l\pm }^R \right| ^2  + \left| g^R_{{\rm b}l\pm } \right| ^2 \right) \label{eq:lesser0}  \\
& + \frac{1-\left| C_{l\pm }^R \right| ^2 -\left| S_{l\pm }^R \right| ^2 }{2i\tau _l^N } \Re \left( g^R_{{\rm t}l\pm } g^A_{{\rm b}l\pm } \right) , \nonumber \\
\mathcal{G}_{1l\pm }=& \frac{i}{2\tau _l^N} \Re \left[ S_{l\pm }^R  \left( g^R_{{\rm t}l\pm }-g^R_{{\rm b}l\pm } \right) \left( g^A_{{\rm t}l\pm }+g^A_{{\rm b}l\pm } \right)  \right] , \\
\mathcal{G}_{2l\pm } =& \frac{i}{2\tau _l^N} \Im \left( C_{l\pm }^R S_{l\pm }^A \right) \left| g^R_{{\rm t}l\pm }-g^R_{{\rm b}l\pm } \right| ^2  , \\
\mathcal{G}_{3l\pm } =& \frac{i}{2\tau _l^N} \Re \left[ C_{l\pm }^R \left( g^R_{{\rm t}l\pm } -g^R_{{\rm b}l\pm } \right) \left( g^A_{{\rm t}l\pm } + g^A_{{\rm b}l\pm } \right) \right] . \label{eq:lesser3}
\end{align}
 The nonequilibrium contribution is rather complicated at this stage where no approximation has been made regarding the strength of $W_{F,N} $. As we shall see, however, in the regime where $\left| \sigma _l^R \right| \ll \Delta _l $ and $ w_F^2 \ll w_N^2 $, one recovers familiar expressions for the tunneling currents.

\section{Tracing over spin and sublattice indices}\label{sec:trace}

Formally, the expressions for the Green's functions (\ref{eq:retarded_block}) and (\ref{eq:lesser0}) -(\ref{eq:lesser3}) are valid for any values of $\sigma _l^R /\Delta _l$. In computing the torques and currents, one can proceed without making any further approximations. To carry out the traces, it is convenient to rewrite $G^<$ as
\begin{eqnarray}
\left( G_a^< \right) _{ll^{\prime }} &=& \delta _{ll^{\prime }} \sum _{\mu =0}^3 \Bigg[ \frac{G^<_{\mu l+}+G^<_{\mu l-}}{2} V_l \sigma _{\mu }\otimes 1_{SL}V_l^{\dagger } \nonumber \\
&& + \frac{G^<_{\mu l+} - G^<_{\mu l-}}{2} V_l \sigma _{\mu } \otimes \tau _3 V_l^{\dagger } \Bigg] .
\end{eqnarray}
The Pauli matrices transform as
\begin{eqnarray}
V_l \sigma _1 \otimes 1_{SL} V_l^{\dagger } &=& -\left( \sigma _1 \cos \chi _l + \sigma _3 \sin \theta _l \sin \chi _l \right) \otimes \tau _3 \nonumber \\
&& +1_{SP}\otimes \left( \tau _1 \cos \phi _l + \tau _2 \sin \phi _l \right) \nonumber \\
&& \times  \cos \theta _l \sin \chi _l , \\
V_l \sigma _2 \otimes 1_{SL} V_l^{\dagger } &=& -\sigma _2 \otimes 1_{SL} \cos \theta _l \cos \chi _l \nonumber \\
&& - \left( \sigma _1 \cos \chi _l \sin \theta _l + \sigma _3 \sin \chi _l \right) \nonumber \\
&& \otimes \left( \tau _1 \sin \phi _l - \tau _2 \cos \phi _l \right) , \\
V_l \sigma _3 \otimes 1_{SL} V_l^{\dagger } &=& 1_{SP} \otimes \left( \tau _1 \cos \phi _l + \tau _2 \sin \phi _l \right) \sin \theta _l \nonumber \\
&& + \sigma _3 \otimes \tau _3  \cos \theta _l , \\
V_l 1_{SP} \otimes \tau _3 V_l^{\dagger } &=& -1_{SP} \otimes \tau _3 \cos \theta _l \cos \chi _l \nonumber \\
&& + \left( \sigma _1 \sin \chi _l - \sigma _3 \sin \theta _l \cos \chi _l \right) \nonumber \\
&& \otimes \left( \tau _1 \cos \phi _l + \tau _2 \sin \phi _l \right) , \\
V_l \sigma _1 \otimes \tau _3 V_l^{\dagger } &=& - \sigma _2 \otimes \left( \tau _1 \sin \phi _l - \tau _2 \cos \phi _l \right) \sin \theta _l \nonumber \\
&& + \sigma _1 \otimes 1_{SL} \cos \theta _l  , \\
V_l \sigma _2 \otimes \tau _3 V_l^{\dagger } &=& \sigma _2 \otimes \tau _3 , \\
V_l \sigma _3 \otimes \tau _3 V_l^{\dagger } &=& \left( \sigma _1 \sin \theta _l \sin \chi _l - \sigma _3 \cos \chi _l \right) \otimes 1_{SL} \nonumber \\
&& +\sigma _2 \otimes \left( \tau _1 \sin \phi _l -\tau _2 \cos \phi _l \right) \nonumber \\
&& \times  \cos \theta _l \sin \chi _l .
\end{eqnarray}
Now the traces in Eqs. (\ref{eq:torquemf}) - (\ref{eq:torquend}) can be carried out immediately, yielding
\begin{eqnarray}
\Gamma ^m_{\rm fl}  &=& -\frac{2\Delta _{\rm ex}}{\sin \theta } \int \frac{d\epsilon }{2\pi i} \sum _l \left( G^<_{1l+}+G^<_{1l-} \right) \cos \chi _l , \\
\Gamma ^m_{\rm dl} &=& \frac{2\Delta _{\rm ex}}{\sin \theta } \int \frac{d\epsilon }{2\pi i} \sum _l \left( G^<_{2l+} - G^<_{2l-} \right) , \\
\Gamma ^n_{\rm fl} &=& \frac{2\Delta _{\rm ex}}{\sin \theta } \int \frac{d\epsilon }{2\pi i} \sum _l \Big[ \left( G^<_{1l+} -G^<_{1l-} \right) \cos \theta _l  \nonumber \\
&& + \left( G^<_{3l+} -G^<_{3l-} \right) \sin \theta _l \sin \chi _l \Big] , \\
\Gamma ^n_{\rm dl} &=& -\frac{2\Delta _{\rm ex}}{\sin \theta }\int \frac{d\epsilon }{2\pi i} \nonumber \\
&& \times \sum _l \left( G^<_{2l+}+ G^<_{2l-} \right) \cos \theta _l \cos \chi _l .
\end{eqnarray}
Plugging (\ref{eq:retarded_block}) and (\ref{eq:lesser0}) - (\ref{eq:lesser3}) in and noting
\begin{equation}
S_{l\pm }^R \cos \theta _l + C_{l\pm }^R \sin \theta _l \sin \chi _l = \frac{\Delta _l \sin ^2 \theta _l  \pm \sigma _l^R c_l}{c_l e_{l\pm }^R}\sin \theta  \label{eq:identityB}
\end{equation}
lead to
\begin{eqnarray}
\Gamma ^m_{\rm fl} &=& -\Delta _{\rm ex}  \int \frac{d\epsilon }{2\pi }\sum _{l,\pm } \left( \pm \cos \theta _l \cos \chi _l \right) \Bigg\{ \frac{n_F -n_N}{\tau _l^N} \nonumber  \\
&& \times \Re \left[ \frac{\sigma _l^R}{e_{l\pm }^R}\left( g^R_{{\rm t}l\pm }-g^R_{{\rm b}l\pm } \right) \left( g^A_{{\rm t}l\pm }+g^A_{{\rm b}l\pm } \right) \right]  \nonumber \\
&& + 2n_F \Im \left[ \frac{\sigma _l^R}{e_{l\pm }^R} \left( g^R_{{\rm t}l\pm } - g^R_{{\rm b}l\pm } \right) \right] \Bigg\}  , \label{eq:torquemf4} \\
\Gamma ^m_{\rm dl} &=& \Delta _{\rm ex} \int \frac{d\epsilon }{2\pi } \sum _{l,\pm } \frac{ \cos \theta _l}{\left| e_{l\pm }^R \right| ^2} \frac{n_F -n_N}{\tau _l^N} \nonumber \\
&& \times \Im \left[ \left( \Delta _l \pm \sigma _l^R c_l \right) \sigma _l^A \right] \left| g^R_{{\rm t}l\pm }-g^R_{{\rm b}l\pm } \right| ^2 ,\\
\Gamma ^n_{\rm fl} &=& 2\Delta _{\rm ex}  \int \frac{d\epsilon }{2\pi } \sum _{l,\pm } \Bigg\{  \Re \Bigg[ \frac{\sigma _l^R c_l \pm \Delta _l \sin ^2 \theta _l }{c_l e^R_{l\pm }} \nonumber \\
&& \times \left( g^R_{{\rm t}l\pm } -g^R_{{\rm b}l\pm } \right) \left( g^A_{{\rm t}l\pm } -g^A_{{\rm b}l\pm } \right) \Bigg] \frac{n_F -n_N}{2\tau _l^N}  \nonumber \\
&& + n_F \Im \left[ \frac{\sigma ^R_l c_l \pm \Delta _l \sin ^2 \theta _l}{c_l e^R_{l\pm }} \left( g^R_{{\rm t}l\pm } -g^R_{{\rm b}l\pm } \right) \right]  \Bigg\} ,\;\;\;\; \label{eq:torquenf4} \\
\Gamma ^n_{\rm dl} &=& - \Delta _{\rm ex} \int \frac{d\epsilon }{2\pi } \sum _{l,\pm } \frac{ \pm \cos ^2 \theta _l \cos \chi _l }{\left| e^R_{l\pm } \right| ^2 }\frac{n_F -n_N}{\tau _l^N}  \nonumber \\
&& \times \Im \left[ \left( \Delta _l \pm \sigma _l^R c_l \right) \sigma _l^A \right] \left| g^R_{{\rm t}l\pm } -g^R_{{\rm b}l\pm } \right| ^2 .
\end{eqnarray}
These are the final results for the torques in their general form. The leading order expressions in $\sigma _l^R /\Delta _l $ (\ref{eq:torquemd2}), (\ref{eq:torquenf2}), (\ref{eq:torquemf2}) and (\ref{eq:torquend2}) are obtained by using
\begin{eqnarray}
\Re \left( e_{l\pm }^R \right) &=& \Delta _l \left[ 1 \mp \frac{2\delta _l c_l}{\Delta _l} + O \left( \left\{ \frac{\sigma _l^R}{\Delta _l} \right\} ^2 \right) \right] , \\
\Im \left( e_{l\pm }^R \right) &=& \frac{\mp 1}{2\tau _l^a} \left[ c_l \pm \frac{\delta _l s_l^2}{\Delta _l} + O \left( \left\{ \frac{\sigma _l^R}{\Delta _l} \right\} ^2 \right) \right] ,
\end{eqnarray}
and expansion in $\delta _l /\Delta _l , 1/2\tau _l^a \Delta _l $. In so doing, we also assumed
\begin{equation}
g^R_{{\rm t}l\pm } = \left( \epsilon -\epsilon _l - \Delta _l +\frac{i}{2\tau _l} \right) ^{-1} \left[ 1 + O \left( \frac{\sigma _l^R}{E_0} \right) \right] ,
\end{equation}
and similarly for $g^R_{{\rm b}l\pm }$ where $E_0$ is a characteristic energy scale of the antiferromagnet. It is not necessarily $\Delta _l$ since perturbative expansions of Green's functions are written as their derivatives, whose estimates depend on how they are integrated over. We also note that in the general results (\ref{eq:torquemf4}), (\ref{eq:torquenf4}), there are equilibrium contributions that do not vanish when $n_F =n_N$. They represent static effective fields arising from the full eigenstates of the antiferromagnetic electrons whose spins do not align with $\bm{S}_{A,B}$ due to the mixing with the ferromagnetic electrons. They are higher order in $\sigma _l^R /\Delta _l $.

Similarly, the trace in Eq. (\ref{eq:current}) for the different currents yields
\begin{eqnarray}
I_N &=& 4 \int \frac{d\epsilon }{2\pi i}  \sum _{l,\pm } \frac{1}{\tau _l^N}\bigg[ \frac{G^<_{0l\pm  } }{2}  \nonumber \\
&& + i n_N \Im \left( g^R_{{\rm t}l\pm } +g^R_{{\rm b}l\pm } \right) \bigg] , \\
J^{z_F}_F &=& 4 \int \frac{d\epsilon }{2\pi i } \sum _{l,\pm } \bigg[ \frac{G^<_{0l\pm }}{2\tau _l^a}   \mp \frac{G^<_{1l\pm } s_l + G^<_{3l\pm }c_l }{2\tau _l^F}  \bigg] , \\
J^x_N &=& -4 \int \frac{d\epsilon }{2\pi i} \sum _{l,\pm } \frac{\pm 1}{\tau _l^N} \bigg\{ i n_N \Im \big[ \left( g^R_{{\rm t}l\pm } -g^R_{{\rm b}l\pm } \right) \nonumber \\
&& \times \left( C_{l\pm }^R \sin \theta _l \sin \chi _l + S^R_{l\pm }\cos \theta _l \right)  \big] \nonumber \\
&&  - \frac{G^<_{1l\pm}\cos \theta _l + G^<_{3l\pm } \sin \theta _l \sin \chi _l}{2} \bigg\}  , \\
J^y_N &=& -4 \int \frac{d\epsilon }{2\pi i}\sum _{l,\pm }\frac{1}{2 \tau _l^N}G^<_{2l\pm } \cos \theta _l \cos \chi _l  , \\
J^z_N &=& -4\int \frac{d\epsilon }{2\pi i} \sum _{l,\pm } \frac{\pm \cos \chi _l}{\tau _l^N} \bigg\{ \frac{G^<_{3l\pm }}{2} \nonumber \\
&& - in_N \Im \left[ C^R_{l\pm }\left( g^R_{{\rm t}l\pm } -g^R_{{\rm b}l\pm } \right) \right] \bigg\} .
\end{eqnarray}
In addition to (\ref{eq:identityB}) one uses
\begin{eqnarray}
&& \Re \left( C^R_{l\pm } c_l + S^R_{l\pm }s_l \right) \nonumber \\
&=& \mp 2\tau _l^a \Im \left( e^R_{l\pm } \right) \frac{1+\left| C^R_{l\pm }\right| ^2 + \left| S^R_{l\pm } \right| ^2 }{2} , \\
&&\Im \left( C^R_{l\pm } c_l + S^R_{l\pm }s_l \right) \nonumber \\
&=& \pm 2\tau _l^a \Re \left( e_{l\pm }^R \right) \frac{1-\left| C^R_{l\pm } \right| ^2 - \left| S^R_{l\pm } \right| ^2 }{2} 
\end{eqnarray}
to derive
\begin{eqnarray}
I_N &=& -2e\int \frac{d\epsilon }{2\pi } \sum _{l,\pm } \frac{n_F -n_N}{\tau _l^N} \Bigg[ 2 \Im \left( g^R_{{\rm t}l\pm } +g^R_{{\rm b}l\pm } \right) \nonumber \\
&& + \frac{1+\left| C^R_{l\pm } \right| ^2 + \left| S^R_{l\pm } \right| ^2 }{2\tau _l^N} \left( \left| g^R_{{\rm t}l\pm } \right| ^2 + \left| g^R_{{\rm b}l\pm } \right| ^2 \right) \nonumber \\
&& + \frac{1-\left| C^R_{l\pm } \right| ^2 -\left| S^R_{l\pm } \right| ^2 }{\tau _l^N} \Re \left( g^R_{{\rm t}l\pm } g^A_{{\rm b}l\pm } \right) \Bigg] , \\
J_F^{z_F} &=& - \int \frac{d\epsilon }{2\pi } \sum _{l,\pm }\frac{n_F -n_N}{\tau _l^N} \Bigg[ \frac{1+\left| C_{l\pm }^R \right| ^2 + \left| S^R_{l\pm } \right| ^2 }{\tau _l^F} \nonumber \\
&& \times \bigg\{ \frac{\tau _l^F}{\tau _l^a} \left( \left| g^R_{{\rm t}l\pm } \right| ^2 + \left| g^R_{{\rm b}l\pm } \right| ^2 \right) \nonumber \\
&& -2\tau _l^a \Im \left( e_{l\pm }^R \right)  \left( \left| g^R_{{\rm t}l\pm } \right| ^2 -\left| g^R_{{\rm b}l\pm } \right| ^2 \right)   \bigg\}  \nonumber \\
&& + \frac{1-\left| C^R_{l\pm } \right| ^2 -\left| S^R_{l\pm } \right| ^2 }{\tau _l^F} \bigg\{ \frac{\tau _l^F}{\tau _l^a} \Re \left( g^R_{{\rm t}l\pm } g^A_{{\rm b}l\pm } \right) \nonumber \\
&&  - 2\tau _l^a \Re \left( e_{l\pm }^R \right) \Im \left( g^R_{{\rm t}l\pm }g^A_{{\rm b}l\pm }  \right) \bigg\} \Bigg] , \\
J^x_N &=& \int \frac{d\epsilon }{2\pi } \sum _{l,\pm } \frac{n_F -n_N }{\tau _l^N}\sin \theta  \Bigg\{ \Re \Bigg[ \frac{\sigma _l^R c_l \pm \Delta _l \sin ^2 \theta _l}{c_l e_{l\pm }^R} \nonumber \\
&& \times \frac{1}{\tau _l^N} \left( g^R_{{\rm t}l\pm } -g^R_{{\rm b}l\pm } \right) \left( g^A_{{\rm t}l\pm } +g^A_{{\rm b}l\pm } \right) \Bigg]  \nonumber \\
&& + \Im \left[ \frac{\sigma _l^R c_l \pm \Delta _l \sin ^2 \theta _l}{c_l e_{l\pm }^R} \left( g^R_{{\rm t}l\pm } -g^R_{{\rm b}l\pm } \right) \right] \Bigg\}   , \\
J^y_N &=& -\int \frac{d\epsilon }{2\pi } \sum _{l,\pm } \frac{n_F -n_N}{\tau _l^N} \frac{\sin \theta \cos \theta _l}{\left| e_{l\pm }^R \right| ^2 }   \nonumber \\
&& \times \frac{\Im \left( \sigma _l^R c_l \pm \Delta _l \right) \sigma _l^A }{\tau _l^N} \left| g^R_{{\rm t}L\pm } -g^R_{{\rm b}l\pm } \right| ^2 , \\
J^z_N &=& -\int \frac{d\epsilon }{2\pi } \sum _{l,\pm } \frac{n_F -n_N}{\tau _l^N} \cos \theta \Bigg\{ \Re \Bigg[ \frac{\sigma _l^R c_l \pm \Delta _l}{c_l e^R_{l\pm }} \nonumber \\
&& \times \frac{1}{\tau _l^N} \left( g^R_{{\rm t}l\pm } -g^R_{{\rm b}l\pm } \right) \left( g^A_{{\rm t}l\pm }+g^A_{{\rm b}l\pm } \right) \Bigg] \nonumber \\
&& + \Im \left[ \frac{\sigma _l^R c_l \pm \Delta _l}{c_l e_{l\pm }^R} \left( g^R_{{\rm t}l\pm } -g^R_{{\rm b}l\pm } \right) \right] \Bigg\} .
\end{eqnarray}
The expansion in $\sigma _l^R /\Delta _l$ is carried out similarly to the torques. It is helpful to note
\begin{equation}
\left| C^R_{l\pm } \right| ^2 + \left| S^R_{l\pm } \right| ^2 = 1 + \frac{s_l^2}{2\left( \tau _l^a \Delta _l \right) ^2} + O \left( \left\{ \frac{\sigma _l^R}{\Delta _l} \right\} ^3 \right) .
\end{equation}

\section{Scattering theory description}\label{sec:scattering}

Scattering theory has been commonly applied to transport problems in spintronics. \cite{Brataas2000,Tserkovnyak2002,Cheng2014c} Based on this approach, it has also been claimed in Ref.~\onlinecite{Nunez2006} that electron transmission coefficients of one-dimensional $PT$ symmetric antiferromagnets are generally spin independent, and thus such systems cannot serve as a spin current filter. It is therefore worth clarifying the relation between scattering theory and the non-equilibrium Green's function approach taken in the present paper. Below we first present a simple toy model of an antiferromagnetic structure that results in spin-dependent electron transmission coefficients. Then we translate our results into the language of scattering theory following Bode et al. (Ref.~\onlinecite{Bode2012a}) and show that spin-dependent transmission coefficients arise, consistent with our spin current computation in Sec. \ref{sec:current}.

Let us consider free electrons scattered by an antiferromagnetic exchange field. For simplicity, we consider a two-dimensional strip of infinite length along $x$. In the $y$ direction, we impose a periodic boundary condition with the period $2\pi d$. We model an antiferromagnet confined in a region near $x=0$ by the Hamiltonian
\begin{eqnarray}
H &=&  -\nabla ^2 + V\left( \bm{r} \right) \sigma _3 ,  \\
V\left( \bm{r} \right) &=& J \left[ e^{-\left( x-a \right) ^2 /2w^2} - e^{-\left( x+a \right) ^2 /2w^2 } \right] \sin \frac{y}{\ell } .
\end{eqnarray}
The antiferromagnetic exchange field $V\left( \bm{r}\right) $ represents two banks of 1D antiferromagnetic chains of period $\ell $ located at $x = \pm a $. The range of the exchange field is given by $w$ and we assume that there exists an integer $N_0 $ such that $N_0 \ell = 2\pi d$. Because of the Gaussian window functions $e^{-\left( x\pm a\right) ^2 /2w^2 }$, the exchange field is practically zero for $\left| x\right| \gg w$ and the eigenstates behave as plane waves in that region. For a given energy $E$, we introduce the bare Green's function $G_0^R \left( \bm{r} \right) $ by
\begin{equation}
\left( E + \nabla ^2 \right) G^R_0 \left( \bm{r} \right) = \delta \left( \bm{r} \right) 
\end{equation}
with an appropriate boundary condition for the retarded propagation. In the present problem it is explicitly written as
\begin{equation}
G_0^R \left( \bm{r} \right) = \frac{1}{2\pi d}\sum _n \int \frac{dk}{2\pi } \frac{e^{ikx + iny/2\pi d}}{E + i\delta - k^2 - \left( n/2\pi d\right) ^2 }.
\end{equation}
We choose an incoming plane wave $\psi _{0kn\pm } = N e^{ikx +iny/2\pi d} $ where $\pm $ denote up and down spin states and $N$ is a normalization constant. To obtain the scattering state, we solve the Lippmann-Schwinger equation \cite{DiVentra2008}
\begin{equation}
\psi _{kn\pm } = \psi _{0kn\pm } \pm \int d\bm{r}^{\prime } G^R_0 \left( \bm{r}-\bm{r}^{\prime } \right) V\left( \bm{r}^{\prime } \right) \psi _{kn\pm }\left( \bm{r}^{\prime } \right) .
\end{equation}
For our purpose of demonstrating spin dependence of the scattering, we replace $\psi _{kn\pm } $ on the right-hand-side by $\psi _{0kn\pm }$. This leads to a solution in the Born approximation given by
\begin{eqnarray}
\psi _{kn\pm } &=& \psi _{0kn\pm } \mp \sqrt{2\pi }w J N \int dk^{\prime } \sin \left[ \left( k-k^{\prime } \right) a \right] \nonumber \\
&& \times \Bigg\{ \frac{e^{iny/2\pi d}}{E+i\delta -k^{\prime 2} -\left[ \left( n+N_0 \right) /2\pi d\right] ^2 }\nonumber \\
&&  - \frac{e^{-iny/2\pi d}}{E+i\delta -k^{\prime 2} -\left[ \left( n-N_0 \right) /2\pi d \right] ^2 } \Bigg\} \nonumber \\
&& \times e^{ik^{\prime }x + iny/2\pi d -w^2 \left( k^{\prime }-k\right) ^2 /2} .
\end{eqnarray}
The spin dependent part does not vanish unless $n=0$. In particular, the transmission coefficients that correspond to the contributions with $k^{\prime } >0$ are spin dependent and lead to non-conservation of the transverse spin current. 

Having demonstrated the spin dependence of transmission through the antiferromagnet in a real space model, we go back to our tunneling junction and construct its scattering theory description. Let us denote the normalized single particle eigenstates in the leads by $\left| \psi ^F_{k\sigma } \right\rangle , \left| \psi ^N_{m\sigma } \right\rangle $. Incoming and outgoing scattering states $\left| \psi ^{F\pm }_{k\sigma } \right\rangle , \left| \psi ^{N\pm }_{m\sigma } \right\rangle $ evolving out of $\left| \psi ^F_{k\sigma } \right\rangle , \left| \psi ^N_{m\sigma } \right\rangle $ respectively are given by
\begin{eqnarray}
\left| \psi ^{F+}_{k\sigma } \right\rangle &=& \left| \psi ^F_{k\sigma } \right\rangle + \sum _{k^{\prime },\sigma ^{\prime }} \left( G^{R}_F W_F G^{R}_a W_F^{\dagger } \right) ^{\sigma ^{\prime }\sigma }_{k^{\prime }k} \left| \psi ^F_{k^{\prime } \sigma ^{\prime }} \right\rangle \nonumber \\
&& + \sum _{m^{\prime },\sigma ^{\prime }} \left( G^R_N W_N G^R_a W_F^{\dagger } \right) ^{\sigma ^{\prime }\sigma }_{m^{\prime }k} \left| \psi ^N_{m^{\prime }\sigma ^{\prime }} \right\rangle , \\
\left| \psi ^{F-}_{k\sigma } \right\rangle &=& \left| \psi ^F_{k\sigma } \right\rangle + \sum _{k^{\prime },\sigma ^{\prime }} \left( G^{A}_F W_F G^{A}_a W_F^{\dagger } \right) ^{\sigma ^{\prime }\sigma }_{k^{\prime }k} \left| \psi ^F_{k^{\prime } \sigma ^{\prime }} \right\rangle \nonumber \\
&& + \sum _{m^{\prime },\sigma ^{\prime }} \left( G^A_N W_N G^A_a W_F^{\dagger } \right) ^{\sigma ^{\prime }\sigma }_{m^{\prime }k} \left| \psi ^N_{m^{\prime }\sigma ^{\prime }} \right\rangle , \\
\left| \psi ^{N+}_{m\sigma } \right\rangle &=& \left| \psi ^N_{m\sigma } \right\rangle + \sum _{m^{\prime },\sigma ^{\prime }} \left( G^{R}_N W_N G^{R}_a W_N^{\dagger } \right) ^{\sigma ^{\prime }\sigma }_{m^{\prime }m} \left| \psi ^N_{m^{\prime } \sigma ^{\prime }} \right\rangle \nonumber \\
&& + \sum _{k^{\prime },\sigma ^{\prime }} \left( G^R_F W_F G^R_a W_N^{\dagger } \right) ^{\sigma ^{\prime }\sigma }_{k^{\prime }m} \left| \psi ^F_{k^{\prime }\sigma ^{\prime }} \right\rangle , \\
\left| \psi ^{N-}_{m\sigma } \right\rangle &=& \left| \psi ^N_{m\sigma } \right\rangle + \sum _{m^{\prime },\sigma ^{\prime }} \left( G^{A}_N W_N G^{A}_a W_N^{\dagger } \right) ^{\sigma ^{\prime }\sigma }_{m^{\prime }m} \left| \psi ^N_{m^{\prime } \sigma ^{\prime }} \right\rangle \nonumber \\
&& + \sum _{k^{\prime },\sigma ^{\prime }} \left( G^A_F W_F G^A_a W_N^{\dagger } \right) ^{\sigma ^{\prime }\sigma }_{k^{\prime }m} \left| \psi ^F_{k^{\prime }\sigma ^{\prime }} \right\rangle . 
\end{eqnarray}
Here the Green's functions $G^{R(A)}_{a,c,d}$ are in the energy representation and the energy argument is set to be the eigenvalue of the reference state, namely $\epsilon =\epsilon _{k\sigma }$ for $\left| \psi ^{F\pm }_{k\sigma }\right\rangle $ and $\epsilon =\epsilon _m $ for $\left| \psi ^{N\pm }_{m\sigma } \right\rangle $. The scattering matrix is defined by
\begin{equation}
\mathcal{S} = \begin{pmatrix}
	\left\langle \psi ^{F-} | \psi ^{F+} \right\rangle & \left\langle \psi ^{F-} | \psi ^{N+} \right\rangle  \\
	\left\langle \psi ^{N-} | \psi ^{F+} \right\rangle & \left\langle \psi ^{N-} |\psi ^{N+} \right\rangle \\
	\end{pmatrix} \equiv \begin{pmatrix}
	r_F & t_F \\
	t_N & r_N \\
	\end{pmatrix} .
\end{equation}
The reflection and transmission matrices are given in terms of the Green's functions as
\begin{eqnarray}
r_F &=& 1 + \left( G^R_F -G^A_F \right)  W_F G^R_a W_F^{\dagger }  , \\
t_F &=& \left( G^R_F -G^A_F \right) W_F G^R_a W_N^{\dagger } , \\
r_N &=& 1 + \left( G^R_N -G^A_N \right) W_N G^R_a W_N^{\dagger } , \\
t_N &=& \left( G^R_N -G^A_N \right) W_N G^R_a W_F^{\dagger } .
\end{eqnarray}
We are interested in spin-dependent scatterings due to the antiferromagnetic order. In order to eliminate any spin-dependent influence of the ferromagnetic lead, we assume the $c$ lead is also a normal metal in this section. It amounts to setting $\sigma _l^R =0$. The antiferromagnetic full Green's function $G^R_a$ then reads
\begin{eqnarray}
 G^R_a  &=& \delta _{ll^{\prime }}\Bigg[  \frac{G^R_{{\rm t}l} + G^R_{{\rm b}l} }{2}  - \frac{G^R_{{\rm t}l} -G^R_{{\rm b}l}}{2} \Big\{ \sigma _3 \otimes \tau _3 \cos \theta _l  \nonumber \\
&& + 1_{SP} \otimes \left( \tau _1 \cos \phi _l + \tau _2 \sin \phi _l \right) \sin \theta _l \Big\} \Bigg]   ,
\end{eqnarray}
where
\begin{eqnarray}
G^R_{{\rm t}l}  &=& \frac{1}{\epsilon - \epsilon _l - \Delta _l  + i/2\tau _l} , \\
 G^R_{{\rm b}l}  &=& \frac{1}{\epsilon -\epsilon _l + \Delta _l + i/2\tau _l } .
\end{eqnarray}
Carrying out the matrix multiplications, one obtains the reflection and transmission coefficients as
\begin{align}
\left( r_F \right) _{kk^{\prime }} =& \delta _{kk^{\prime }} 1_{SP} \\
& -2\pi i \delta \left( \epsilon _{k} -\epsilon _{k^{\prime }} \right) \left( \mathcal{T}_{kk^{\prime }}1_{SP} + \mathcal{T}_{kk^{\prime }}^s \sigma _3 \right)  , \nonumber  \\
\left( t_F \right) _{km^{\prime }} =& -2\pi i \delta \left( \epsilon _{k } -\epsilon _{m^{\prime }} \right)  \left( \mathcal{T}_{km^{\prime }}1_{SP} + \mathcal{T}^s_{km^{\prime }} \sigma _3 \right) , \\
\left( r_N \right) _{mm^{\prime }} =& \delta _{mm^{\prime }} 1_{SP} \\
& -2\pi i \delta \left( \epsilon _{m} -\epsilon _{m^{\prime }} \right) \left( \mathcal{T}_{mm^{\prime }}1_{SP} + \mathcal{T}_{mm^{\prime }}^s \sigma _3 \right)  , \nonumber  \\
\left( t_N \right) _{mk^{\prime }} =& -2\pi i \delta \left( \epsilon _{m } -\epsilon _{k^{\prime }} \right)  \left( \mathcal{T}_{mk^{\prime }}1_{SP} + \mathcal{T}^s_{mk^{\prime }} \sigma _3 \right) .
 \end{align}
 $\mathcal{T}$ and $\mathcal{T}^s$ are spin independent and dependent parts of the $T$ matrix given by
 \begin{eqnarray}
 \mathcal{T}_{kk^{\prime }} &=& \sum _l \bigg[  \frac{G^R_{{\rm t}l}\left( \epsilon _k \right)+G^R_{{\rm b}l}\left( \epsilon _k \right)}{2}  \sum _{\alpha =A,B}\left( W^{\alpha }_F \right) _{kl} \overline{\left( W^{\alpha }_F \right) _{k^{\prime }l}}   \nonumber \\
 && -\frac{G^R_{{\rm t}l}\left( \epsilon _k \right) -G^R_{{\rm b}l}\left( \epsilon _k \right) }{2} \bigg\{ e^{i\phi _l} \left( W^B_F \right) _{kl} \overline{\left( W^A_F \right) _{k^{\prime }l}} \nonumber \\
 && + e^{-i\phi _l} \left( W^A_F \right) _{kl} \overline{\left( W^B_F \right) _{k^{\prime }l}} \bigg\} \sin \theta _l \bigg] , \\
 \mathcal{T}_{km^{\prime }} &=&  \sum _l \bigg[  \frac{G^R_{{\rm t}l}\left( \epsilon _k \right)+G^R_{{\rm b}l}\left( \epsilon _k \right) }{2}  \sum _{\alpha =A,B}\left( W^{\alpha }_F \right) _{kl} \overline{\left( W^{\alpha }_N \right) _{m^{\prime }l}}   \nonumber \\
 && -\frac{G^R_{{\rm t}l}\left( \epsilon _k \right) -G^R_{{\rm b}l}\left( \epsilon _k \right)}{2} \bigg\{ e^{i\phi _l} \left( W^B_F \right) _{kl} \overline{\left( W^A_N \right) _{m^{\prime }l}} \nonumber \\
 && + e^{-i\phi _l} \left( W^A_F \right) _{kl} \overline{\left( W^B_N \right) _{m^{\prime }l}} \bigg\} \sin \theta _l \bigg] , \\
  \mathcal{T}_{mk^{\prime }} &=& \sum _l \bigg[  \frac{G^R_{{\rm t}l}\left( \epsilon _m \right) +G^R_{{\rm b}l}\left( \epsilon _m \right) }{2}  \sum _{\alpha =A,B}\left( W^{\alpha }_N \right) _{ml} \overline{\left( W^{\alpha }_F \right) _{k^{\prime }l}}   \nonumber \\
 && -\frac{G^R_{{\rm t}l}\left( \epsilon _m \right) -G^R_{{\rm b}l}\left( \epsilon _m \right) }{2} \bigg\{ e^{i\phi _l} \left( W^B_N \right) _{ml} \overline{\left( W^A_F \right) _{k^{\prime }l}} \nonumber \\
 && + e^{-i\phi _l} \left( W^A_N \right) _{ml} \overline{\left( W^B_F \right) _{k^{\prime }l}} \bigg\} \sin \theta _l \bigg] , \\
   \mathcal{T}_{mm^{\prime }} &=& \sum _l \bigg[  \frac{G^R_{{\rm t}l}\left( \epsilon _m \right) +G^R_{{\rm b}l} \left( \epsilon _m \right) }{2}  \sum _{\alpha =A,B}\left( W^{\alpha }_N \right) _{ml} \overline{\left( W^{\alpha }_N \right) _{m^{\prime }l}}   \nonumber \\
 && -\frac{G^R_{{\rm t}l}\left( \epsilon _m \right) -G^R_{{\rm b}l}\left( \epsilon _m \right)  }{2} \bigg\{ e^{i\phi _l} \left( W^B_N \right) _{ml} \overline{\left( W^A_N \right) _{m^{\prime }l}} \nonumber \\
 && + e^{-i\phi _l} \left( W^A_N \right) _{ml} \overline{\left( W^B_N \right) _{m^{\prime }l}} \bigg\} \sin \theta _l \bigg] ,\\
\mathcal{T}^s_{kk^{\prime }} &=& - \sum _{l}  \frac{G^R_{{\rm t}l}\left( \epsilon _k \right) - G^R_{{\rm b}l}\left( \epsilon _k \right)}{2} \bigg[ \left( W^A_F \right) _{kl} \overline{\left( W^A_F \right) _{k^{\prime } l}} \nonumber \\
&& - \left( W^B_F \right) _{kl} \overline{\left( W^B_F \right) _{k^{\prime } l}} \bigg]  \cos \theta _l , \\
\mathcal{T}^s_{km^{\prime }} &=& - \sum _{l}  \frac{G^R_{{\rm t}l}\left( \epsilon _k \right) - G^R_{{\rm b}l}\left( \epsilon _k \right)}{2} \bigg[ \left( W^A_F \right) _{kl} \overline{\left( W^A_N \right) _{m^{\prime } l}} \nonumber \\
&& - \left( W^B_F \right) _{kl} \overline{\left( W^B_N \right) _{m^{\prime } l}} \bigg] \cos \theta _l  , \\
\mathcal{T}^s_{mk^{\prime }} &=& - \sum _{l}  \frac{G^R_{{\rm t}l}\left( \epsilon _m \right) - G^R_{{\rm b}l}\left( \epsilon _m \right)}{2} \bigg[ \left( W^A_N \right) _{ml} \overline{\left( W^A_F \right) _{k^{\prime } l}} \nonumber \\
&& - \left( W^B_N \right) _{ml} \overline{\left( W^B_F \right) _{k^{\prime } l}}  \bigg] \cos \theta _l  ,  \\
\mathcal{T}^s_{mm^{\prime }} &=& - \sum _{l}  \frac{G^R_{{\rm t}l}\left( \epsilon _m \right) - G^R_{{\rm b}l}\left( \epsilon _m \right)}{2} \bigg[ \left( W^A_N \right) _{ml} \overline{\left( W^A_N \right) _{m^{\prime } l}} \nonumber \\
&& - \left( W^B_N \right) _{ml} \overline{\left( W^B_N \right) _{m^{\prime } l}} \bigg] \cos \theta _l .
\end{eqnarray}
We note that at this stage we cannot yet impose the statistical sublattice symmetry of the tunneling matrices $W^{A,B}_{F,N}$ since the self-averaging has been assumed when summations over $k,m$ are taken. 

Let us now consider specifically an incoming electron at energy $\epsilon _k$ in the $c$ lead with spin polarization along the positive $x$ direction. As before, the antiferromagnetic quantization axis is along $z$. Upon scattering by the antiferromagnet, the spin expectation value of the injected electron changes. The reflected and transmitted spin expectation values are given by
\begin{eqnarray}
\left\langle s^x_k \right\rangle _r &=& \frac{1}{2} \sum _{k^{\prime }} \left[ \overline{\left( r_F \right) ^{\uparrow \uparrow }_{k^{\prime }k}} \left( r_F \right) ^{\downarrow \downarrow }_{k^{\prime }k} +  \overline{\left( r_F \right) ^{\downarrow \downarrow }_{k^{\prime }k}} \left( r_F \right) ^{\uparrow \uparrow }_{k^{\prime }k} \right] , \\
\left\langle s^x_k \right\rangle _t &=&   \frac{1}{2} \sum _{m^{\prime }} \left[ \overline{\left( t_N \right) ^{\uparrow \uparrow }_{m^{\prime }k}} \left( t_N \right) ^{\downarrow \downarrow }_{m^{\prime }k} +  \overline{\left( t_N \right) ^{\downarrow \downarrow }_{m^{\prime }k}} \left( t_N \right) ^{\uparrow \uparrow }_{m^{\prime }k} \right] , \qquad\\
\left\langle s^y_k \right\rangle _r &=& \frac{1}{2i} \sum _{k^{\prime }} \left[ \overline{\left( r_F \right) ^{\uparrow \uparrow }_{k^{\prime }k}} \left( r_F \right) ^{\downarrow \downarrow }_{k^{\prime }k} -  \overline{\left( r_F \right) ^{\downarrow \downarrow }_{k^{\prime }k}} \left( r_F \right) ^{\uparrow \uparrow }_{k^{\prime }k} \right] , \\
\left\langle s^y_k \right\rangle _t &=&   \frac{1}{2i} \sum _{m^{\prime }} \left[ \overline{\left( t_N \right) ^{\uparrow \uparrow }_{m^{\prime }k}} \left( t_N \right) ^{\downarrow \downarrow }_{m^{\prime }k} -  \overline{\left( t_N \right) ^{\downarrow \downarrow }_{m^{\prime }k}} \left( t_N \right) ^{\uparrow \uparrow }_{m^{\prime }k} \right] .
\end{eqnarray}
We remark that the reflected spin expectation values are essentially real and imaginary parts of the spin-mixing conductance \cite{Brataas2000} of the antiferromagnetic junction. We focus on the transmission here and derive
\begin{align}
\left\langle s^x_k \right\rangle _t =& 4\pi ^2 D_F  \sum _{m}D_N \left( \left| \mathcal{T}_{mk} \right| ^2 - \left| \mathcal{T}^s_{mk} \right| ^2 \right) \nonumber \\
=& 4\pi ^2 D_F D_N \sum _{ll^{\prime }, \alpha \beta }  \Lambda ^{\alpha \beta }_{ll^{\prime }} \overline{\left( W^{\alpha }_F \right) _{kl}} \left( W^{\beta }_F \right) _{kl^{\prime }}  \\
\left\langle s^y_k \right\rangle _t =& 4\pi ^2 i D_F \sum _{m} D_N \left( \overline{\mathcal{T}_{mk}}\mathcal{T}^s_{mk} -\overline{\mathcal{T}^s_{mk}} \mathcal{T}_{mk} \right) \nonumber \\
=& 4\pi ^2 D_F D_N \sum _{ll^{\prime },\alpha \beta } \Pi ^{\alpha \beta }_{ll^{\prime }} \overline{\left( W^{\alpha }_F \right) _{kl}} \left( W^{\beta }_F \right) _{kl^{\prime }} , 
\end{align}
where $D_{F,N}$ are the density of states of the leads taken to be constant for simplicity, and 
\begin{widetext}
\begin{eqnarray}
\Lambda ^{\alpha \beta }_{ll^{\prime }} &=& \sum _m \Bigg[ \left\{ \frac{G^R_{{\rm t}l}\left( \epsilon _m \right) + G^R_{{\rm b}l}\left( \epsilon _m \right) }{2}\left( W^{\alpha }_N \right) _{ml} - \frac{G^R_{{\rm t}l}\left( \epsilon _m \right) - G^R_{{\rm b}l}\left( \epsilon _m \right) }{2} \left( W^{\overline{\alpha }}_N \right) _{ml} e^{i\nu _{\alpha } \phi _l} \sin \theta _l  \right\} \nonumber \\
&& \times \left\{ \frac{G^A_{{\rm t}l^{\prime }}\left( \epsilon _m \right) + G^A_{{\rm b}l^{\prime }}\left( \epsilon _m \right) }{2} \overline{\left( W^{\beta }_N \right) _{ml^{\prime }}} - \frac{G^A_{{\rm t}l^{\prime }}\left( \epsilon _m \right) - G^A_{{\rm b}l^{\prime }}\left( \epsilon _m \right) }{2} \overline{\left( W^{\overline{\beta }}_N \right) _{ml^{\prime }}} e^{-i\nu _{\beta } \phi _{l^{\prime }}} \sin \theta _{l^{\prime }}  \right\} \nonumber \\
&& - \nu _{\alpha }\nu _{\beta } \frac{G^R_{{\rm t}l}\left( \epsilon _m \right) - G^R_{{\rm b}l}\left( \epsilon _m \right)}{2} \frac{G^A_{{\rm t}l^{\prime }}\left( \epsilon _n \right) - G^A_{{\rm b}l^{\prime }}\left( \epsilon _m \right) }{2} \left( W^{\alpha }_N \right) _{ml} \overline{\left( W^{\beta }_N \right) _{ml^{\prime }}} \cos \theta _l \cos \theta _{l^{\prime }} \Bigg]  ,\\
\Pi ^{\alpha \beta }_{ll^{\prime }} &=& i \sum _m \Bigg[  \left( W^{\alpha }_N \right) _{ml} \overline{\left( W^{\beta }_N \right) _{ml^{\prime }}}  \bigg\{ \frac{G^R_{{\rm t}l}\left( \epsilon _m \right) + G^R_{{\rm b}l}\left( \epsilon _m \right) }{2} \frac{G^A_{{\rm t}l^{\prime }}\left( \epsilon _m \right) - G^A_{{\rm b}l^{\prime }}\left( \epsilon _m \right) }{2}  \nu _{\beta } \cos \theta _{l^{\prime }} \nonumber \\
&& -  \frac{G^R_{{\rm t}l}\left( \epsilon _m \right) - G^R_{{\rm b}l}\left( \epsilon _m \right) }{2} \frac{G^A_{{\rm t}l^{\prime }}\left( \epsilon _m \right) + G^A_{{\rm b}l^{\prime }}\left( \epsilon _m \right) }{2}  \nu _{\alpha } \cos \theta _{l} \bigg\}  + \frac{G^R_{{\rm t}l}\left( \epsilon _m \right) - G^R_{{\rm b}l}\left( \epsilon _m \right) }{2} \frac{G^A_{{\rm t}l^{\prime }}\left( \epsilon _m \right) - G^A_{{\rm b}l^{\prime }}\left( \epsilon _m \right) }{2} \nonumber \\
&& \times \left\{ \left( W^{\alpha }_N \right) _{ml} \overline{\left( W^{\beta }_N \right) _{ml^{\prime }}} \nu _{\alpha } e^{i\nu _{\beta }\phi _{l^{\prime }}} \cos \theta _l \sin \theta _{l^{\prime }} +\left( W^{\overline{\alpha }}_N \right) _{ml} \overline{\left( W^{\overline{\beta }}_N \right) _{ml^{\prime }}}  \nu _{\beta } e^{i\nu _{\alpha } \phi _l} \cos \theta _{l^{\prime }}\sin \theta _l \right\} \Bigg] .
\end{eqnarray}
\end{widetext} 
We have introduced the notation $\overline{A} = B, \overline{B} =A $ and the signature function $\nu _A = +1 , \nu _B = -1$. Now that $\Lambda ^{\alpha \beta }_{ll^{\prime }} ,\Pi ^{\alpha \beta }_{ll^{\prime }}$ contain a summation over $m$, one can use the statistical properties of $W_{F,N}$ to estimate their elements. We first of all drop the off-diagonal terms in $ll^{\prime }$ that are suppressed by $1/\sqrt{N_N} $. Further, one can discard products $W^{\alpha } W^{\overline{\alpha }}$ by the same reasoning. This leaves us with
\begin{eqnarray}
\Lambda ^{\alpha \alpha }_{ll^{\prime }} & \approx & \delta _{ll^{\prime }} \sum _m \Bigg[ \left| \left( W^{\alpha }_d \right) _{ml}\right| ^2  \left| \frac{G^R_{{\rm t}l}\left( \epsilon _m \right) + G^R_{{\rm b}l}\left( \epsilon _m \right)}{2} \right| ^2 \nonumber \\
&& -\left\{ \left| \left( W^{\alpha }_d \right) _{ml}\right| ^2 \cos ^2 \theta _l -   \left| \left( W^{\overline{\alpha }}_d \right) _{ml} \right| ^2 \sin ^2 \theta _l \right\} \nonumber \\
&& \times   \left| \frac{G^R_{{\rm t}l}\left( \epsilon _m \right) - G^R_{{\rm b}l}\left( \epsilon _m \right) }{2} \right| ^2 \Bigg] , \\
\Lambda ^{\alpha \overline{\alpha }}_{ll^{\prime }} &\approx &-  \delta _{ll^{\prime }} \sum _m e^{i\nu _{\alpha }\phi _l }\sin ^2 \theta _l \Bigg[ \frac{G^R_{{\rm t}l}\left( \epsilon _m \right) + G^R_{{\rm b}l}\left( \epsilon _m \right) }{2} \nonumber \\
&& \times \frac{G^A_{{\rm t}l}\left( \epsilon _m \right) - G^A_{{\rm b}l}\left( \epsilon _m \right)}{2}\left| \left( W^{\alpha }_d \right) _{ml} \right| ^2 \nonumber \\
&& +  \frac{G^R_{{\rm t}l}\left( \epsilon _m \right) - G^R_{{\rm b}l}\left( \epsilon _m \right) }{2} \nonumber \\ 
&& \times \frac{G^A_{{\rm t}l}\left( \epsilon _m \right) + G^A_{{\rm b}l}\left( \epsilon _m \right)}{2}\left| \left( W^{\overline{\alpha }}_d \right) _{ml} \right| ^2  \Bigg] , \\
\Pi ^{\alpha \beta }_{ll^{\prime }} & \approx & i\delta _{ll^{\prime }}\delta _{\alpha \beta } \nu _{\alpha } \sum _m \Bigg[ \left| \left( W^{\alpha }_d \right) _{ml} \right| ^2 \cos \theta _l \nonumber \\
&& \times \left\{ G^R_{{\rm b}l}\left( \epsilon _m \right) G^A_{{\rm t}l}\left( \epsilon _m \right) - G^R_{{\rm t}l}\left( \epsilon _m \right) G^A_{{\rm b}l}\left( \epsilon _m \right) \right\}  \nonumber \\
&& + \left\{ \left| \left( W^{\alpha }_d \right) _{ml} \right| ^2 + \left| \left( W^{\overline{\alpha }}_d \right) _{ml} \right| ^2 \right\} e^{i\nu _{\alpha }\phi _l}  \nonumber \\
&& \times \left| \frac{G^R_{{\rm t}l}\left( \epsilon _m \right) - G^R_{{\rm b}l}\left( \epsilon _m \right) }{2} \right| ^2 \cos \theta _l \sin \theta _l \Bigg] .
\end{eqnarray}
Finally, we also take advantage of the sublattice symmetry so that $\left| \left( W^{\alpha }_d \right) _{ml} \right| ^2 = \left| \left( W^{\overline{\alpha }}_d \right) _{ml} \right| ^2 $ under the summation over $m$. Then one can see that if $\sin \theta _l =0$, namely if there is no intersublattice overlap, both $\Lambda ^{\alpha \beta }_{ll^{\prime }}$ and $\Pi ^{\alpha \beta }_{ll^{\prime }}$ are proportional to the interband combinations of the Green's functions $G^R_{{\rm b}l}G^A_{{\rm t}l} , G^R_{{\rm t}l} G^A_{{\rm b}l}$ that are suppressed by a factor of $1/\tau _l \Delta _l $ compared to the intraband products $\left| G^R_{{\rm t,b}l} \right| ^2 $. Although not presented here, the charge current transmission is proportional to $\left| G^R_{{\rm t,b}l}\right| ^2 $. Therefore, one concludes that both the $x$ and $y$ components of the transmitted spin per transmitted particle is small by a factor $\sim 1/\tau _l \Delta _l $. This suppression represents the result of dephasing extensively discussed in the main text. When there is a finite intersublattice overlap, the transmitted $x$ and $y$ components are proportional to $\left| t_l \right| ^2 /\Delta _l^2 , \left| t_l \right| /\Delta _l $ respectively. This is also consistent with the result of the Keldysh approach as $t_l$ opens up a channel of transport through the antiferromagnet in which the transverse spin is conserved. Regardless of the strength of the intersublattice overlap, this calculation shows that scatterings by an antiferromagnet are strongly spin dependent and the transmitted spin per transmitted particle is generally reduced from what is injected.

\section{F/F/N and AF/AF/N junctions}\label{sec:AFMlead}

The Keldysh formalism we have developed can be easily applied to F/F/N or AF/AF/N junctions instead of F/AF/N studied in the main text. In this section, we first present the spin expectation values of F/F/N junction to clarify the role of N. It adds a contribution to the known leading-order tunneling expressions. \cite{Chudnovskiy2008} Second, we demonstrate that AF/AF/N junction in the present formalism does not lead to any spin dependence of the electron transport. We further discuss how the spin-transfer torques in continuous antiferromagnetic textures, studied in Ref.~\onlinecite{Yamane2016a}, could be understood in our framework.

For F/F/N junction, one uses the same Hamiltonian (\ref{eq:total_Hamiltonian}), but $H_{0,c,d}$ are replaced by
\begin{eqnarray}
H_0 &=& \sum _{l,\sigma } \epsilon _l a_{l\sigma }^{\dagger }a_{l\sigma } - \frac{\Delta _{\rm ex}}{S} \sum _{l,\sigma \sigma ^{\prime }} \bm{\sigma }^{\sigma \sigma ^{\prime }}\cdot \bm{S}_F a_{l\sigma }^{\dagger } a_{l\sigma ^{\prime }} , \\
H_F &=& \sum _{kl,\sigma \sigma ^{\prime }} \left[ c_{k\sigma }^{F\dagger } \left( W_F \right) _{kl} R^{\sigma \sigma ^{\prime }}_{SP} a_{l\sigma ^{\prime }} + {\rm h.c.} \right] , \\
H_N &= & \sum _{lm,\sigma }\left[ c^{N\dagger }_{m\sigma } \left( W_N \right) _{ml} a_{l\sigma } +{\rm h.c.} \right] .
\end{eqnarray}
The procedure is identical to F/AF/N junction case. One first obtains the full retarded Green's function
\begin{align}
\left( G^R_a\right) _{ll^{\prime }} = \delta _{ll^{\prime }}\bigg[ \frac{g^R_{{\rm t}l} +g^R_{{\rm b}l}}{2} + \frac{g^R_{{\rm t}l} -g^R_{{\rm b}l}}{2}  \left( C_l^R \sigma _3 +S_l^R \sigma _1 \right) \bigg] ,
\end{align}
where
\begin{eqnarray}
\sigma _l^R &=& \sum _k \left| \left( W_F \right) _{kl} \right| ^2 \frac{1}{2}\left( \frac{1}{\epsilon - \epsilon _{k\uparrow }+i\delta }-\frac{1}{\epsilon -\epsilon _{k\downarrow }+i\delta }\right) ,\nonumber \\
&&\\
e_l^R &=& \sqrt{\Delta _{\rm ex}^2 + 2\Delta _{\rm ex}\sigma _l^R \cos \theta + \left( \sigma _l^R \right) ^2 } , \\
g^R_{{\rm t,b}l} &=& \frac{1}{\epsilon +i/2\tau _l -\epsilon _l \mp e_l^R } , \\
C^R_l &=& \frac{\Delta _{\rm ex} +\sigma _l^R \cos \theta }{e_l^R} , \quad S_l^R = -\frac{\sigma _l^R \sin \theta }{e_l^R}.
\end{eqnarray}
The definitions of $1/\tau _l^{F,N},1/\tau _l $ are also accordingly modified. The lesser Green's function yields
\begin{eqnarray}
\left( G^<_a \right) _{ll^{\prime }} &=& \delta _{ll^{\prime }} \sum _{\mu } G^<_{\mu l } \sigma _{\mu } , \\
G^<_{0l} &=& i\frac{n_N  -n_F}{\tau _l^N} \Bigg[ \frac{1+\left| C^R_l \right| ^2 + \left| S^R_l \right| ^2 }{2} \frac{\left| g^R_{{\rm t}l}\right| ^2 + \left| g^R_{{\rm b}l}\right| ^2 }{2} \nonumber \\
&& + \frac{1- \left| C^R_l \right| ^2 - \left| S^R_l \right| ^2 }{2} \Re \left( g^R_{{\rm t}l}g^A_{{\rm b}l} \right) \Bigg] \nonumber \\
&& -in_F \Im \left( g^R_{{\rm t}l} + g^R_{{\rm b}l}\right) , \\
G^<_{1l} &=& i\frac{n_N -n_F}{2\tau _l^N} \Re \left[ S^R_l \left( g^R_{{\rm t}l} - g^R_{{\rm b}l}\right) \left( g^A_{{\rm t}l}+g^A_{{\rm b}l}\right) \right]  \nonumber \\
&& + in_F \Im \left[ S^R_l \left( g^R_{{\rm t}l} -g^R_{{\rm b}l} \right) \right]  ,  \\
G^<_{2l} &=& -i\frac{n_N -n_F}{2\tau _l^N} \Im \left( C^R_l S^R_l \right) \left| g_{{\rm t}l}^R -g^R_{{\rm b}l}\right| ^2  , \\
G^<_{3l} &=& i\frac{n_N -n_F}{2\tau _l^N} \Re \left[ C^R_l \left( g^R_{{\rm t}l} -g^R_{{\rm b}l} \right) \left( g^A_{{\rm t}l} -g^A_{{\rm b}l} \right) \right] \nonumber \\
&& -n_F \Im \left[ C^R_l \left( g^R_{{\rm t}l}-g^R_{{\rm b}l} \right) \right] .
\end{eqnarray}
The similarity with the antiferromagnetic case, Eqs. (\ref{eq:lesser0}) - (\ref{eq:lesser3}), is clear. Indeed, the antiferromagnetic Green's function can be regarded as two copies of the ferromagnetic ones above with different dispersions characterized by $e_{l\pm }^R$. The electron spin expectation values at the leading order in $\sigma _l^R / \Delta _{\rm ex}$ read
\begin{eqnarray}
\left\langle s_x \right\rangle & \approx & -\int \frac{d\epsilon }{2\pi } \left( n_F -n_N \right) \sum _l \frac{2\delta _l \sin \theta }{\Delta _{\rm ex}\tau _l^N}\left\{ \frac{1}{\tau _l^2} -\frac{\cos ^2 \theta }{\left( \tau _l^a \right) ^2 }\right\} ^{-1} \nonumber \\
&& \times \left[ \frac{A_{{\rm t}l} -A_{{\rm b}l}}{\tau _l} - \left( A_{{\rm t}l}+A_{{\rm b}l}\right) \frac{\cos \theta }{\tau _l^a}\right] , \\
\left\langle s_y \right\rangle & \approx & \int \frac{d\epsilon }{2\pi } \left( n_F -n_N \right) \sum _l \frac{\sin \theta }{\Delta _{\rm ex} \tau _l^N \tau _l^a} \left\{ \frac{1}{\tau _l^2} -\frac{\cos ^2 \theta }{\left( \tau _l^a \right) ^2 } \right\} ^{-1}  \nonumber \\
&& \times \left[ \frac{A_{{\rm t}l} +A_{{\rm b}l}}{\tau _l} - \left( A_{{\rm t}l}-A_{{\rm b}l}\right) \frac{\cos \theta }{\tau _l^a}\right] .
\end{eqnarray}
We have introduced the real $\delta _l$ and imaginary $-1/2\tau _l^a $ parts of $\sigma _l^R $ and the spectral functions $A_{{\rm t,b}l}$ as before. If $\tau _l /\tau _l^a $ is ignored compared to unity, the above expressions become equivalent to the results in Ref.~\onlinecite{Chudnovskiy2008}. The modifications are thus related to the higher-order tunneling processes between two Fs. Treating those higher order processes requires inclusion of an additional source of relaxation: Otherwise, the repeated tunneling would lead to an equilibrium between Fs and a trivial result would follow. As explained in Sec. \ref{sec:torque}, these higher-order terms are also crucial for deriving $\Gamma ^n_{\rm dl} $ and $\Gamma ^m_{\rm fl}$.

The AF/AF/N case can be handled similarly. We keep $H_0$ as in (\ref{eq:AFM_Hamiltonian}) and modify the $c$ lead and the associated tunneling as
\begin{eqnarray}
\epsilon _{k\sigma }c^{\dagger }_{k\sigma }c_{k\sigma } &\rightarrow & \begin{pmatrix}
	c_{Ak\sigma }^{\dagger } & c_{Bk\sigma }^{\dagger } \\
	\end{pmatrix} \bigg[ \begin{pmatrix}
	\epsilon _k & \overline{t_k} \\
	t_k & \epsilon _k  \\
	\end{pmatrix} \nonumber \\
&& -  \sigma \tilde{\Delta }_{\rm ex} \tau _3 \bigg] \begin{pmatrix}
	c_{Ak\sigma } \\
	c_{Bk\sigma } \\
	\end{pmatrix} , \\
H_F &=& \sum _{kl,\sigma \sigma ^{\prime }} \bigg[ R^{\sigma \sigma ^{\prime }}\begin{pmatrix}
	c_{Ak\sigma }^{\dagger } & c_{Bk\sigma }^{\dagger } \\ 
	\end{pmatrix} \nonumber \\
&& \times \begin{pmatrix}
	W^{Aa}_F & W^{Ab}_F \\
	W^{Ba}_F & W^{Bb}_F \\
	\end{pmatrix}_{kl} \begin{pmatrix}
	a_{l\sigma ^{\prime }} \\
	b_{l\sigma ^{\prime }} \\
	\end{pmatrix} + {\rm h.c.} \bigg] .
\end{eqnarray}
We keep the subscript $F$ for the lead despite it being antiferromagnetic. The diagonal basis for the antiferromagnetic lead is given by
\begin{eqnarray}
\begin{pmatrix}
	\alpha _{k\sigma } \\
	\beta _{k\sigma } \\
	\end{pmatrix} &=& U_{k\sigma} \begin{pmatrix}
	c_{Ak\sigma } \\
	c_{Bk\sigma } \\
	\end{pmatrix} , 
\end{eqnarray}
where
\begin{align}
U_{k\uparrow } &= \begin{pmatrix}
	e^{-i\phi _k /2}\cos \left( \theta _k /2\right) & -e^{-i\phi _k /2}\sin \left( \theta _k /2\right) \\
	e^{i\phi _k /2}\sin \left( \theta _k /2\right) & e^{i\phi _k /2}\cos \left( \theta _k /2\right) \\
	\end{pmatrix} , \\
U_{k\downarrow } &=  \begin{pmatrix}
	e^{i\phi _k /2}\cos \left( \theta _k /2\right) & e^{i\phi _k /2}\sin \left( \theta _k /2\right) \\
	e^{-i\phi _k /2}\sin \left( \theta _k /2\right) & -e^{-i\phi _k /2}\cos \left( \theta _k /2\right) \\
	\end{pmatrix} .
\end{align}
We have defined
\begin{eqnarray}
\gamma _k &=& \sqrt{\tilde{\Delta }_{\rm ex}^2 + \left| t_k \right| ^2 } ,  \\
\cos \theta _k &=& \frac{\tilde{\Delta }_{\rm ex}}{\gamma _k } , \quad e^{i\phi _k} \sin \theta _k = - \frac{t_k}{\gamma _k} .
\end{eqnarray}
Any modification to the lead appears through the associated self-energy $\Sigma ^R_F$. It is formally written as
\begin{eqnarray}
\Sigma ^R_F &=& R^{\dagger }_{SP} \sum _k \begin{pmatrix}
	W^{Aa}_F & W^{Ab}_F \\
	W^{Bb}_F & W^{Bb}_F \\
	\end{pmatrix} _{SL}^{\dagger } \begin{pmatrix}
	U_{k\uparrow }& 0\\
	0 &U_{k\downarrow } \\
	\end{pmatrix} ^{\dagger } G^R_F \nonumber \\
&& \times \begin{pmatrix}
	U_{k\uparrow } & 0 \\
	 0 & U_{k\downarrow } \\
	 \end{pmatrix} \begin{pmatrix}
	 W^{Aa}_F & W^{Ab}_F \\
	W^{Bb}_F & W^{Bb}_F \\
	\end{pmatrix} _{SL} R_{SP} ,
\end{eqnarray}
where the lead Green's function $G^R_F$ for the antiferromagnetic case is given by
\begin{equation}
\left( G^R_F \right) _{kk^{\prime }} = \delta _{kk^{\prime }} \begin{pmatrix}
	G^R_{k+} & 0 & 0 & 0 \\
	0 & G^R_{k-} & 0 & 0 \\
	0 & 0 & G^R_{k-} & 0 \\
	0 & 0 & 0 & G^R_{k+} \\
	\end{pmatrix} 
\end{equation}
with $G^R_{k\pm } = \left( \epsilon -\epsilon _k \pm \gamma _k + i\delta \right) ^{-1}$. Introducing matrices in the sublattice space by
\begin{eqnarray}
\left( K_0 \right) ^{\alpha \beta }_{ll^{\prime }} &=& \sum _k \frac{G^R_{k+}+G^R_{k-}}{2}  \\
&& \times \left[ \overline{\left( W^{A\alpha }_F \right) _{kl}}\left( W^{A\beta }_F \right) _{kl^{\prime }}  + \overline{\left( W^{B\alpha }_F \right) _{kl}} \left( W^{B\beta }_F \right) _{kl^{\prime }} \right] \nonumber , \\
\left( K_1 \right) ^{\alpha \beta }_{ll^{\prime }} &=& \sum _k \frac{G^R_{k+}-G^R_{k-}}{2} \sin \theta _k  \\
&& \times \left[ \overline{\left( W^{A\alpha }_F \right) _{kl}}\left( W^{B\beta }_F \right) _{kl^{\prime }}  + \overline{\left( W^{B\alpha }_F \right) _{kl}} \left( W^{A\beta }_F \right) _{kl^{\prime }} \right] \nonumber ,\\
\left( K_3 \right) ^{\alpha \beta }_{ll^{\prime }} &=& \sum _k \frac{G^R_{k+}-G^R_{k-}}{2}\cos \theta _k  \\
&& \times \left[ \overline{\left( W^{A\alpha }_F \right) _{kl}}\left( W^{A\beta }_F \right) _{kl^{\prime }}  - \overline{\left( W^{B\alpha }_F \right) _{kl}} \left( W^{B\beta }_F \right) _{kl^{\prime }} \right] \nonumber ,
\end{eqnarray}
where $\alpha ,\beta =a,b$ are the sublattice indices, the self-energy can be written as
\begin{equation}
\Sigma _F^R = 1_{SP} \otimes \left( K_0 - K_1 \right) + S_{SP}\left( \theta \right) \otimes K_3 . \label{eq:self_AFM-AFM}
\end{equation}
As before, we invoke the randomness of the tunneling matrix elements for varying $k$ and estimate the matrices $K_{0,1,3}$. Since $K_1$ involves products of tunnelings with $A$ and $B$ channels of the antiferromagnetic lead, it is suppressed by $1/\sqrt{N_F}$ compared to $K_0$ as long as the tunnelings through the two channels are uncorrelated. Similarly, appealing to the statistical sublattice symmetry of the tunneling, one can estimate $K_3 $ to be again smaller than $K_0$ by a factor of $1/\sqrt{N_F}$. This implies that at the leading order in the number of lead channels $N_F$, the self-energy is independent of spin. Therefore there is neither spin torque nor spin-dependent current under these assumptions. 

This conclusion may appear unsatisfactory given that there are a number of theoretical studies reporting the existence of a spin torque in continuous textures inside antiferromagnets \cite{Yamane2016a}. Although the antiferromagnetic orders in the lead and the dynamical antiferromagnetic free-layer individually do not vary in space in our model, the difference in the orientation between the lead and dynamical antiferromagnetic moments can be considered as an extreme limit of spatial texture. Hence it is reasonable to expect a qualitative connection between the present approach and the continuum theories. There is no immediate contradiction, however, due to our assumption on the statistical distributions of the tunneling matrices. In a crystalline antiferromagnet with a spatially varying order parameter, one can expect that the electron at one sublattice site has different probabilities of moving to a next site in the same or the other sublattice. For instance, when the exchange splitting is strong, an electron at $A$ site will be more likely to move to another $A$ site than a $B$ site. Qualitatively, the corresponding situation in our model would be having different variances for $W_F^{Aa}$ and $W_F^{Ba}$. In deriving the above estimate, however, we assumed the two variances are the same. If we assume otherwise, $K_3$ will be of the same order in $N_F$ as $K_0$ so that we will see some spin-dependent effects. In the context of the tunneling junction, we believe that taking $W_F^{Aa}$ and $W_F^{Ba}$  to be equivalent is more reasonable and do not pursue this direction in the present study.

%

\end{document}